\documentclass[%
longbibliography,
reprint,
superscriptaddress,
amsmath,amssymb,
aps,
floatfix, 
]{revtex4-1}
\usepackage[english]{babel}
\usepackage{comment}
\usepackage{natbib}
\usepackage{graphicx} 
\usepackage{dcolumn} 
\usepackage{bm} 
\usepackage[dvipsnames]{xcolor}
\usepackage{amsmath}
\usepackage{listings}
\usepackage{mathtools}
\usepackage{xcolor}
\usepackage{physics}
\usepackage{lipsum}
\usepackage{cancel}
\usepackage{url}
\usepackage{esvect}
\usepackage[colorlinks=true,
	linkcolor=blue,
	urlcolor=blue,
	citecolor=blue,
	anchorcolor=blue,
	breaklinks=true
	]{hyperref}
\usepackage{breakurl} 
\usepackage{colortbl}
\usepackage{array}
\usepackage{microtype} 
\usepackage{ragged2e} 
\begin{document}
\preprint{APS/123-QED}
\title{Strained topological insulator spin-orbit torque random access memory (STI-SOTRAM) bit cell for energy-efficient Processing in Memory}

\author{Md Golam Morshed}
\email{mm8by@virginia.edu}
\affiliation{Department of Electrical and Computer Engineering, University of Virginia, Charlottesville, VA 22904, USA\looseness=-1}
\author{Hamed Vakili}
\affiliation{Department of Physics and Astronomy and Nebraska Center for Materials and Nanoscience, University of Nebraska, Lincoln, NE 68588, USA\looseness=-1}
\author{Mohammad Nazmus Sakib}
\affiliation{Department of Electrical and Computer Engineering, University of Virginia, Charlottesville, VA 22904, USA\looseness=-1}
\author{Samiran Ganguly}
\affiliation{Department of Electrical and Computer Engineering, Virginia Commonwealth University, Richmond, VA 23284, USA\looseness=-1}
\author{Mircea R. Stan}
\affiliation{Department of Electrical and Computer Engineering, University of Virginia, Charlottesville, VA 22904, USA\looseness=-1}
\author{Avik W. Ghosh}
\affiliation{Department of Electrical and Computer Engineering, University of Virginia, Charlottesville, VA 22904, USA\looseness=-1}
\affiliation{Department of Physics, University of Virginia, Charlottesville, VA, 22904, USA\looseness=-1}

\date{\today}

             
\begin{abstract}
We present a novel design of a strained topological insulator spin-orbit torque random access memory (STI-SOTRAM) bit cell comprising a piezoelectric/magnet (gating)/topological insulator (TI)/magnet (storage) heterostructure that leverages the TI's high charge-to-spin conversion efficiency coupled with the piezo-induced strain-based gating mechanism for low-power in-memory computing. The piezo-induced strain effectively modulates the conductivity of the topological surface state (TSS) by altering the gating magnet's magnetization from out-to-in-plane, facilitating the storage magnet's spin-orbit torque (SOT) switching. Through comprehensive coupled stochastic Landau-Lifshitz-Gilbert (LLG) simulations, we explore the device dynamics, anisotropy-stress phase space for switching, and write conditions and demonstrate a significant reduction in energy dissipation compared to conventional heavy metal (HM)-based SOT switching. Additionally, we project the energy consumption for in-memory Boolean operations (AND and OR). Our findings suggest the promise of the STI-SOTRAM for low-power, high-performance edge computing.
\end{abstract}

\pacs{Valid PACS appear here}

\maketitle 


\section{Introduction}\label{sec:intro}
Electronics in the edge computing era poses unique challenges that may need heterogeneous combinations of materials with diverse attributes, and devices and circuits that capitalize on those attributes \cite{edge_computing,news-article}. Along this path, the intersection of magnetism, topology, and strain offers a highly complementary suite of symmetries and tunable properties that could play off of each other strategically. For instance, strain can alter the energy landscape~\cite{Roy2011Aug} of a magnet and rotate it by $90^{\circ}$ ~\cite{Biswas2017Jun}. Rotating a magnet out of plane can modulate spin conductivity of a 3D topological insulator (TI) by breaking time-reversal symmetry and gapping its topological surface states (TSS)  ~\cite{fm_ti_ncstate,Rojas-Sanchez2016Mar,Nikolic}. Conversely, a TI can act on the magnet when its spin-momentum locked TSS write spin information onto the free layer of a magnetic tunnel junction (MTJ) with a current driven spin-orbit torque (SOT), its spin Hall angle (SHA) exceeding unity~\cite{ti_ralph,mag_doped_ti_kang_wang,ti_TbCo_Liu,Pham1}. Integrating these properties into a single stack may allow us to naturally integrate multiple functionalities into a compact, vertical monolithic structure. 

\begin{figure*}[!ht]
    \centering
    \includegraphics[width=0.98\textwidth]{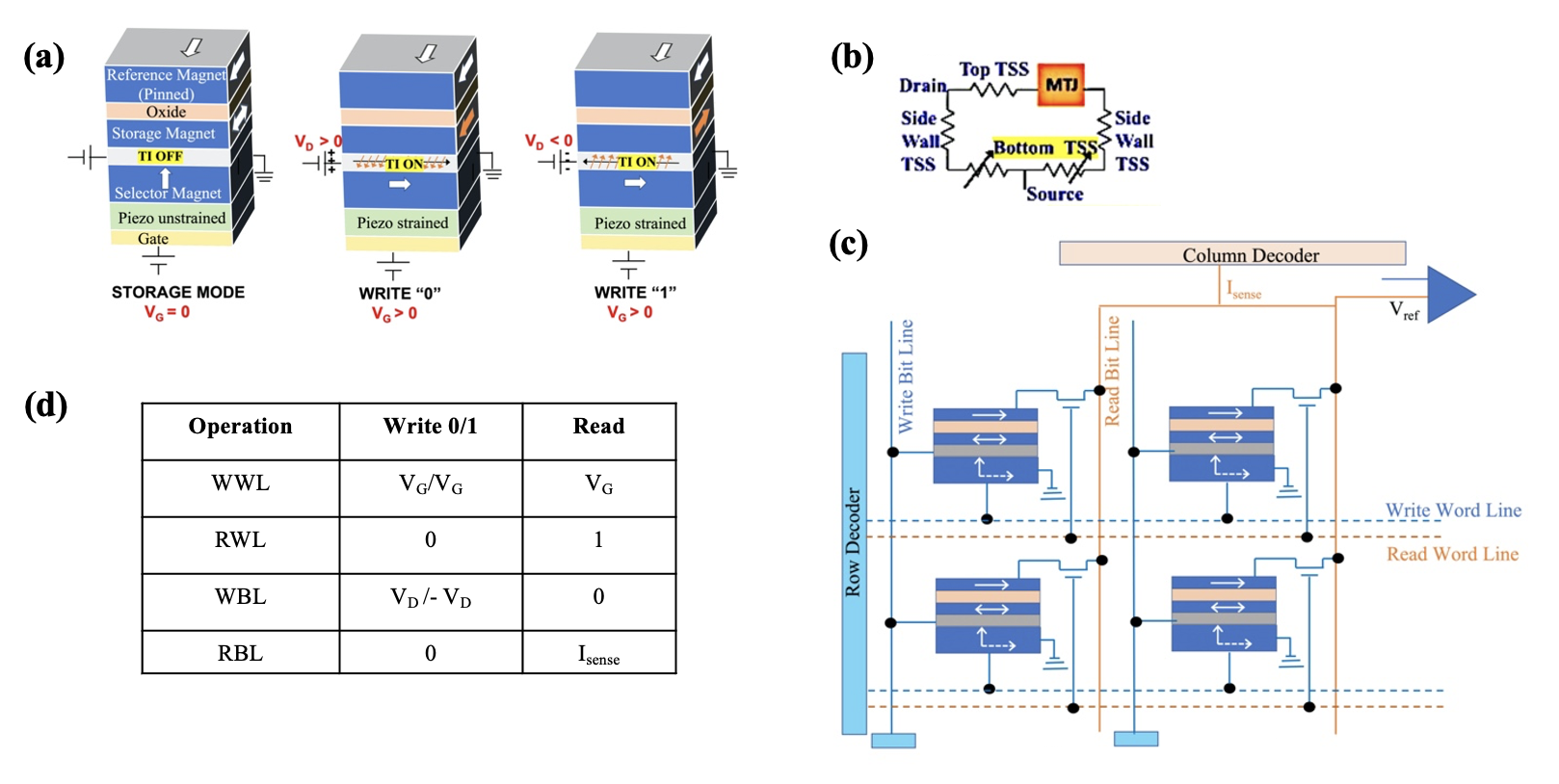}
    \caption{(a) Schematic of a compact, low-power STI-SOTRAM. An out-of-plane gating (selector) magnet gaps the bottom surface states of the TI and places the upper magnet in storage mode (left fig.). The strain generated by the piezoelectric rotates the selector magnet from out-of-plane to in-plane, restoring the bottom layer surface states and activating the TI. The drain polarity $V_D$ sets the storage bit with $m_{2y}$ (bit `$0$') or $-m_{2y}$ (bit `$1$') using high SHA SOT (middle and right figs, respectively). The fixed layer of the MTJ reads the state as `$0$' (parallel) or `$1$' (antiparallel). (b) Equivalent resistance model of the TI. An out-of-plane gating magnet breaks the current path in the bottom TSS. (c) Schematic of a PiM crossbar architecture where each activated stack (bit cell) can be selected by activating the row column. Two selected bit cells can feed to a sense amplifier that processes the local inputs by comparing with a reference voltage/current and performing Boolean logic operations (e.g., AND, OR, etc.), thereby processing local data from the magnetic memory. (d) Baising condition for read-write operations. WWL: Write Word Line, RWL: Read Word Line, WBL: Write Bit Line, and RBL: Read Bit Line.}
    \label{fig:ti_fig1}
\end{figure*}

In this paper, we will discuss an example stack, a multifunctional device composed out of a four-layer piezo/magnet/TI/magnet structure (Fig.~\ref{fig:ti_fig1}), which exploits the above attributes to achieve both logic and memory functions in a single bit cell. Here strain actuates a bottom magnet in and out of the TI plane with a voltage-gated piezo, which modulates the TSS and drives a top storage magnet into a write mode, a read mode, or a logic mode using a sense amplifier. 
Since selector and storage magnets are co-located in a vertical geometry, the structure is scalable and naturally suited for a Processing-in-Memory (PiM) architecture.
In contrast to traditional Von Neumann architecture (separate memory and logic cores with data transfer latency)~\cite{memory_wall,zou2021breaking}, key processing tasks are performed locally within a PiM ~\cite{early_pim,pim_mutlu,pim_princeton,pim_survey}, allowing the pre-processed data to be transferred to the primary processing unit with reduced latency~\cite{Mutlu2019Jun}. The individual memory cells in a PiM architecture (bit cells) are often organized in a crossbar layout. Selectors control each row and column of the crossbar grid, enabling the bit cells for read or write operations. Utilizing sense amplifiers, the entire row of the crossbar can be read by comparing the state of the bit cells with a known reference voltage or current to perform logic operations within the memory~\cite{Parveen,Boolean_1,sparse_matrix}. 

The specific design of the stack depends on the convenience of fabrication (e.g., here we assume the thick piezo layer sits at the bottom, the bottom magnet on it to share the strain, then the TI to feel its proximal magnetization, and finally, the top magnet to get a write operation on it). Other lateral or flipped structures may well serve as better designs, if only for ease of fabrication. The core functionality, however, is employing SOT from a single channel material, in this case, the TI, that delivers both high SHA and simultaneously gate tunability. Let us now discuss these two features separately. 

 Magnetic random access memory (MRAM) is a leading contender for a PiM bit cell due to its non-volatility, high speed, low power consumption, high endurance, scalability, and excellent compatibility with complementary metal-oxide-semiconductor (CMOS) process technology~\cite{pim_spintronics,sot_pim_fan,pim_survey}. The fundamental building block for MRAM devices is an MTJ, which consists of a thin insulator sandwiched between two magnetic layers—a “pinned layer” whose magnetization is fixed and a “free layer” whose magnetization can be reoriented by a spin current. Conventional MRAM uses current in the overhead to apply a magnetic field to write data on the MTJ free layer. Recently, current-induced SOT mechanism, resulting from either the spin Hall effect~\cite{Liu_sot_1,Liu_sot_2} or the Rashba effect~\cite{Miron2011Aug}, has emerged as a promising approach for energy-efficient switching of the MTJ's free layer. SOT-based switching addresses several limitations of its already commercialized counterpart, spin-transfer torque (STT)-based switching, such as the need for high write voltages, shared read-write paths creating read-disturbs, and degradation of the insulating layer~\cite{Prenat2015,Ahmed2017Oct,Shao2021May}. SOT-based MRAM (SOTRAM) is a three-terminal device featuring a decoupled read-write path and in-plane charge current flow in the non-magnetic layer underneath the MTJ free layer with a high charge-to-spin conversion efficiency, yielding a low write current and infinite endurance~\cite{Liu_sot_1,Liu_sot_2}. SOTRAM is attractive as a high density last level-embedded cache~\cite{SOT_cache,Oboril2015Jan}, with distinct advantages for PiM-like edge-based motion detection and cross-modality object recognition~\cite{sot_pim_fan,sot_matrix}. They do suffer from the higher footprint associated with the third terminal, a higher charge current needed to supply the spin current orthogonal to it, and the need for a field-assist when the free layer spin stagnates onto a symmetry direction in the magnet/heavy metal (HM) interfacial plane.  
 Consequently, achieving a significantly high charge-to-spin conversion efficiency (large SHA) is essential to bypass these challenges for energy-efficient applications. 

Traditional SOT devices typically utilize HMs such as Pt~\cite{Liu_sot_2,Pt_sot}, Ta~\cite{Liu_sot_1}, and W~\cite{sot_w} as spin current injectors; however, these HMs exhibit low SHA (Pt: 0.08, Ta: 0.15, and W: 0.4~\cite{Pham2}), leading to low charge-to-spin conversion efficiency~\cite{Pham1}. Recently, TIs have garnered attention as an attractive spin source for SOT switching~\cite{ti_ralph,mag_doped_ti_kang_wang,ti_TbCo_Liu,ti_current_ratio,ti_Jian_Ping}. TIs are characterized by their spin-momentum-locked TSS, insulating bulk states, and SHA greater than unity~\cite{zahid_hasan_princeton,ti_review,ti_review_2}. The conducting TSS arise from band inversion at the surface, mediated by strong spin-orbit coupling, and are topologically protected by time-reversal symmetry~\cite{ti_tss_1,ti_tss_2,ti_tss_3}. {More encouragingly, sputtered TIs such as Bi$_2$Se$_3$ show very high ($>$ 10) SHAs, primarily because of the reduction in bulk conductivity that would otherwise shunt current away from the TSS, as well as better defect control that suppresses unwanted surface scattering.}
These attributes make TIs highly efficient spin current injectors for SOT switching, facilitating energy-efficient information writing~\cite{ti_memory}. 

Conversely, a magnet's ability to modulate the spin conductivity of the TSS of a TI, when in proximity with the TI, based on its magnetization direction, offers a pathway for an intrinsic gating mechanism to control the TI surface current~\cite{fm_ti_ncstate,hamed_fm_ti}. Switching the magnet (gating magnet hereafter) from in-plane to out-of-plane orientations, gaps in the TSS can be opened, thereby modulating the spin conductivity~\cite{Rojas-Sanchez2016Mar,Nikolic}. This $90^{\circ}$ switching of the gating magnet can be achieved through various mechanisms, such as strain~\cite{strain_2010,strain_2011,strain_2016,sb_review}, voltage control magnetic anisotropy~\cite{vcma}, and changing the anisotropy by an applied voltage~\cite{voltage,fm_ti_ncstate}. Nonetheless, strain-induced $90^{\circ}$ switching of magnets is very energy-efficient~\cite{Roy2011Aug,Roy2012Jul,strain_2016,sb_review}, where a piezoelectric material induces electrical strain in response to an applied gate voltage to facilitate switching from in-plane to out-of-plane and vice versa. 


In this paper, we model the aforementioned potentially compact and energy-efficient four-layer piezoelectric/magnet/TI/magnet (MTJ) stack, which we refer to as a strained topological insulator spin-orbit torque random access memory (STI-SOTRAM). We argue that this structure is naturally suited as a compact bit cell in an in-memory computing architecture. We employ stochastic Landau-Lifshitz-Gilbert (LLG) simulation to analyze the device dynamics. We present the device's functionality, the required phase space for stress generated by the piezo-induced strain and the gating magnet's anisotropy, and the writing condition's phase space (writing voltage vs. time/delay) of the MTJ. We estimate the energy cost for gating and writing mechanisms and find that energy dissipation is significantly reduced compared to HM-based SOT switching, which indicates the benefits of utilizing TI as a spin source and an intrinsic gating mechanism as opposed to an access transistor traditionally used. Furthermore, we project the energy cost for 2-bit AND and OR operations, which shows lower energy costs than traditional SOT switching. Finally, we show the impact of various material parameters on the device metric.

\section{STI-SOTRAM Bit Cell Design for PiM}\label{sec:structure}

Figure~\ref{fig:ti_fig1}(a) depicts the schematic of the compact four-layer structure of the device comprising a piezoelectric/gating magnet/TI/MTJ stack in the vertical direction. The device contains a gating (selector) magnet at the bottom whose magnetization can be switched in and out of the plane of the TI in nanoseconds with a voltage-gated piezoelectric at a low energy cost ($\sim 10~\mathrm{aJ}$)~\cite{Fashami2011Mar,Roy2012Jul,Biswas2017Jun}. An out-of-plane selector magnet gaps the bottom surface states of the TI, leading to breaking the current path (Fig.~\ref{fig:ti_fig1}(b)) and places the upper MTJ free layer in storage mode (Fig.~\ref{fig:ti_fig1}(a), left). Activating the TI bottom surface with applied voltage restores the current path and drives the top MTJ free layer into one of three modes – (\textit{i}) altering its magnetization with drain bias for data writing (Fig.~\ref{fig:ti_fig1}(a), middle and right), (\textit{ii}) discharging its stored magnetization state for data reading, the output set by its low (parallel)/high (antiparallel) resistance relative to the MTJ pinned layer; (\textit{iii}) or execute a logic operation (e.g., bitwise AND, OR, etc.) using a sense amplifier (Fig.~\ref{fig:ti_fig1}(c)). Two selected bit cells can feed their voltage/current to a sense amplifier that processes the local inputs by comparing them with a reference voltage/current and performing Boolean logic operations. For example, for current-based sensing, if the current of the bit cells is $I_c$, AND and OR operations can be performed by selecting a reference current of $1.5I_c$ and $0.5I_c$, respectively. Similarly, we can choose different reference voltages for a voltage-based sensing scheme (discussed later). Since selector and storage magnets are co-located in a vertical geometry, the structure is scalable and naturally suited for a PiM architecture that pre-processes stored data locally, all with the same vertically integrated, compact bit cell as shown in Fig.~\ref{fig:ti_fig1}(c). Figure~\ref{fig:ti_fig1}(d) shows the biasing condition for various operations.

From Mars rovers to mine robots, edge computing avoids cloud access that is either unavailable or unreliable. Energy and latency are its main concerns. Our device gating energy is low because the strain is remarkably energy-efficient, while our write energy is low because TIs offer a very high SHA. Our results (Fig.~\ref{fig:ti_fig5}) show that at the individual bit cell level, our device requires much lower energy ($\sim2–6\times$) than 45-nm Dynamic and Static Random Access Memory (DRAM, SRAM) and HM-based SOTRAM with a lower/comparable speed ($2~\mathrm{ns}$), further improvable with larger bandgap TIs. Moreover, at a PiM architecture level, we significantly outperform DRAM on both non-volatility and energy cost by rolling memory and logic into a single unit, reducing data transit latency from $\sim$mm-cm data bus lengths between separate memory-logic cores to just $\sim~\mu$m-mm array dimensions, as the sense amplifier processes suitably segmented data from locally accessed storage magnets. At data transfer rates of $\sim$Mbps-Gbps, this enables a $2–3\times$ speed-up over traditional Von Neumann architecture, in addition to the energy gain. Our device exploits the TI as a unique semiconducting channel that delivers high SOT without field assist for low energy write and voltage gating with strain for low power row-column select in PiM architecture.

\section{Methods}\label{sec:method}
To characterize the magnetization dynamics in our device, we solve a coupled stochastic LLG equation in the macrospin limit using the fourth order-Runge Kutta Method. As explained in detail in section~\ref{sec:structure}, the device consists of a piezoelectric/gating magnet/TI/MTJ heterostructure. We simultaneously solve the magnetization dynamics of the gating magnet and the MTJ free layer. In the case of the gating magnet, the LLG equation is described as:

\begin{equation}
\begin{split}
        \frac{1+\alpha_1^2}{\gamma}\cdot\pdv{\vb{m_1}}{t} &= - \mu_0\cdot(\vb{m_1} \cp \vb{H}_\mathrm{eff1}) \\
        & - \alpha_1 \mu_0\cdot\vb{m_1} \cp (\vb{m_1} \cp \vb{H}_\mathrm{eff1}),
\end{split}
\label{eq:llg1}
\end{equation}
where $\vb{m_1} = \vb{M_1}/M_{s1}$ is the normalized magnetization and $M_{s1}$ is the saturation magnetization of the gating magnet. $\alpha_1$, $\mu_0$, and $\gamma$ are magnetic damping coefficient, permeability of free space, and gyromagnetic ratio, respectively. We consider $\vb{H}_\mathrm{eff1} = \vb{H}_{k1} + \vb{H}_\mathrm{stress} + \vb{H}_\mathrm{th}$, where $\vb{H}_{k1}$ and $\vb{H}_{stress}$ are the effective anisotropy field and stress field of the gating magnet, respectively~\cite{Winters2019Sep,Mishra2023Sep}. $H_{k1} = \frac{2K_{u1}}{\mu_0 M_{s1}} - M_{s1}$ ($K_{u1}$ is the uniaxial anisotropy of the gating magnet). $H_{stress} = \frac{3 \lambda_s \sigma_s}{\mu_0M_{s1}}$ ($\lambda_s$ is the magnetostriction coefficient of the gating magnet, and $\sigma_s$ is the stress generated by the electrical strain induced by the piezoelectric). 

For the MTJ free layer, the switching is facilitated by the SOT arising from the TI, and the LLG equation takes the form: 
\begin{equation}
\begin{split}
        \frac{1+\alpha_2^2}{\gamma}\cdot\pdv{\vb{m_2}}{t} &= - \mu_0\cdot (\vb{m_2} \cp \vb{H}_\mathrm{eff2}) \\
         & - \alpha_2 \mu_0\cdot\vb{m_2} \cp (\vb{m_2} \cp \vb{H}_\mathrm{eff2})\\ 
         & - \frac{\hbar}{2e}\cdot\frac{\theta_{sh}^{eff} J}{M_{s2}t_{f2}}\cdot\vb{m_2} \cp (\vb{m_2} \cp \boldsymbol{\sigma}_{p}),
\end{split}
\label{eq:llg2}
\end{equation}
where $e$ is the elementary charge, $\hbar$ is the reduced Plank constant, $\theta_{sh}^{eff}$ is the effective SHA of the TI, $t_{f2}$ is the thickness of the MTJ free layer, $J$ is the surface current density of the TI, and $\boldsymbol{\sigma}_{p}$ is the unit vector along the spin polarization direction. The other variables and constants have the same meaning as defined in Eq.~(\ref{eq:llg1}), and a subscript of $2$ represents the parameter for the MTJ free layer. $\theta_{sh}^{eff} = \theta_{sh}(1-sech(t_{TI}/\lambda))$, where $\theta_{sh}$, $t_{TI}$, and $\lambda$ are the SHA, thickness, and spin diffusion length of the TI. In Eq.~(\ref{eq:llg2}), $\vb{H}_\mathrm{eff2} = \vb{H}_{d} + \vb{H}_\mathrm{th}$. $\vb{H_d}$ is the demagnetization field of the in-plane MTJ free layer and expressed as $H_d = - M_{s2}[N_{dx}~N_{dy}~N_{dz}]$, where $N_{dx}$, $N_{dy}$, $N_{dz}$ are the demagnetization factors of the free layer along $x$, $y$, $z$ axis, respectively.  

In both Eqs.~(\ref{eq:llg1}) and (\ref{eq:llg2}), $\vb{H}_\mathrm{th}$ is a random thermal field with zero mean ($\mu = 0$) and standard deviation
\begin{equation}
 \text{SD}  = \sqrt{\frac{2 \alpha k_BT}{\mu_0^2 \gamma M_s V \Delta t}},
\label{eq:equ3}
\end{equation}
where $\alpha$, $M_s$, and $V$ are the damping coefficient, saturation magnetization, and volume of the respective magnets. $\Delta t$ is the simulation time step. $\mu_0$, $\gamma$, $k_B$, and $T$ have their usual meanings. 

Throughout our study, we use TbCo as both the gating magnet and the MTJ free layer magnet, Bi$_2$Se$_3$ as the TI, and Pb(Zr,Ti)O$_3$ (PZT) as the piezoelectric materials unless otherwise specified. The used parameters are listed in Table~\ref{tab:parameters}.

The modulation of the surface current density J of the TI by the gating magnet is modeled separately with an empirical function, described next. 


\section{Results and Discussion}\label{sec:results}
The first functional block of the entire stack is the piezoelectric/gating magnet heterostructure. 
Applying a voltage to the piezoelectric (PZT), electrical strain is generated and transferred to the adjacent gating magnet (TbCo) via the magnetostriction effect. This strain is then converted into stress via the elastic modulus of the gating magnet, which counteracts the uniaxial anisotropy of the gating magnet. When the stress is sufficient, it can switch the gating magnet from out-of-plane to in-plane with minuscule energy consumption in the order of $\sim 10~\mathrm{aJ}$~\cite{Fashami2011Mar,Roy2012Jul,Biswas2017Jun} (energy cost estimation for our device is shown later). The strain-induced switching of the gating magnet facilitates the opening and closing of the band gap in the TSS of the TI (Bi$_2$Se$_3$), depending on the out-of-plane and in-plane magnetization orientation of the gating magnet, respectively.

\begin{figure*}[!htbp]
    \centering
    \includegraphics[width=0.95\textwidth]{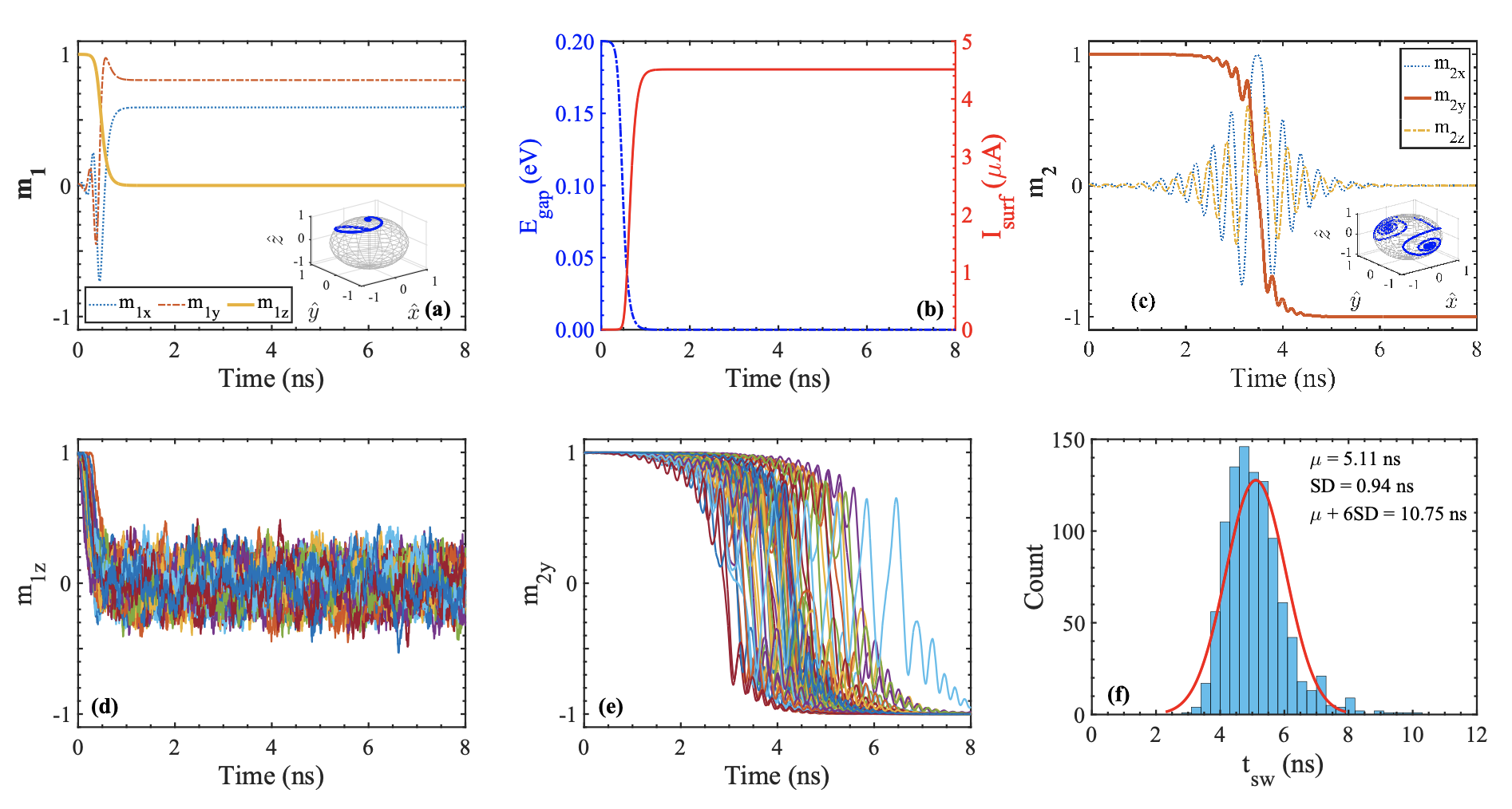}
    \caption{Coupled LLG simulations showing gating magnet switching with strain (a), turning ON TSS (closing band gap) and hence delivering surface current (b), and switching MTJ free layer with SOT (c) without the thermal field. (d)-(e) Magnetization dynamics for the gating magnet and the MTJ free layer, respectively, influenced by the stochastic thermal field ($50$ simulations). (f) Histogram of switching time from $1000$ stochastic simulations to estimate the switching time corresponding to $\text{WER}=10^{-9}$. The switching time is calculated as $t_{sw} = \mu + 6\text{SD}$ from the histogram. The red curve is the Gaussian fit. The results are generated using $I_{0,surf}=6I_{c,surf}$.}  
    \label{fig:ti_fig2}
\end{figure*}

Figure~\ref{fig:ti_fig2}(a) shows the magnetization dynamics of the gating magnet under the influence of the stress induced by the piezoelectric strain. We started with an out-of-plane magnetization ($m_{1z} = + 1$) for the gating magnet, which corresponds to the OFF state of the device as it opens a band gap in the TSS and, hence, no current flow. 
Under the influence of stress, it takes a very short time, $\sim 1~\mathrm{ns}$ for the $90^{\circ}$ switching ($m_{1z} = 0$) of the gating magnet for a stress $\sigma_s = 100~\mathrm{MPa}$. The switching delay depends on the strength of $\sigma_s$ and the material parameters of the gating magnet. The out-of-plane magnetization component ($m_{1z}$) is fed to the next block (TI) dynamically, which modulates the band gap in the TSS and, hence, the surface current. We consider a Dirac Hamiltonian form to model the surface state of the TI. In the presence of a magnet in proximity to the TI, the Hamiltonian can be expressed as $H = \hbar v_F(\boldsymbol{\sigma} \cross \vb{k}) \cdot \hat{z} + M_{0} \vb{M_1} \cdot \boldsymbol{\sigma}$~\cite{sci_dirac,fm_ti_ncstate}, where $v_F$ is the Fermi velocity, $\vb{k}$ is the wavevector,  $\boldsymbol{\sigma}$ is the Pauli spin matrices, $M_{0}$ is the exchange strength between the gating magnet and TI, and $\vb{M_1}$ is magnetization of the gating magnet. From simple algebra, for the case of an out-of-plane magnetization orientation of the gating magnet, we can show the energy dispersion takes the form $E = \pm\sqrt{\hbar^2v_F^2|k|^2 + (M_0m_{1z})^2}$, where $|k|=\sqrt{k_x^2 + k_y^2}$. This gives rise to a band gap $E_{gap} = \sqrt{(2M_0m_{1z})^2}$ at $k=0$. The band gap will modulate the surface current of the TI as $I_{surf} = I_{0,surf}e^{-E_{gap}/k_BT}$, where $I_{0,surf}$ represents the surface current needed to switch the MTJ free layer for a specific switching delay. Figure~\ref{fig:ti_fig2}(b) shows the evolution of the $E_{gap}$ and the $I_{surf}$ with time in response to the magnetization dynamics of the gating magnet (Fig.~\ref{fig:ti_fig2}(a)). Initially, when $m_{1z} = +1$, a band gap of $2M_0$ opened, and we get vanishing surface current. As the gating magnet switches ($m_{1z} = 0$) by the stress, the conductivity of the TSS is restored since the band gap is closed. We use $M_0 = 0.1~\mathrm{eV}$, a typical value for Bi$_2$Se$_3$~\cite{fm_ti_ncstate,hamed_fm_ti}. Note that we cannot open an infinite band gap as it is limited by the bulk band gap of the TI ($0.3~\mathrm{eV}$ for Bi$_2$Se$_3$~\cite{Cho2011May,yunkun_ti}). We use $I_{0,surf} = 6I_{c,surf}$ while generating Fig.~\ref{fig:ti_fig2}(b), where $I_{c,surf}$ is the critical surface current needed to switch the MTJ free layer. 

\begin{figure*}[!htbp]
    \centering
    \includegraphics[width=0.45\textwidth]{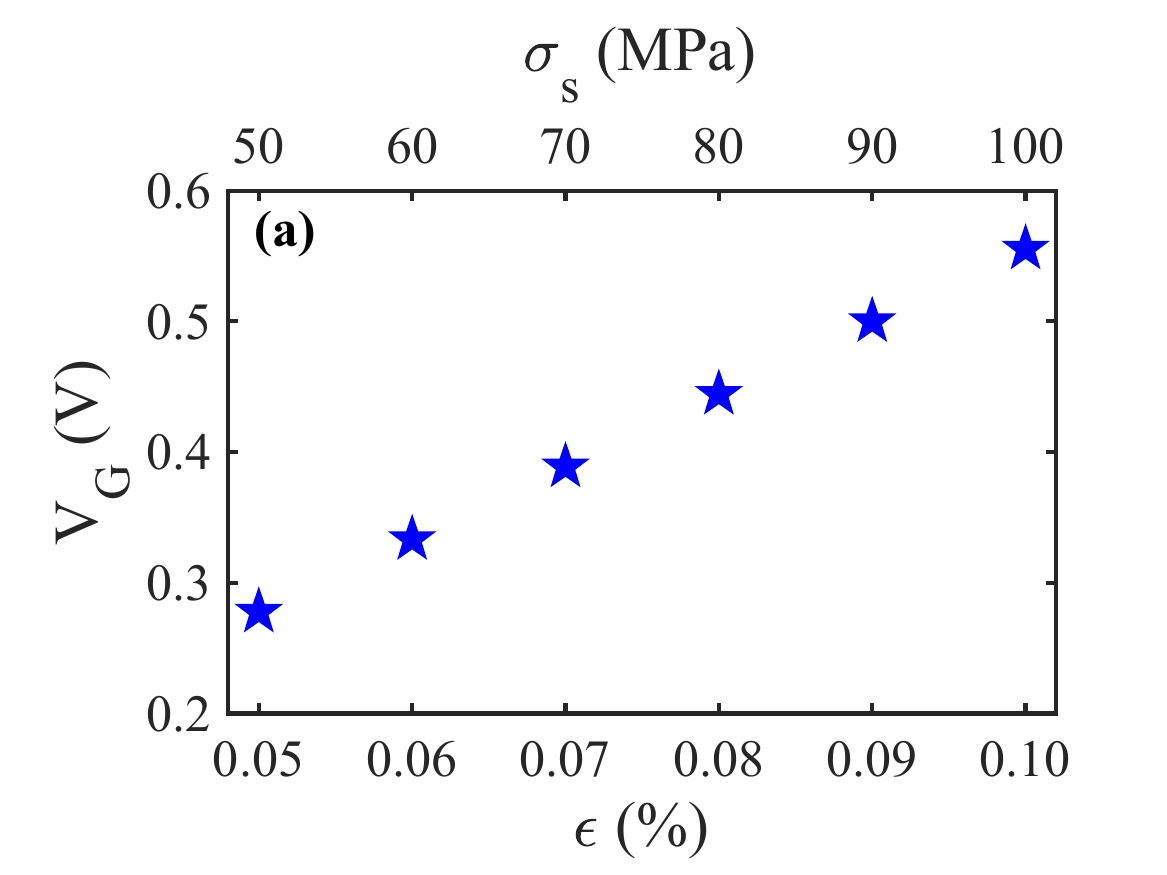}
    \includegraphics[width=0.45\textwidth]{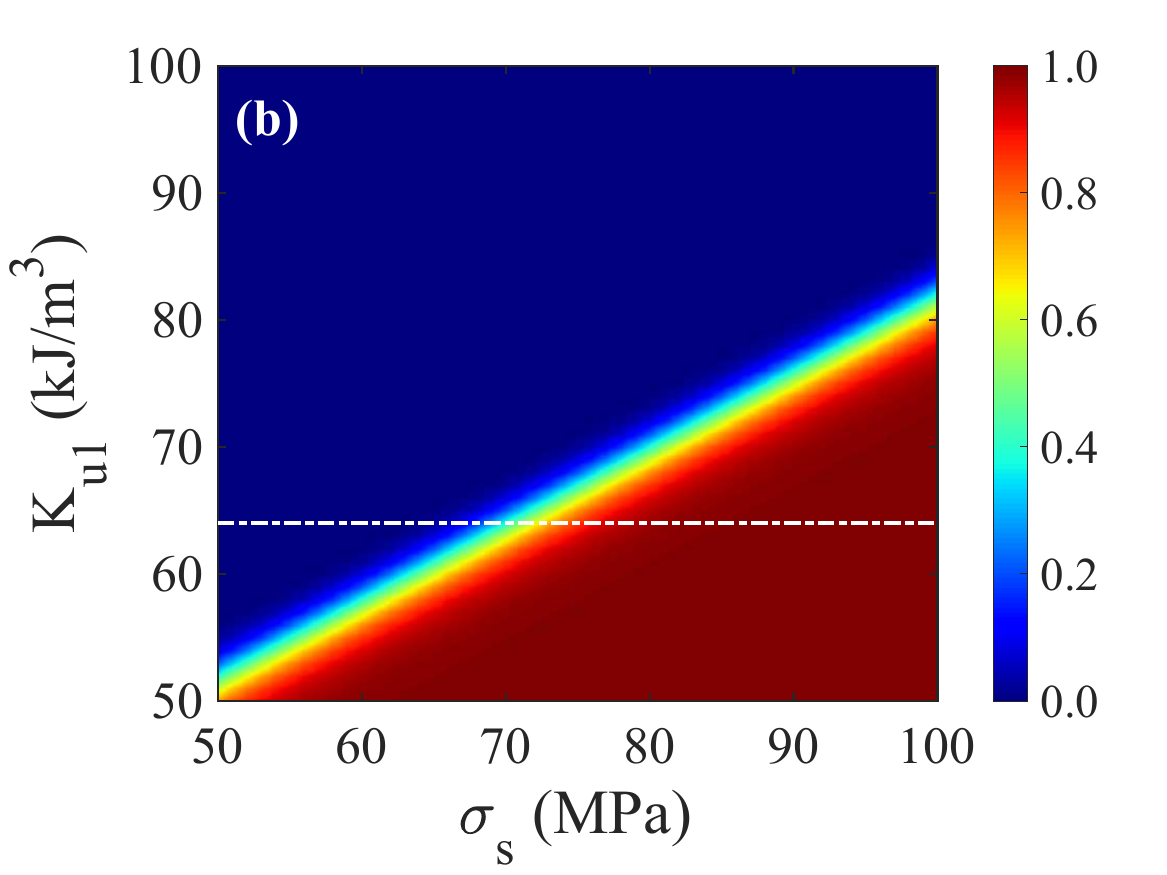}
    \caption{(a) Gate voltage requirement in the PZT to generate electrical strain (stress). The capacity to generate strain in the PZT limits the maximum achievable stress. (b) Gating magnet's uniaxial anisotropy vs. stress phase-space for the switching probability (colorplot) of the MTJ free layer. The probability of switching is calculated from $10^5$ stochastic simulations. The dash-dotted line represents the TbCo uniaxial anisotropy ($K_{u1} = 64~\mathrm{kJ/m^3}$).}
    \label{fig:ti_fig3}
\end{figure*}
\begin{figure}[!htbp]
    \centering
    \includegraphics[width=0.90\linewidth]{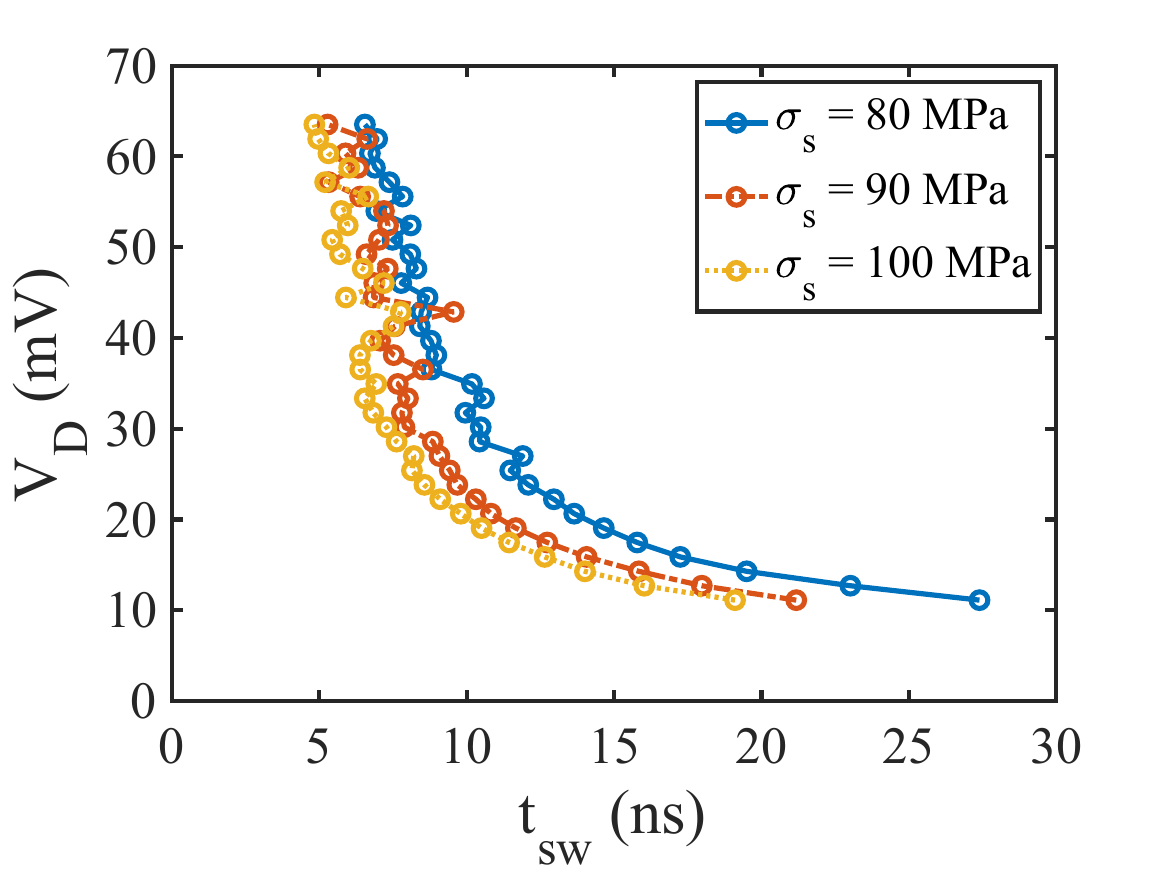}
    \caption{Supplied drain voltage to the TI for write operation. We use $I_{surf}/I_{tot} = 15\%$.
    }
    \label{fig:ti_fig4}
\end{figure}

\begin{figure*}[!htbp]
    \centering
    \includegraphics[width=0.32\textwidth]{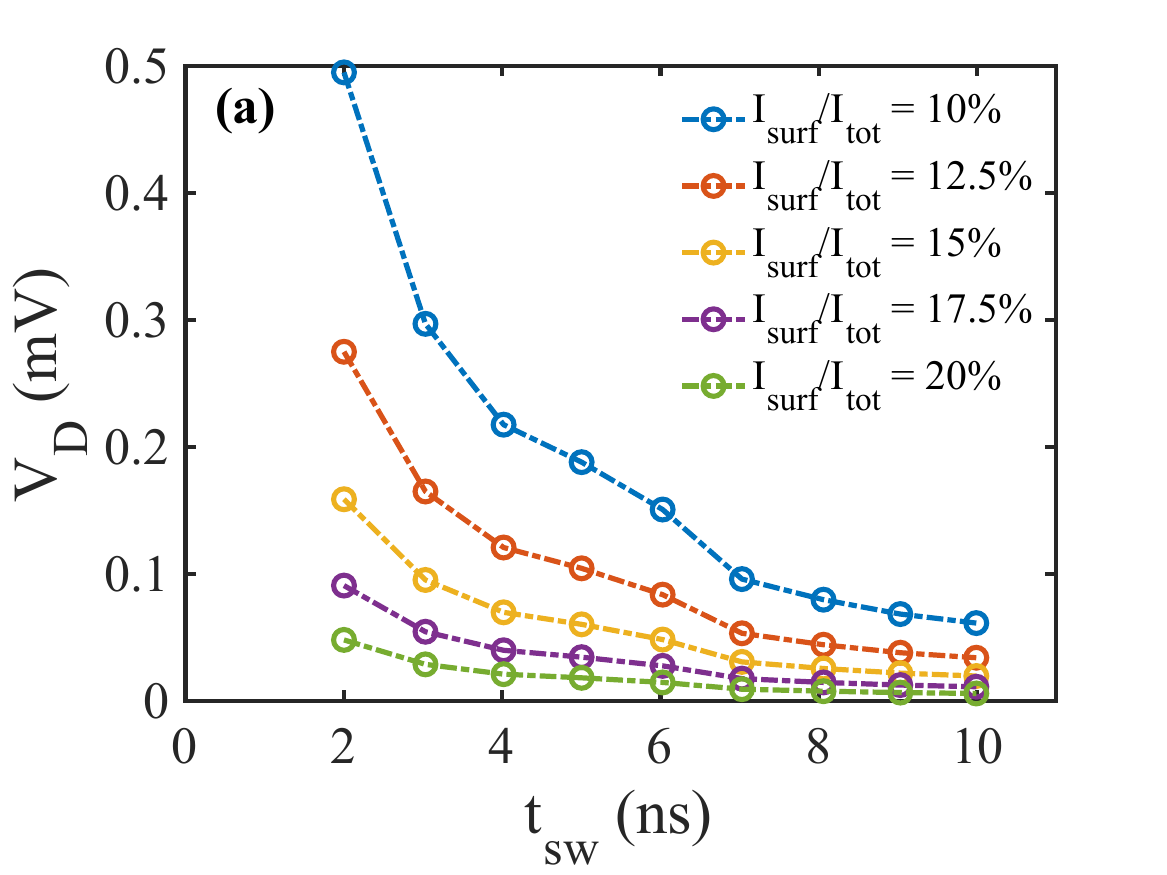}
    \includegraphics[width=0.32\textwidth]{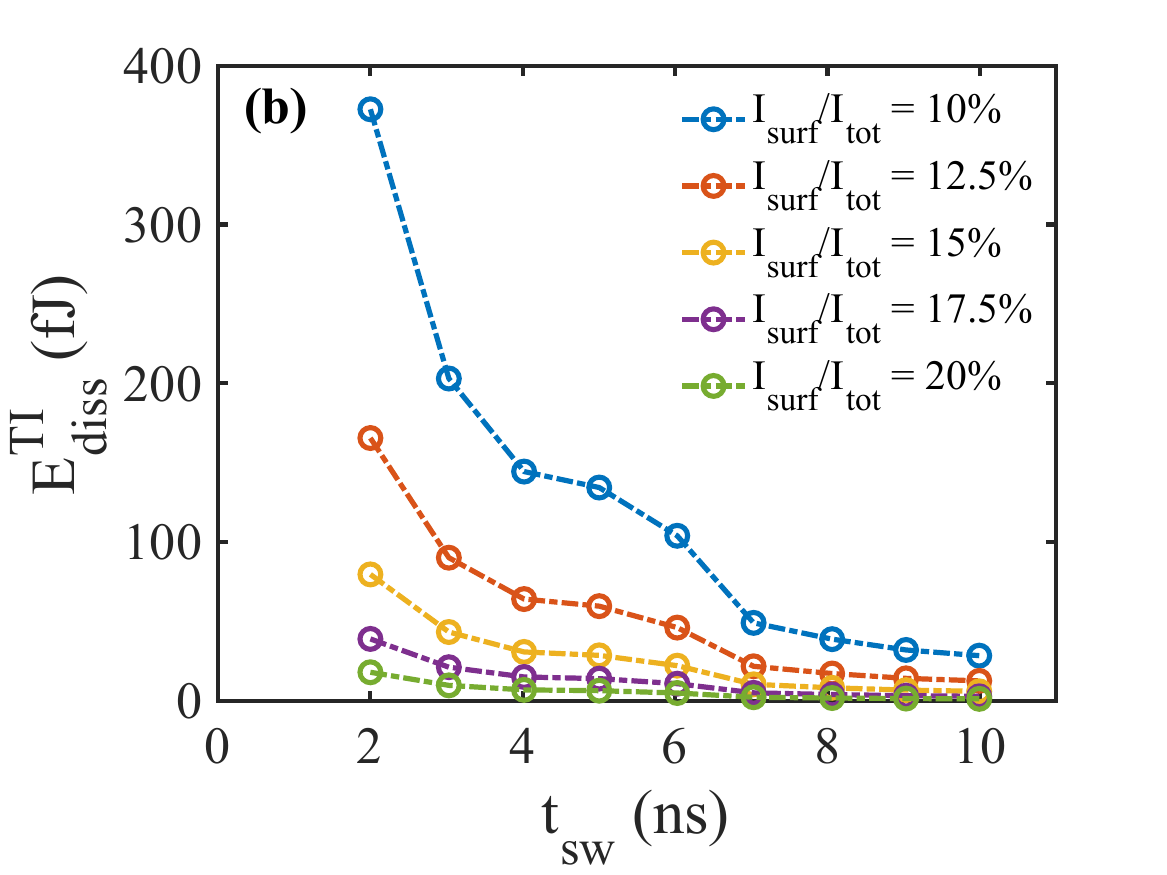}
    \includegraphics[width=0.32\textwidth]{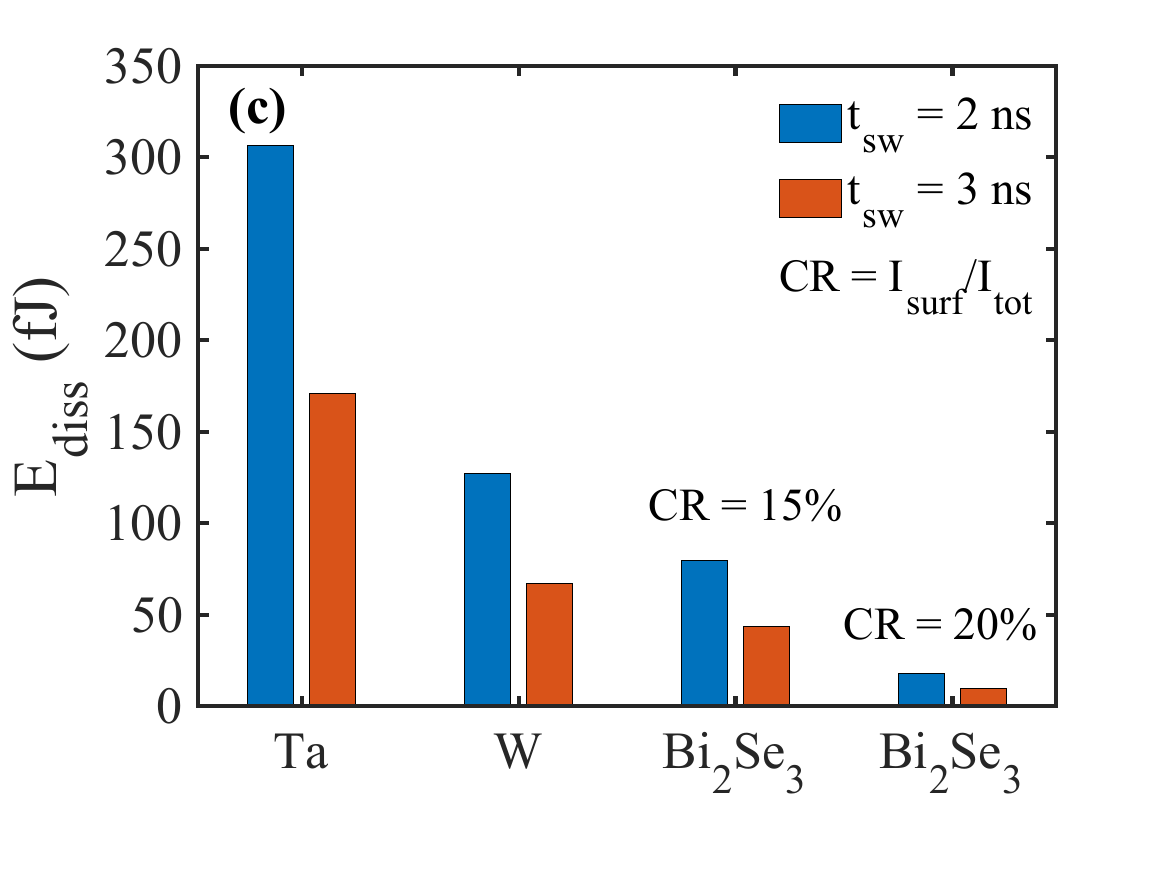}
    \caption{(a) Drain voltage requirement for high speed switching ($2-10~\mathrm{ns}$). (b) Energy dissipation in the TI during the writing process. (c) Comparison of energy dissipation between traditional HM- and TI-based SOT mechanism. TI consumes orders of magnitude less energy than the HM.}
    \label{fig:ti_fig5}
\end{figure*}

We utilize an in-plane magnet-based MTJ for our device for primarily two reasons: (\textit{i}) SOT can provide deterministic switching without an external field assist only for an in-plane magnet because of in-plane spin polarization (orthogonal to the charge current)~\cite{Liu_sot_1,ti_ralph,Fukami2016Jul}, and (\textit{ii}) we are limited by the interaction between the TI and the MTJ free layer as such an out-of-plane free layer will open gap in the TSS and hence no current conduction~\cite{fm_ti_ncstate,Rojas-Sanchez2016Mar,Nikolic}. While utilizing an in-plane magnet occupies more area than a perpendicular magnetic anisotropy (PMA) magnet, our design integrates the gating and storage magnets within the same stack. This configuration offers both area and energy efficiency advantages, along with the added benefit of field-free switching. Extensive research is underway for field-free switching of PMA magnets. However, these require a graded structure or an added magnet that provides the symmetry-breaking field and adds structural complexity~\cite{field-free1,field-free2}. We use an in-plane magnet of ``type y'' (easy axis in y-direction). The critical current density for switching $J_c$ for this type of MTJ is expressed as~\cite{Katine2000Apr,Lee2013Mar,Fukami2016Jul}:

\begin{equation}
    J_c = \frac{2e\alpha_2\mu_0M_{s2}t_{f2}}{\hbar\theta_{sh}^{eff}}\left(H_{in}+\frac{H_{out}}{2}\right)
    \label{eq:crit_J}
\end{equation}
$H_{in}$ is the in-plane shape anisotropy of the free layer, and $H_{out}$ is the out-of-plane demagnetization component. The other variables and constants have the same meaning defined in section~\ref{sec:method}. Using the parameters listed in Table~\ref{tab:parameters}, we find $J_c = 1.88 \times 10^{10}~\mathrm{A/m^2}$, which is at least one order of magnitude less than conventional HM-based SOT switching~\cite{Fukami2016Jul,Pham2} and laid the foundation of energy-efficient switching. This amount of critical current density needs to be supplied by the top surface states of TI for the switching of MTJ free layer, from which we can estimate the critical surface current $I_{c,surf} = J_cWt_{surf}=0.75~\mathrm{\mu A}$ ($W$ is the device width, $t_{surf} = 1~\mathrm{nm}$ is the thickness of the TI top surface~\cite{ti_current_ratio}). 

Figure~\ref{fig:ti_fig2}(c) shows the magnetization dynamics of the MTJ free layer in response to the SOT generated by the TI surface current. We can clearly see the magnetization switching ($m_{2y} = +1~\text{to}~m_{2y} = -1$) with a switching delay of $\sim 4~\mathrm{ns}$ for a surface current of $6I_{c,surf}$. Note that Figs.~\ref{fig:ti_fig2}(a)-(c) are generated in the absence of a thermal field. Figures~\ref{fig:ti_fig2}(d) and (e) show the magnetization dynamics of the gating magnet and the MTJ free layer, respectively, under the influence of the thermal field for $50$ simulation runs for the same current as without thermal case, and we can see the stochastic behavior of the magnetization dynamics. Figure~\ref{fig:ti_fig2}(f) shows the histogram from $1000$ simulations runs, and we find a switching time of $10.75~\mathrm{ns}$, calculated by $t_{sw} = \mu + 6\text{SD}$, corresponding to a write error rate $\text{WER} = 10^{-9}$, which is the industry standard for memory applications~\cite{Ahmed2017Oct}. In the subsequent part of the paper, we calculate $t_{sw}$ similarly unless otherwise specified.

Figure~\ref{fig:ti_fig2} demonstrates the functionality of the device. We will now delve into the details and constraints of the individual blocks. Beginning from the bottom to up, the piezoelectric-gating magnet stack enables the strain-driven out-plane to in-plane switching. As discussed earlier, the switching is governed by the competition between the uniaxial anisotropy energy and the stress energy induced by the strain. PZT typically can generate an electrical strain of $0.05\% - 0.1\%~(500 -1000~\mathrm{ppm})$ in response to an applied voltage~\cite{Roy2012Jul}, which sets the achievable stress energy limit to counteract the anisotropy energy. The relation between the applied voltage and strain is $V_G/t_{piezo} = \epsilon/d_{31}$~\cite{Roy2011Aug,Roy2012Jul}, where $V_G$, $t_{piezo}$, $\epsilon$, $d_{31}$ are the applied gate voltage, thickness of piezoelectric, strain, and piezoelectric constant, respectively. The typical $d_{31}$ constant for PZT is $1.8 \times 10^{-10}$~\cite{Roy2011Aug}. TbCo has Young's modulus of $100~\mathrm{GPa}$~\cite{Youngs_mod} and $\lambda_s=400 \times 10^{-6}$~\cite{magnetostriction}, which gives a range of stress $\sigma_s$ from $50$ to $100~\mathrm{MPa}$. Figure~\ref{fig:ti_fig3}(a) shows the gate voltage required for generating the strain (stress) for a $100~\mathrm{nm}$ thick PZT. The critical question is whether this stress is sufficient to overcome the anisotropy barrier. TbCo has a low uniaxial anisotropy $K_{u1}$ of $64~\mathrm{kJ/m^3}$~\cite{TbCo_ku_ms,ti_TbCo_Liu}. Considering the demagnetization effect, the effecting anisotropy becomes $38.9~\mathrm{kJ/m^3}$ ($K_{eff} = K_{u1} - \frac{1}{2}\mu_0M_{s1}^2$). Conversely, the stress energy $E_{stress} = \frac{3}{2}\lambda_s\sigma_s$~\cite{Winters2019Sep,Mishra2023Sep} ranges from $30~\mathrm{kJ/m^3}$ to $60~\mathrm{kJ/m^3}$ depending on the amount of strain (stress) shown in Fig.~\ref{fig:ti_fig3}(a). Figure~\ref{fig:ti_fig3}(b) presents the phase space for anisotropy and stress for the switching probability of the MTJ-free layer (colormap). The $P_{sw}$ is calculated from $10^5$ stochastic LLG runs~\cite{ankit} for $10~\mathrm{ns}$ with a surface current of $6I_{c,surf}$. The figure illustrates the reciprocal relation between the $K_{u1}$ and $\sigma_s$, and the range of $K_{u1}$ and $\sigma_s$ for a working device. This phase space is important because $K_{u1}$ can be tuned by adjusting the composition of TbCo and alloying~\cite{Zhao2015Mar,TbCo_param_Liu}, while $\sigma_s$ is adjustable through applied gate voltage.

Next, we consider the voltage requirement for the MTJ write operation (switching the MTJ free layer). For the write operation, three important quantities are the switching time, switching current/voltage, and WER. We calculate the switching voltage vs. switching time for various stress values in Fig.~\ref{fig:ti_fig4}, showing the typical reciprocal relation for spin-torque switching~\cite{Bedau2010Jan,strain_sot,golam_mbm}. The switching voltage shown in Fig.~\ref{fig:ti_fig4}(a) correspond to $\text{WER} = 10^{-9}$. As $\sigma_s$ increases, the out-to-in-plane switching time of the gating magnet is reduced, and it requires less time for the band gap to close, which is attributed to the lower overall switching time of the device. We vary the surface current from $3.5$ to $20$ times the critical surface current and let the magnetization evolve until it reaches $95\%$ of its final value ($m_{2y} = -0.95$)~\cite{Roy2016Jun}. We denote the time as the switching time for the individual switching event, and from the switching time distribution of $1000$ simulations, we calculate the $t_{sw}$ corresponding to $\text{WER} = 10^{-9}$. While calculating the total current requirement for the required surface current, we consider a ${I_{surf}}/{I_{tot}} = 15\%$ (see more discussion later). 

\begin{figure*}[!htbp]
    \centering
    \includegraphics[width=0.42\textwidth]{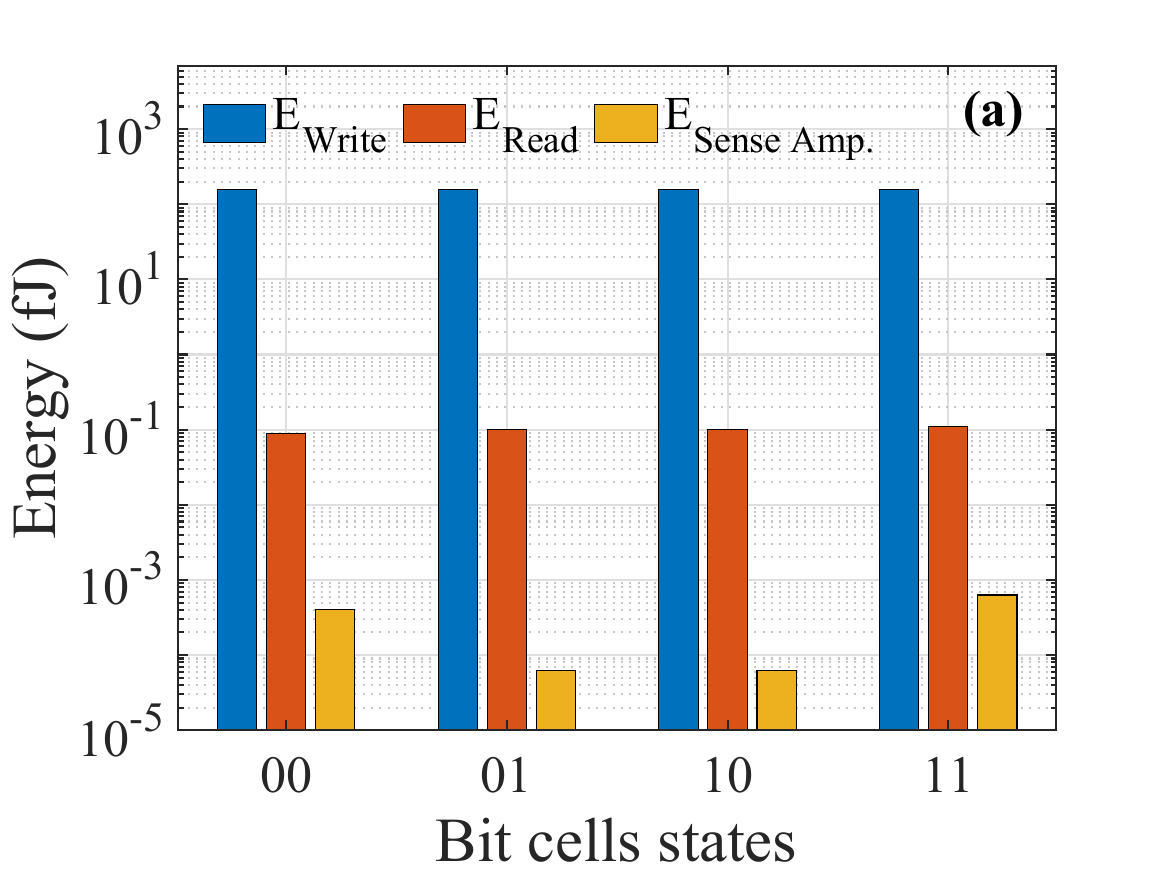}
    \includegraphics[width=0.42\textwidth]{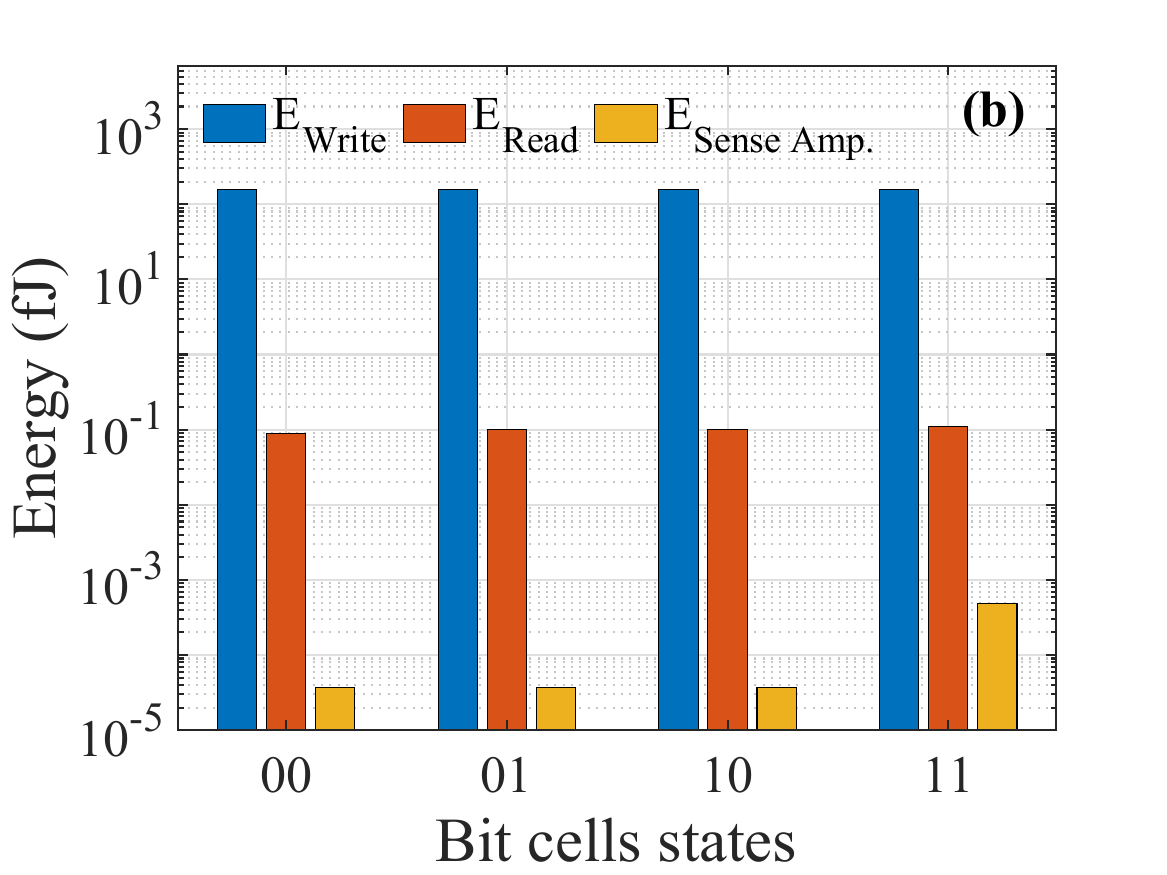}
    \caption{Energy consumption for in-memory (a) AND and (b) OR operations. $E_{write}$ and $E_{read}$ are the combined energy for both the bit cells, while $E_{Sense~Amp.}$ represents the energy associated with a single sense amplifier.}
    \label{fig:ti_fig6}
\end{figure*}

\begin{figure*}[!htbp]
    \centering
    \includegraphics[width=0.32\textwidth]{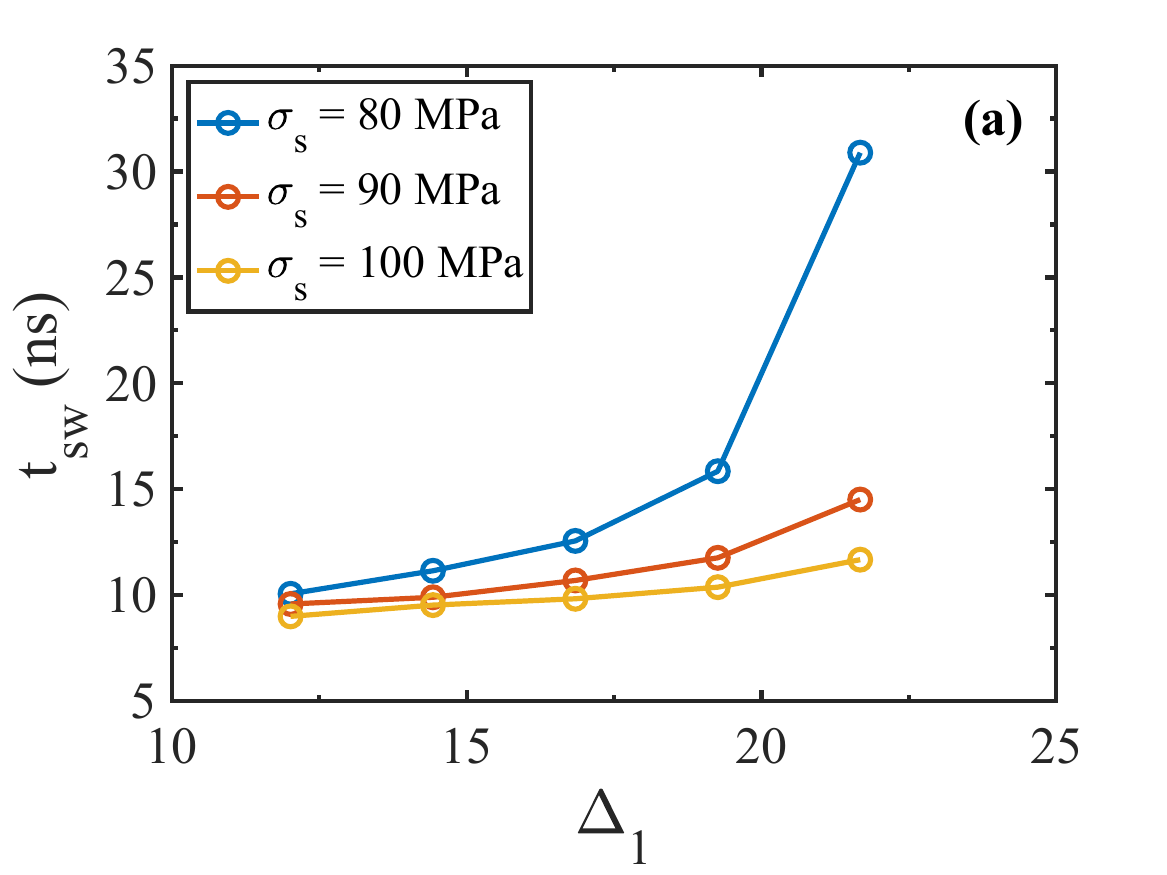}
    \includegraphics[width=0.32\textwidth]{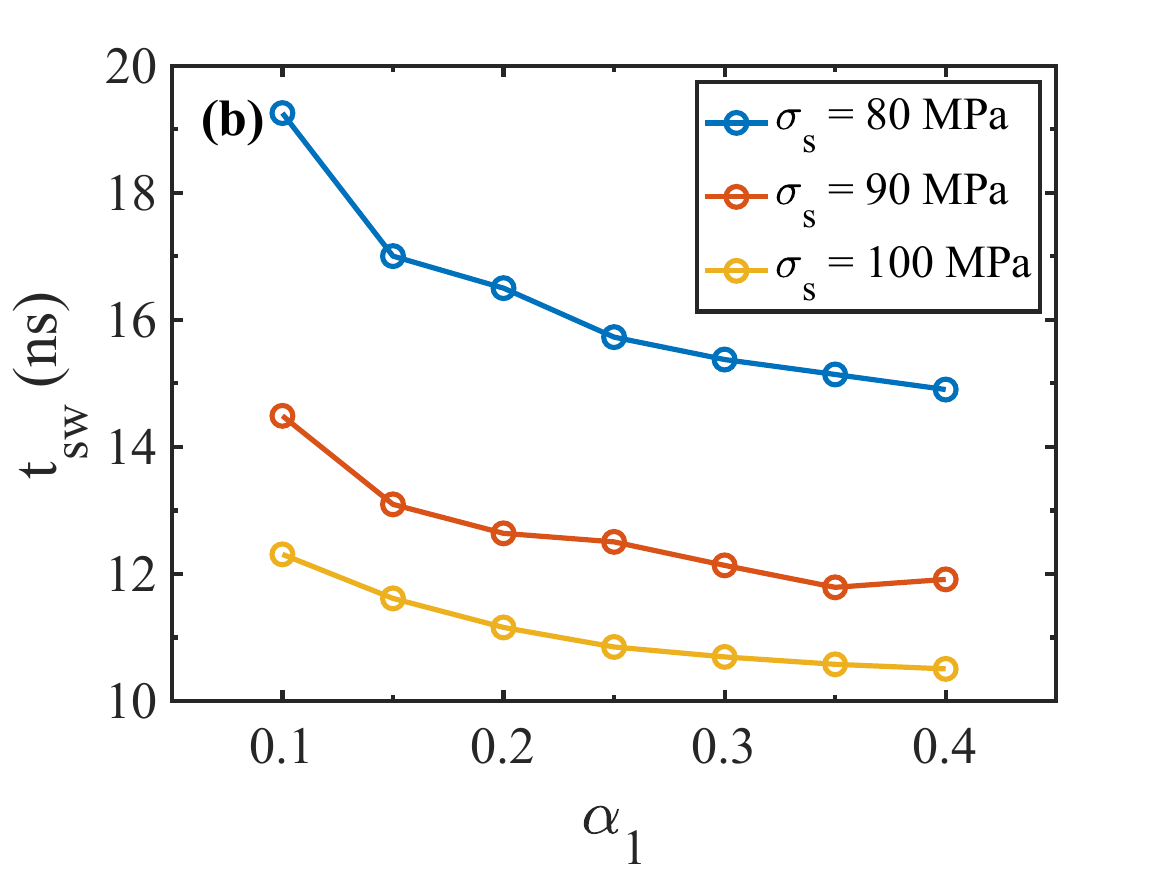}
    \includegraphics[width=0.32\textwidth]{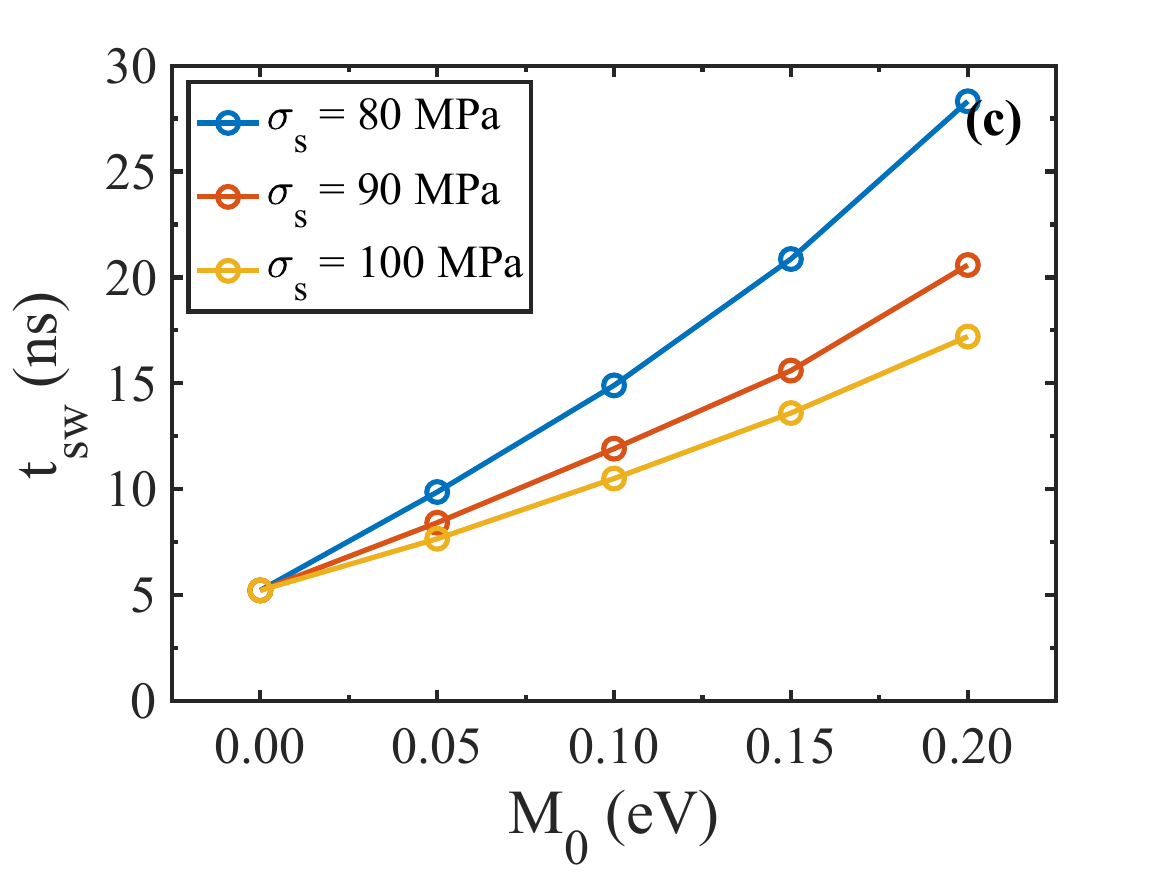}
    \includegraphics[width=0.32\textwidth]{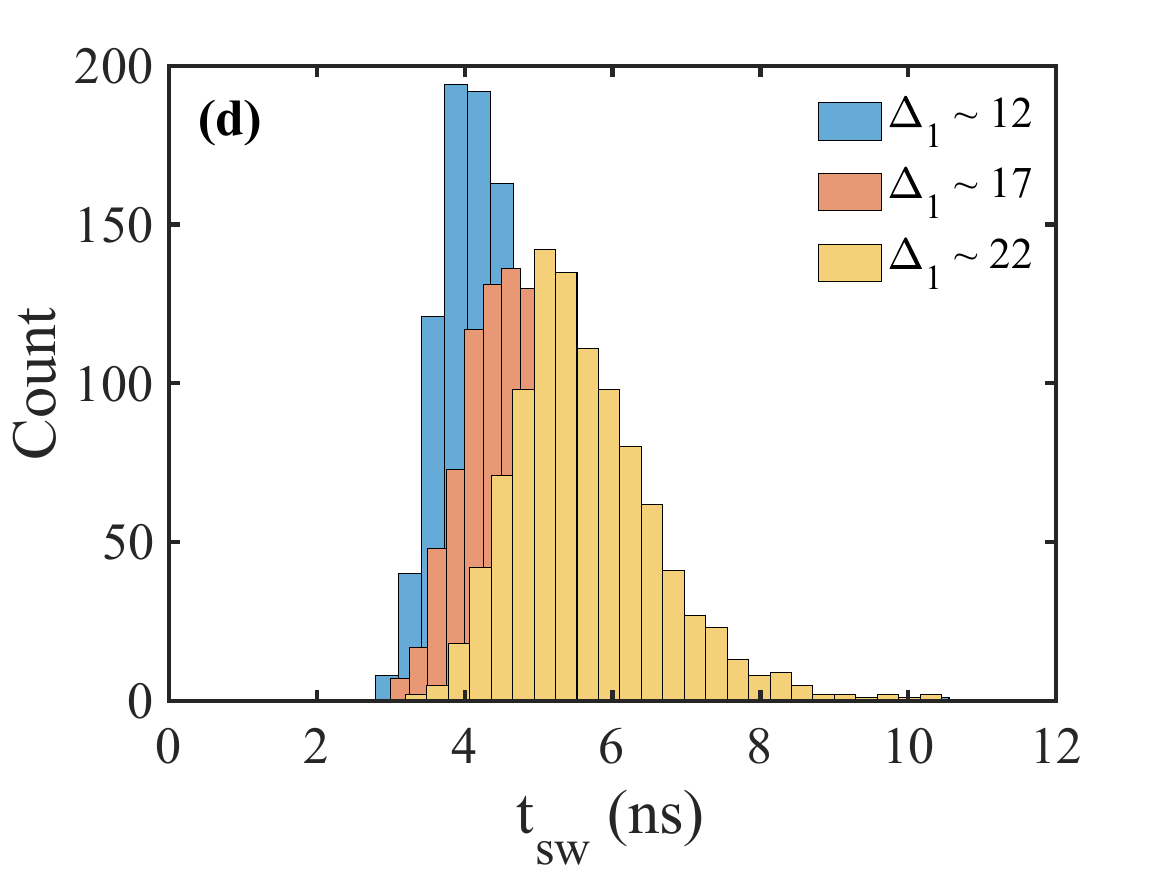}
    \includegraphics[width=0.32\textwidth]{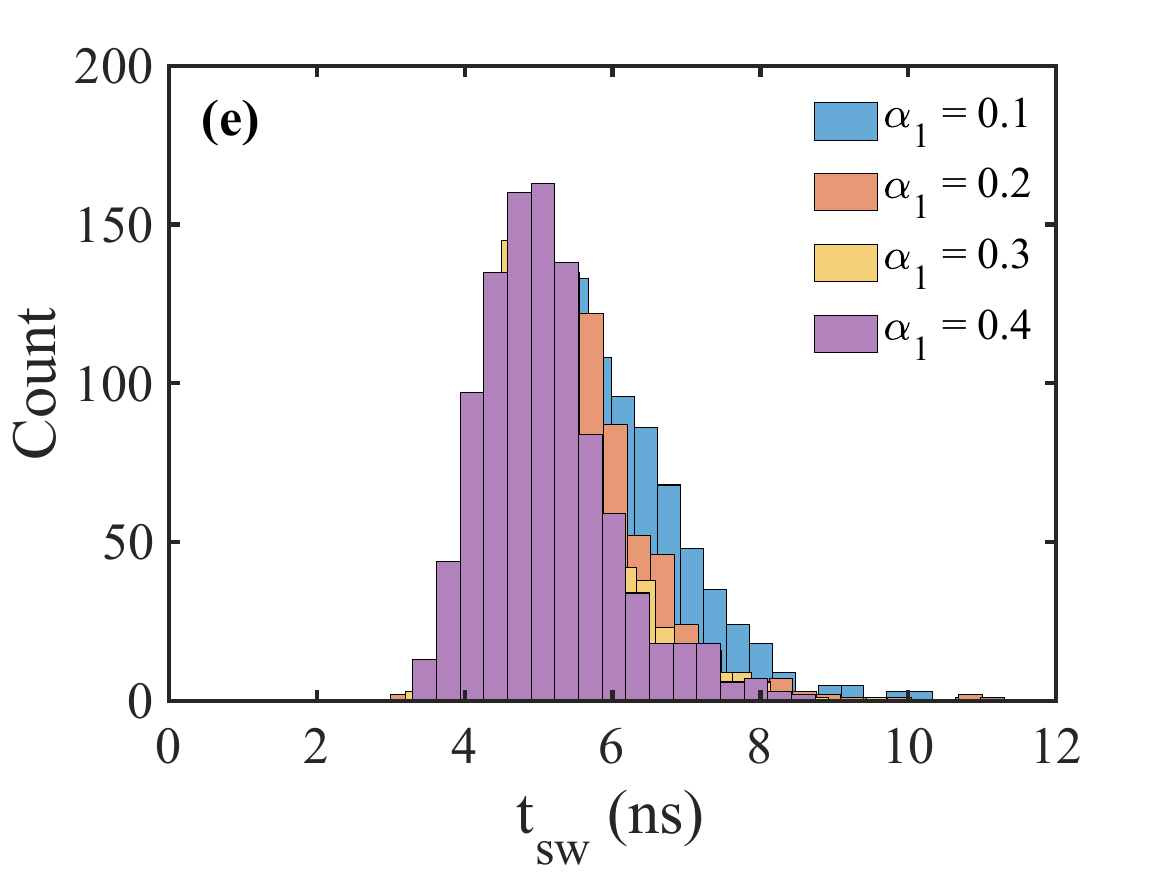}
    \includegraphics[width=0.32\textwidth]{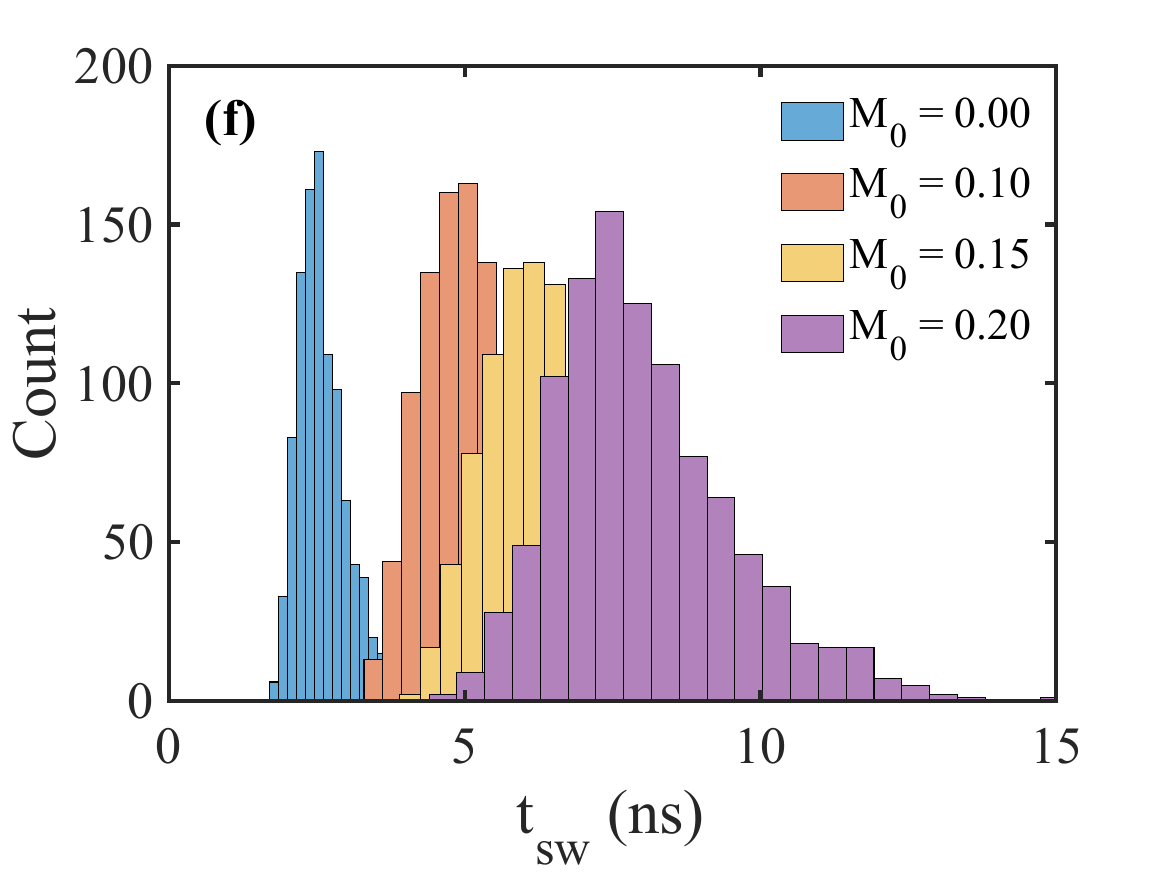}
    \caption{Variation in switching time with respect to (a) the thermal stability factor of the gating magnet, (b) the damping coefficient of the gating magnet, and (c) the exchange strength between the TI and the gating magnet. (d)-(f) Histograms of the switching time from $1000$ stochastic LLG simulations corresponding to (a)-(c), respectively. We use $I_{0,surf}=6I_{c,surf}$ while calculating the $t_{sw}$.}
    \label{fig:ti_fig7}
\end{figure*}

Furthermore, we estimate the energy consumption for the switching process. The energy has two components: the gating energy and the switching energy. The energy associated with the gating mechanism comes from the applied voltage in the piezoelectric and can be estimated as $E_{piezo} = \frac{1}{2}C_pV_G^2$~\cite{Roy2011Aug,Roy2012Jul}, where $C_p$ is the capacitance of the PZT, and $V_G$ is the applied gate voltage. For the case of $0.1\%$ strain, we get $V_G=0.56~\mathrm{V}$ and $C_p = {\epsilon_r \epsilon_0 W L}/{t_{piezo}} = 0.071 \times 10^{-15}~\mathrm{F}$ ($L$ is the length of the device, $\epsilon_0$ and $\epsilon_r = 1000$~\cite{Roy2011Aug} are the permittivity of free space and the relative permittivity of PZT, respectively). We obtain $E_{piezo} = 11.13~\mathrm{aJ}$, which is minuscule and consistent with previous studies~\cite{Fashami2011Mar,Roy2012Jul,Biswas2017Jun}. The other part of the energy is the $I^2R$ loss in the TI. TI is typically modeled as parallel surface and bulk channels (bulk channel accounts for the shunting through the bulk states)~\cite{ti_current_ratio,yunkun_ti}. The bulk resistance is $R_{bulk}=L/\sigma_cWt_{bulk} = 1.46~\mathrm{k\Omega}$, where $\sigma_c$ is the average conductivity of the TI ($\sigma_c = 5.7 \times 10^{4}~\Omega^{-1}m^{-1}$ for Bi$_2$Se$_3$~\cite{Pham2}), and $t_{bulk}$ is the thickness of the bulk state of TI. We use $t_{bulk} = 6~\mathrm{nm}$ since the TI is $8~\mathrm{nm}$ thick and the thickness of the top and the bottom surface is $1~\mathrm{nm}$ each~\cite{ti_current_ratio}. We estimate $R_{surf}$ based on the current distribution in the top surface channel, which is found to be $30\%$ of the total current as previously reported in both theory~\cite{yunkun_ti} and experiments~\cite{ti_current_ratio}. This makes $R_{surf} = 1.94~\mathrm{k\Omega}$. Nonetheless, the bottom surface needs to be grounded through the gating magnet for the gating mechanism to work for our vertical structure (Fig.~\ref{fig:ti_fig1}(b)). Therefore, we divide the bottom surface resistance equally with a ground in the middle and estimate the equivalent resistance $R_{eq} = 633.5~\Omega$, and eventually, we achieve $15\%$ current through the top surface channel of the TI for the resistance configuration shown in Fig.~\ref{fig:ti_fig1}(b). Shunting through the bulk in TI is an open question, and it requires materials with a higher band gap to get more current in the surface~\cite{yunkun_ti}. For high-speed applications, we want the switching time in single-digit nanoseconds~\cite{Oboril2015Jan,Yu2016Jun,Ahmed2017Oct}. We estimate the energy used in this operation regime. 
Figure~\ref{fig:ti_fig5}(a) shows the voltage requirement for $t_{sw}$ for $2-10~\mathrm{ns}$ ($\text{WER} = 10^{-9}$), while Fig.~\ref{fig:ti_fig5}(b) shows the energy consumption for the writing operation for a range of $I_{surf}/I_{tot}$. $I_{surf}/I_{tot}$ can be varied by tuning the conductivity, bulk band gap, and thickness of the TI~\cite{ti_current_ratio}. Our analysis suggests energy consumption in the TI is in the order of $<100~\mathrm{fJ}$, even with a low current ratio of $15\%$ in the top surface, which is energy efficient compared to HM-based SOT switching (Fig.~\ref{fig:ti_fig5}(c)), consistent with previous studies~\cite{ti_TbCo_Liu,Pham2}. If we increase the top surface current, the energy consumption will be decreased further. Fig~\ref{fig:ti_fig5}(c) shows the energy consumption comparison between HM and TI-based switching, which shows an order of magnitude reduction in energy consumption for TI. Although the conductivity of HM is very high, the current dictates the $I^2R$ loss, and TI requires a significantly low current because of its higher SHA~\cite{Pham2}. 

The total energy consumption for writing operation $E_{write} = E_{piezo} + E_{TI}$, which is dominated by the $E_{TI}$ because of the minuscule energy in the gating mechanism. This suggests the advantage of the stain-based intrinsic gating mechanism because it eliminates the need for an access transistor for the write operation, costs negligible energy, and has a reduced footprint compared to a CMOS access transistor. The energy consumption in TI will be much lower if we incorporate other TIs with higher SHAs. For example, for BiSb (SHA = $10$ and $\sigma_c = 1.5 \times 10^5~\mathrm{\Omega^{-1}m^{-1}}$~\cite{Pham2}), the energy consumption is $\sim 0.9~\mathrm{fJ}$ and $\sim 0.2~\mathrm{fJ}$ for a current ratio of $15\%$ and $20\%$, respectively, which is $\sim 100 \times$ lower than Bi$_2$Se$_3$ and $\sim 1000 \times$ lower than traditional HM. 

\begin{table}[t!]
\caption{Material parameters}
\centering
\renewcommand{\arraystretch}{1.5}
\setlength{\tabcolsep}{2pt}
\begin{tabular}{|c|c|c|}
\hline
\multicolumn{3}{|c|}{\bf{Dimension}} \\ \hline
Length & L & 20 nm \\ \hline
Width & W & 40 nm \\ \hline
\multicolumn{3}{|c|}{\bf{Gating Magnet: TbCo}} \\ \hline
Thickness & $t_{f1}$ & 2.5 nm \\ \hline
Saturation magnetization & $M_{s1}$ & $200 \times 10^3$ A/m~\cite{TbCo_ku_ms,TbCo_param_Liu} \\ \hline
Damping & $\alpha_1$ & 0.4 \\ \hline
Uniaxial anisotropy & $K_{u1}$ & $64 \times 10^3$ J/m$^3$~\cite{TbCo_ku_ms,ti_TbCo_Liu} \\ \hline
Magnetostrictive coefficient & $\lambda_s$ & $400 \times 10^{-6}$~\cite{magnetostriction} \\ \hline
Young's modulus & Y & $100 \times 10^9$ Pa~\cite{Youngs_mod} \\ \hline
\multicolumn{3}{|c|}{\bf{MTJ Free Layer: TbCo}} \\ \hline
Thickness & $t_{f2}$ & 12.5 nm \\ \hline
Saturation magnetization & $M_{s2}$ & $400 \times 10^3$ A/m~\cite{TbCo_highMs,TbCo_param_Liu} \\ \hline
Damping & $\alpha_2$ & 0.01 \\ \hline
\multicolumn{3}{|c|}{\bf{Topological Insulator: Bi$_2$Se$_3$}} \\ \hline
Thickness & $t_{TI}$ & 8 nm~\cite{ti_TbCo_Liu,ti_current_ratio} \\ \hline
Spin Hall angle & $\theta_{sh}$ & 3.5~\cite{Pham2} \\ \hline
Spin diffusion length & $\lambda$ & 6.2 nm~\cite{spin_diffusion} \\ \hline
Conductivity & $\sigma_c$ & $5.7 \times 10^4$ $\Omega^{-1}$m$^{-1}$~\cite{Pham2} \\ \hline
\multicolumn{3}{|c|}{\bf{Piezoelctric: PZT}} \\ \hline
Thickness & $t_{piezo}$ & 100 nm \\ \hline
Piezoelectric constant & $d_{31}$ & $1.8 \times 10^{-10}$ m/V~\cite{Roy2011Aug} \\ \hline
Max. strain & $\epsilon$ & 1000 ppm ($0.1\%$)~\cite{Roy2012Jul}\\ \hline
Relative dielectric constant & $\epsilon_r$ & 1000~\cite{Roy2011Aug} \\ \hline
\end{tabular}
\label{tab:parameters}
\end{table}

The above-estimated write energy is at the bit cell level. Now, we project the energy cost for a PiM array from the bit cell energy. In particular, we show the projection for a $2-$bit Boolean AND and OR operation using the PiM crossbar array (Fig.~\ref{fig:ti_fig1}(c)). Before doing the projection, we need to estimate the read energy cost. For reading operation, we use a small sense current $I_{sense} = 1~\mathrm{\mu A}$, significantly less than the write current, eliminating the risk of an accidental switch. We assume a resistance area product for the MTJ $R_P A = 2~\mathrm{\Omega \mu m^2}$ and a $\text{TMR} = 100\%$~\cite{laura_smart}, which makes $R_P = 2.5~\mathrm{k \Omega}$ and $R_{AP} = 5~\mathrm{k \Omega}$, respectively ($R_P$ and $R_{AP}$ are the resistance of MTJ in parallel and anti-parallel state, respectively). We consider a $16~\mathrm{nm}$ PTM HP model for the read access transistor with a $R_{ON} = 5~\mathrm{k \Omega}$. We need to keep the gate voltage ON for the read operation so that the TI surface state is conductive, adding the gating energy during the read operation. Considering this, we found a read energy $E_{read}$ of $~\sim 45~\mathrm{aJ}$ and $~\sim 55~\mathrm{aJ}$ for parallel (bit `$0$') and antiparallel (bit `$1$') states at a read speed of $4~
\mathrm{ns}$~\cite{Wei}. For in-memory $2-$bit AND and OR operations, we need to select two bit cells simultaneously, and the PiM operation is done by comparing the sense voltage from the bit cells with a reference voltage $V_{ref}$ through a sense amplifier (see Fig.~\ref{fig:ti_fig1}(b)), where $V_{ref}$ is set to a value depending on the operation we want to perform. For two bit cells, we will have $\{R_{AP},~R_{AP}\},~\{R_{AP},~R_{P}\},~\{R_{P},~R_{AP}\},~\{R_{P},~R_{P}\}$ combinations in the resistance states of the MTJ, corresponding to sense voltages $\{V_{AP},~V_{AP}\},~\{V_{AP},~V_{P}\},~\{V_{P},~V_{AP}\},~\{V_{P},~V_{P}\}$. Setting a $V_{ref} = (V_{AP,AP} + V_{AP,P})/2$ ensures the sense amplifier will produce a logic `$1$' output only when both the bit cells are in the antiparallel state (logic AND). Similarly, for logic OR, $V_{ref} = (V_{AP,P} + V_{P,P})/2$ will serve the purpose. For the resistance and sense current used in our study, we get $V_{AP,AP} = 5~\mathrm{mV}$, $V_{AP,P}/V_{P,AP} = 4.29~\mathrm{mV}$, $V_{P,P} = 3.75~\mathrm{mV}$, which give $V_{ref, \text{AND}}=4.65~\mathrm{mV}$; $V_{ref, \text{OR}}=4.02~\mathrm{mV}$. We can set the reference voltage by tuning the reference resistance of the sense amplifier, and therefore, we can perform re-configurable in-memory computing by tuning the reference voltage. 
The energy components for a full cycle (write-read) in-memory Boolean computing are summarized in Fig.~\ref{fig:ti_fig6}, which shows that $E_{write}$ is the dominant contribution. This suggests the energy efficiency of the STI-SOTRAM bit cell for in-memory computing because the write energy of STI-SOTRAM is much smaller than that of a conventional HM-based device, as shown in Fig.~\ref{fig:ti_fig5} (see the comparison with other emerging technology in Table~\ref{tab:comparison}). Note that we assume a symmetric write operation~\cite{symmetric_current} for both bits `$0$' and `$1$', and the energy associated with the sense amplifier is estimated as $\frac{1}{2}C \Delta V^2$, where $C$ is the capacitance of the sense amplifier (typically $1~\mathrm{pF}$~\cite{Tozer1992May,Conte2005Jan}) and $\Delta V$ is the difference between the sense voltage and the reference voltage. The estimated total area for the Boolean operation is $6720~\mathrm{nm^2}$, which includes the area of the bit cells and the read access transistor (W/L = 10:1).  

Finally, we present the effect of various key parameters on device switching time/delay $t_{sw}$. We show the effect of the thermal stability factor $\Delta_1$, exchange constant $M_0$, and damping coefficient $\alpha_1$ of the gating magnet on $t_{sw}$ for various $\sigma_s$ values. In Fig.~\ref{fig:ti_fig7}, the top panel shows the $t_{sw}$ while the bottom panel shows the histogram of $t_{sw}$ from $1000$ simulations. For the case of varying $\Delta_1$, as it increases, the switching time of the gating magnet will increase for a specific $\sigma_s$, which increases the overall switching time of the device. While for a larger $\alpha_1$, $t_{sw}$ reduces as the gating magnet switching reduces with a large damping coefficient because the gating magnet dynamics is governed by precessional and damping torques. Lastly, if we increase $M_0$, $t_{sw}$ increases because the gap opening is larger, and it takes longer to provide sufficient surface state current in the TI. Note that even if we get a large $M_0$, the TSS band gap cannot be infinitely large as we are limited by the bulk bandgap of the TI. In contrast, we need a larger value of $M_0$ for a high ON/OFF ratio, which suggests the need for a TI with larger bulk bandgap~\cite{hamed_fm_ti}.

\begin{table*}
\caption{Stoplight chart for existing vs. emerging memory technology~\cite{sptlgt_chat_1,Oboril2015Jan,sptlgt_chat_2,Yu2016Jun,sptlgt_chat_3,pim_survey}. Our device uses non-volatile magnetic SOT technology, albeit with a superior TI underlayer with both a large SHA ($> 10$) for low power write and a gate tunable bandgap for energy efficient ($\sim 10$s of aJ) row-column select. The resulting vertical structure of the selector-storage stack (Fig.~\ref{fig:ti_fig1}) is well suited for compact, low latency in-memory edge computing.}
\centering
\renewcommand{\arraystretch}{1.5}
\setlength{\tabcolsep}{0pt}
\begin{tabular}{|c|c|c|c|c|c|c|c|c|c|c|}
    \hline
    & SRAM & DRAM & Flash & Flash & FeRAM & ReRAM & PCRAM & STTRAM & SOTRAM & STI- \\
    & (45 nm) & & (NAND) & (NOR) & & & & & & SOTRAM\\
    \hline
    Memory & Charge & Charge & Charge & Charge & FE & Oxide- & VO$_2$, GST & Tunnel & Tunnel & Tunnel\\
    type & & & & & capacitor & charge & Bi$_2$O$_2$Se & junction & junction & junction \\
    \hline
    Endurance & \cellcolor{YellowGreen}$10^{16}$ & \cellcolor{YellowGreen}$10^{16}$ & \cellcolor{red!50}$10^{4}$ & \cellcolor{red!50}$10^{6}$ & \cellcolor{yellow}$10^{10}$ & \cellcolor{yellow}$10^5-10^8$ & \cellcolor{yellow}$10^{4}-10^9$ & \cellcolor{YellowGreen} $> 10^{15}$& 
    \cellcolor{YellowGreen} $> 10^{15}$& 
    \cellcolor{red!50} $> 10^{6}$\\
    \hline
    Read time & \cellcolor{YellowGreen} 1-100 ns & \cellcolor{YellowGreen} 30 ns & \cellcolor{red!50} $\sim$1 $\mu$s & \cellcolor{red!50} $\sim$50 ns & \cellcolor{YellowGreen} $< 10$ ns &\cellcolor{YellowGreen} $< 10$ ns &\cellcolor{YellowGreen} $< 10$ ns  &\cellcolor{YellowGreen} $< 10$ ns &\cellcolor{YellowGreen} $< 10$ ns &\cellcolor{YellowGreen} $< 10$ ns\\
   Write time & \cellcolor{YellowGreen} 1-100 ns & \cellcolor{YellowGreen} 50 ns &  \cellcolor{red!50} 0.1-1 ms & \cellcolor{red!50} 1-10 $\mu$s & \cellcolor{YellowGreen} $\sim$ 30 ns & \cellcolor{YellowGreen} $<$ 10 ns & \cellcolor{YellowGreen} $\sim$ 50 ns &\cellcolor{YellowGreen} $< 10$ ns &\cellcolor{YellowGreen} $< 10$ ns &\cellcolor{YellowGreen} $< 10$ ns\\
   Erase time& \cellcolor{YellowGreen} & \cellcolor{YellowGreen} 50 ns&\cellcolor{red!50} 0.1 ms & \cellcolor{red!50} 10 ms & \cellcolor{YellowGreen} &\cellcolor{YellowGreen} &\cellcolor{YellowGreen} &\cellcolor{YellowGreen} &\cellcolor{YellowGreen} &\cellcolor{YellowGreen} \\
   \hline
   Write energy & \cellcolor{YellowGreen}$\sim$ fJ & \cellcolor{YellowGreen}$\sim$ 10 fJ & \cellcolor{YellowGreen}$\sim$ 10 fJ & \cellcolor{red!50}$\sim$ 100 pJ & \cellcolor{YellowGreen}$\sim$ 100 fJ & 
   \cellcolor{yellow}0.1-1 pJ & \cellcolor{yellow}10 pJ & \cellcolor{YellowGreen}$\sim$ 100 fJ & \cellcolor{YellowGreen}$<$ 100 fJ & \cellcolor{YellowGreen}1-100 fJ \\
    \hline
    Size F$^2$& \cellcolor{red!50}50-120 & \cellcolor{YellowGreen}6-10 & \cellcolor{YellowGreen}10 & \cellcolor{YellowGreen}5 & \cellcolor{yellow}15-34 &\cellcolor{YellowGreen}6-10 &\cellcolor{YellowGreen}4-10 &\cellcolor{YellowGreen}6-20 &\cellcolor{YellowGreen}6-20 &\cellcolor{yellow}15-25\\
    \hline
    Standby & \cellcolor{red!50}Current & \cellcolor{red!50}Refresh & \cellcolor{YellowGreen}None & \cellcolor{YellowGreen}None & \cellcolor{YellowGreen}None & \cellcolor{YellowGreen}None & \cellcolor{YellowGreen}None & \cellcolor{YellowGreen}None & \cellcolor{red!50}Field assist & \cellcolor{YellowGreen}No assist\\
     & \cellcolor{red!50}Leakage & \cellcolor{red!50}current & \cellcolor{YellowGreen} & \cellcolor{YellowGreen} & \cellcolor{YellowGreen} & \cellcolor{YellowGreen} & \cellcolor{YellowGreen} & \cellcolor{YellowGreen} & \cellcolor{red!50} & \cellcolor{YellowGreen} \\
    \hline 
    High voltage& \cellcolor{YellowGreen}No & \cellcolor{Yellow}2 V & \cellcolor{red!50}10 V & \cellcolor{red!50}$<$10 V & \cellcolor{Yellow}$<$3 V & \cellcolor{Yellow}$<$3 V & \cellcolor{Yellow}$<$3 V & \cellcolor{Yellow}3 V & \cellcolor{YellowGreen}$<$1.5 V & \cellcolor{YellowGreen}$<$1V \\
    \hline 
    Non-volatile& \cellcolor{red!50}No & \cellcolor{red!50}No & \cellcolor{YellowGreen}Yes & \cellcolor{YellowGreen}Yes & \cellcolor{YellowGreen}Yes & \cellcolor{YellowGreen}Yes & \cellcolor{YellowGreen}Yes & \cellcolor{YellowGreen}Yes & \cellcolor{YellowGreen}Yes & \cellcolor{YellowGreen}Yes \\
    \hline 
\end{tabular}
\label{tab:comparison}  
\end{table*}
\section{Challenges and Opportunities }\label{sec:challenge} 
The above results show excellent promise for utilizing the STI-SOTRAM for an energy-efficient bit-storage device. However, it is worth discussing the challenges associated with implementing STI-SOTRAM devices. We will separate these discussions into two categories - material issues vs. device issues.
\subsection{Material Challenges}
There exist material issues~\cite{Schenk2020Jun} since we need three separate interfacial processes with input-output isolation -- (\textit{a}) strain rotating the bottom selector magnet, (\textit{b}) modulating the TSS with such rotation, and (\textit{c}) TSS writing information onto the top storage magnet. Let us discuss these three processes one by one.

{\bf{{(a) Can strain switch a magnet by $90^{\circ}$?}}}
The magnetocrystalline anisotropy density $K_u$ is often larger than stress energy. We will need to compositionally tune sputtered magnets like amorphous TbCo ~\cite{Zhao2015Mar,TbCo_param_Liu,Anuniwat2013Jan}, or tailor film thickness of epitaxially grown CoFe to approach a small $K_u$ near the out-of-plane to in-plane crossover ~\cite{Mandal2018Dec}. The resulting low $K_u$ can be overcome with stress, while being large enough to avoid spontaneous thermal fluctuations. The condition for switching is $(3/2)\lambda_s \sigma_s > K_u$, where $\sigma_s$ is the uniaxial stress. The magnets referenced above have magnetostriction coefficient $\lambda_s \sim 200-400 \times 10^{-6}$~\cite{magnetostriction,Serizawa2019May} with elastic moduli $\sim 100~\mathrm{GPa}$~\cite{Youngs_mod} and strain $\sim 0.1\%$, which gives (3/2)$\lambda_s\sigma_s \sim 30-60$ kJ/m$^3$ and suggests successful switching of the magnetization if we can keep $K_u$ lower.

{\bf{{(b) Can a rotated magnet kill the TSS?}}}
Theory~\cite{Nikolic} and experiments~\cite{Rojas-Sanchez2016Mar} show that an interfacing out-of-plane magnet opens a TSS bandgap, leaving only chiral quantum anomalous edge states with significantly diminished spin surface current. A 3D tight binding-based study suggests the same, even accounting for bands along the patterned TI side walls~\cite{hamed_fm_ti}. Critically, the magnet must adjoin the TI, providing at least $50~\mathrm{mV}$ exchange coupling. The g-factor for the Zeeman gap is highly material dependent ($\sim 18$ for Bi$_2$Se$_3$, $\sim 6$ for Sb$_2$Te$_2$Se~\cite{Fu2016Feb}), while that depends on bulk bandgap due to the magnetic proximity effect ($\sim 20~\mathrm{meV}$ for EuS/Bi$_2$Se$_3$~\cite{Wang2023Aug}, $\sim 100~\mathrm{meV}$ for MnBi$_2$Se$_4$/Bi$_2$Se$_3$~\cite{Kaveev2021Dec}).

{\bf{{(c) Can an activated TI write efficiently on the top magnet, accounting for small TI bandgap and current shunting?}}}
SHA, even in sputtered TI, can be large ($\sim 10-20$, about $100 \times$ greater than conventional HMs, e.g., Pt), switching magnets very efficiently~\cite{Pham2,Pham3}. TIs have small bulk bandgaps $\sim 100-300~\mathrm{meV}$ that shunt current into the bulk and sidewalls. However, previous studies suggest that even for a modest bandgap TI, an ON-OFF ratio of $\sim 10$ might be able to flip spins of a magnet with a low  WER ~\cite{hamed_fm_ti}.  This self-correction happens through internal anisotropy fields~\cite{Rehm2024Jan}, an increased bandgap with thin film quantization~\cite{Pham2,Chang2012Dec}, and reduced charge current with gap opening.  To avoid current shunting into the storage magnet, a thin layer of insulating NiO~\cite{Wang2019Nov} or MgO~\cite{Khanal2021Dec} can be grown that transmits magnon torque between the magnet/TI, or else use an insulating BaFe$_{12}$O$_{19}$ as the storage magnet’s free layer~\cite{Li2019Aug}.

{\bf{(d) Can we grow and pattern the heterogeneous stack?}}
To grow the stack with compatible processing temperatures, we will need to sputter the Bi$_2$Se$_3$ directly onto a magnet on a piezo such as Pb(Zr,Ti)O$_3$(PZT)/Pb(Mg$_{1/3}$Nb$_{2/3}$)O$_3$-PbTiO$_3$ (PMNT). 
Previous experiments show that amorphous TI retains bandgap, topological protection, and high SHA~\cite{Dc2019Aug,Pham2,Lu2022Mar}, even after patterning with wet etching~\cite{Barton2019Sep}. Besides mechanical decoupling, the stacks are also electrically decoupled, as the strains on a common piezo substrate stay localized around the contacts~\cite{Biswas2017Jun,Karki2023Nov}.

\subsection{Device/circuit/architecture issues}

Table~\ref{tab:comparison} shows projected performance metrics for our device vs. competing memory technology, especially for PiM/embedded applications. 

{\bf{(a) Key application space.}} Our device is similar to a DRAM crossbar, complementary to non-volatile memory in the low power/embedded space, and competitive in NAND space with similar trade benefits as MRAM/SOTRAM. It capitalizes on SOT technology, which is attractive as last level embedded cache for digital AI and stand-alone off chip cache for analog AI. The device uses a superior TI channel that is both spin-selective with high SOT efficiency and gate tunable with low power strain gating. For in-memory computing in embedded applications (image processing, combinatorial optimization) for edge intelligence, error tolerance and retention needs are low while energy and latency are premium. Our energy budget is low with write voltage $<100$ mV for thin film ($\sim$ 1 $\mu$m on conducting subtrate)/2-D monolayer piezos. Our device projects excellent metrics with $\sim$ 2-3X improvement over DRAM in energy-delay products. 

{\bf{(b) Critical current/voltage.}} Our projected current density $\sim 10^5-10^6$A/cm$^2$ is typical of STTRAMs and can be further reduced with higher SHA. Since the TSS are metallic in ON-state (conductivity of BiSe $\sim$ 8,000-60,000 $\Omega^{-1}m^{-1}$), our room-temperature write voltage is projected to be $< 100$ mV. 

{\bf{(c) Back-end-of-the-line (BEOL) integration.}} TIs can be sputtered and wet etched while retaining high SHA, so usual BEOL metallization steps should work. Both TI and magnets should survive 85$^\circ$-100$^\circ$C processing temperatures. Thermal degradation of TI does not usually occur below 300$^\circ$C. 

{\bf{(d) Endurance, stability and device-to-device variations.}} While HM-based SOTRAM endurance is  $\sim 10^{15}$, piezo fatigue and TI endurance limit our STI-SOTRAM device to $\sim 5 \times 10^6$ cycles~\cite{Zhu2013Apr,Jang2020Nov}. While this is significantly lower endurance, we believe this would suffice for hyperdense, nonvolatile persistent cells in PiM applications, where the probability of data overwrite per cell is very low ($\sim 10^{-3}$, easily error corrected~\cite{Hey2023May}).

{\bf{(e) Data retention, read disturb tolerance and scalability.}} The lower selector magnet needs small out-of-plane anisotropy for strain gating, while the upper storage magnet needs higher barrier consistent with SOTRAM technology with 10 year retention. Like SOT, we avoid read disturbs with separate read-write paths (unlike SOT, we do not need field-assist to circumvent stagnation, as our TI spins and storage magnet are in-plane). The top MTJ requires ion-milling and mesa etching, which the TSS are known to survive.  

{\bf{(f) Density.}} Our device footprint is similar to three terminal SOT, slightly bigger than STTRAM, DRAM, and flash. We compete on density (and latency) by rolling storage and processing into one bit cell, along with non-volatile magnetic storage and ultra-low energy field-free write using strain gating ($\sim$ 10 aJ), and high SHA ($\sim$ 10), key metrics for embedded applications and edge intelligence.

\section{Conclusion}\label{sec:conclusion} 
In summary, we demonstrated a novel four-layer piezoelectric/magnet/TI/MTJ bit cell design with reduced energy cost and footprint for in-memory computing. Eliminating an access transistor with a built-in strain-based gating mechanism has proved energy-efficient, yielding an overall reduced energy cost. High-speed operation is achieved with significantly less energy than the traditional SOT metals at the device and array level. Our results suggest this heterogeneous stack may provide a compact and energy-efficient design for low-power, high-speed in-situ applications.

The challenges at this time are the yet-to-be-demonstrated ability to fabricate the entire stack, the lower endurance associated with TIs (which is a lesser issue with PiM as the data overwrite per cell is low), and the established challenges with SOT, such as higher footprint (again, PiM saves at the circuit level by rolling memory and logic into one unit), and higher current (but lower voltage since the HM has lower resistance). The in-plane spins help with field assist but hurt with scaling. Out-of-plane field-free switching has been demonstrated with SOTs and with Weyl semi-metals~\cite{WSM1,WSM2,WSM3}. 


\section{Acknowledgments}\label{sec:ack}
We thank S. Joseph Poon at the University of Virginia, Supriyo Bandyopadhyay at Virginia Commonwealth University, Joseph A. Hagmann at Mitre, Patrick Taylor, George de Coster, and Mahesh R Neupane at Army Research Laboratories, Andrew D. Kent at New York University, Daniel B. Gopman at the National Institute of Standards at Technology, and Steve Kramer at Micron for insightful discussions. This work was supported in part by the NSF I/UCRC on Multi-functional Integrated System Technology (MIST) Center; IIP-1439644, IIP-1439680, IIP-1738752, IIP-1939009, IIP-1939050, and IIP-1939012. The calculations were performed using the computational resources from High-Performance Computing systems at the University of Virginia (Rivanna).

\bibliography{main,supporting}

\begin{thebibliography}{107}%
\makeatletter
\providecommand \@ifxundefined [1]{%
 \@ifx{#1\undefined}
}%
\providecommand \@ifnum [1]{%
 \ifnum #1\expandafter \@firstoftwo
 \else \expandafter \@secondoftwo
 \fi
}%
\providecommand \@ifx [1]{%
 \ifx #1\expandafter \@firstoftwo
 \else \expandafter \@secondoftwo
 \fi
}%
\providecommand \natexlab [1]{#1}%
\providecommand \enquote  [1]{``#1''}%
\providecommand \bibnamefont  [1]{#1}%
\providecommand \bibfnamefont [1]{#1}%
\providecommand \citenamefont [1]{#1}%
\providecommand \href@noop [0]{\@secondoftwo}%
\providecommand \href [0]{\begingroup \@sanitize@url \@href}%
\providecommand \@href[1]{\@@startlink{#1}\@@href}%
\providecommand \@@href[1]{\endgroup#1\@@endlink}%
\providecommand \@sanitize@url [0]{\catcode `\\12\catcode `\$12\catcode `\&12\catcode `\#12\catcode `\^12\catcode `\_12\catcode `\%12\relax}%
\providecommand \@@startlink[1]{}%
\providecommand \@@endlink[0]{}%
\providecommand \url  [0]{\begingroup\@sanitize@url \@url }%
\providecommand \@url [1]{\endgroup\@href {#1}{\urlprefix }}%
\providecommand \urlprefix  [0]{URL }%
\providecommand \Eprint [0]{\href }%
\providecommand \doibase [0]{http://dx.doi.org/}%
\providecommand \selectlanguage [0]{\@gobble}%
\providecommand \bibinfo  [0]{\@secondoftwo}%
\providecommand \bibfield  [0]{\@secondoftwo}%
\providecommand \translation [1]{[#1]}%
\providecommand \BibitemOpen [0]{}%
\providecommand \bibitemStop [0]{}%
\providecommand \bibitemNoStop [0]{.\EOS\space}%
\providecommand \EOS [0]{\spacefactor3000\relax}%
\providecommand \BibitemShut  [1]{\csname bibitem#1\endcsname}%
\let\auto@bib@innerbib\@empty
\bibitem [{\citenamefont {Roohi}\ \emph {et~al.}(2022)\citenamefont {Roohi}, \citenamefont {Angizi},\ and\ \citenamefont {Fan}}]{edge_computing}%
  \BibitemOpen
  \bibfield  {author} {\bibinfo {author} {\bibfnamefont {A.}~\bibnamefont {Roohi}}, \bibinfo {author} {\bibfnamefont {S.}~\bibnamefont {Angizi}}, \ and\ \bibinfo {author} {\bibfnamefont {D.}~\bibnamefont {Fan}},\ }in\ \href {\doibase 10.1007/978-3-031-16344-9_11} {\emph {\bibinfo {booktitle} {{Frontiers of Quality Electronic Design (QED):AI, IoT and Hardware Security}}}}\ (\bibinfo  {publisher} {Springer},\ \bibinfo {address} {Cham, Switzerland},\ \bibinfo {year} {2022})\ pp.\ \bibinfo {pages} {415--464}\BibitemShut {NoStop}%
\bibitem [{new(2024)}]{news-article}%
  \BibitemOpen
  \href {https://semi.org/en/blogs/technology-and-trends/maximizing-edge-intelligence-requires-more-than-computing} {\enquote {\bibinfo {title} {{Maximizing Edge Intelligence Requires More Than Computing {$\vert$} SEMI}},}\ } (\bibinfo {year} {2024}),\ \bibinfo {note} {[Online; accessed 7. Jun. 2024]}\BibitemShut {NoStop}%
\bibitem [{\citenamefont {Roy}\ \emph {et~al.}(2011{\natexlab{a}})\citenamefont {Roy}, \citenamefont {Bandyopadhyay},\ and\ \citenamefont {Atulasimha}}]{Roy2011Aug}%
  \BibitemOpen
  \bibfield  {author} {\bibinfo {author} {\bibfnamefont {K.}~\bibnamefont {Roy}}, \bibinfo {author} {\bibfnamefont {S.}~\bibnamefont {Bandyopadhyay}}, \ and\ \bibinfo {author} {\bibfnamefont {J.}~\bibnamefont {Atulasimha}},\ }\href {\doibase 10.1063/1.3624900} {\bibfield  {journal} {\bibinfo  {journal} {Appl. Phys. Lett.}\ }\textbf {\bibinfo {volume} {99}} (\bibinfo {year} {2011}{\natexlab{a}}),\ 10.1063/1.3624900}\BibitemShut {NoStop}%
\bibitem [{\citenamefont {Biswas}\ \emph {et~al.}(2017)\citenamefont {Biswas}, \citenamefont {Ahmad}, \citenamefont {Atulasimha},\ and\ \citenamefont {Bandyopadhyay}}]{Biswas2017Jun}%
  \BibitemOpen
  \bibfield  {author} {\bibinfo {author} {\bibfnamefont {A.~K.}\ \bibnamefont {Biswas}}, \bibinfo {author} {\bibfnamefont {H.}~\bibnamefont {Ahmad}}, \bibinfo {author} {\bibfnamefont {J.}~\bibnamefont {Atulasimha}}, \ and\ \bibinfo {author} {\bibfnamefont {S.}~\bibnamefont {Bandyopadhyay}},\ }\href {\doibase 10.1021/acs.nanolett.7b00439} {\bibfield  {journal} {\bibinfo  {journal} {Nano Lett.}\ }\textbf {\bibinfo {volume} {17}},\ \bibinfo {pages} {3478} (\bibinfo {year} {2017})}\BibitemShut {NoStop}%
\bibitem [{\citenamefont {Duan}\ \emph {et~al.}(2015)\citenamefont {Duan}, \citenamefont {Li}, \citenamefont {Li}, \citenamefont {Semenov},\ and\ \citenamefont {Kim}}]{fm_ti_ncstate}%
  \BibitemOpen
  \bibfield  {author} {\bibinfo {author} {\bibfnamefont {X.}~\bibnamefont {Duan}}, \bibinfo {author} {\bibfnamefont {X.-L.}\ \bibnamefont {Li}}, \bibinfo {author} {\bibfnamefont {X.}~\bibnamefont {Li}}, \bibinfo {author} {\bibfnamefont {Y.~G.}\ \bibnamefont {Semenov}}, \ and\ \bibinfo {author} {\bibfnamefont {K.~W.}\ \bibnamefont {Kim}},\ }\href {\doibase 10.1063/1.4937407} {\bibfield  {journal} {\bibinfo  {journal} {J. Appl. Phys.}\ }\textbf {\bibinfo {volume} {118}} (\bibinfo {year} {2015}),\ 10.1063/1.4937407}\BibitemShut {NoStop}%
\bibitem [{\citenamefont {Rojas-S{\ifmmode\acute{a}\else\'{a}\fi}nchez}\ \emph {et~al.}(2016)\citenamefont {Rojas-S{\ifmmode\acute{a}\else\'{a}\fi}nchez}, \citenamefont {Oyarz{\ifmmode\acute{u}\else\'{u}\fi}n}, \citenamefont {Fu}, \citenamefont {Marty}, \citenamefont {Vergnaud}, \citenamefont {Gambarelli}, \citenamefont {Vila}, \citenamefont {Jamet}, \citenamefont {Ohtsubo}, \citenamefont {Taleb-Ibrahimi}, \citenamefont {Le~F{\ifmmode\grave{e}\else\`{e}\fi}vre}, \citenamefont {Bertran}, \citenamefont {Reyren}, \citenamefont {George},\ and\ \citenamefont {Fert}}]{Rojas-Sanchez2016Mar}%
  \BibitemOpen
  \bibfield  {author} {\bibinfo {author} {\bibfnamefont {J.-C.}\ \bibnamefont {Rojas-S{\ifmmode\acute{a}\else\'{a}\fi}nchez}}, \bibinfo {author} {\bibfnamefont {S.}~\bibnamefont {Oyarz{\ifmmode\acute{u}\else\'{u}\fi}n}}, \bibinfo {author} {\bibfnamefont {Y.}~\bibnamefont {Fu}}, \bibinfo {author} {\bibfnamefont {A.}~\bibnamefont {Marty}}, \bibinfo {author} {\bibfnamefont {C.}~\bibnamefont {Vergnaud}}, \bibinfo {author} {\bibfnamefont {S.}~\bibnamefont {Gambarelli}}, \bibinfo {author} {\bibfnamefont {L.}~\bibnamefont {Vila}}, \bibinfo {author} {\bibfnamefont {M.}~\bibnamefont {Jamet}}, \bibinfo {author} {\bibfnamefont {Y.}~\bibnamefont {Ohtsubo}}, \bibinfo {author} {\bibfnamefont {A.}~\bibnamefont {Taleb-Ibrahimi}}, \bibinfo {author} {\bibfnamefont {P.}~\bibnamefont {Le~F{\ifmmode\grave{e}\else\`{e}\fi}vre}}, \bibinfo {author} {\bibfnamefont {F.}~\bibnamefont {Bertran}}, \bibinfo {author} {\bibfnamefont {N.}~\bibnamefont {Reyren}}, \bibinfo {author} {\bibfnamefont {J.-M.}\ \bibnamefont {George}}, \ and\
  \bibinfo {author} {\bibfnamefont {A.}~\bibnamefont {Fert}},\ }\href {\doibase 10.1103/PhysRevLett.116.096602} {\bibfield  {journal} {\bibinfo  {journal} {Phys. Rev. Lett.}\ }\textbf {\bibinfo {volume} {116}},\ \bibinfo {pages} {096602} (\bibinfo {year} {2016})}\BibitemShut {NoStop}%
\bibitem [{\citenamefont {Marmolejo-Tejada}\ \emph {et~al.}(2017)\citenamefont {Marmolejo-Tejada}, \citenamefont {Dolui}, \citenamefont {Lazi{\ifmmode\acute{c}\else\'{c}\fi}}, \citenamefont {Chang}, \citenamefont {Smidstrup}, \citenamefont {Stradi}, \citenamefont {Stokbro},\ and\ \citenamefont {Nikoli{\ifmmode\acute{c}\else\'{c}\fi}}}]{Nikolic}%
  \BibitemOpen
  \bibfield  {author} {\bibinfo {author} {\bibfnamefont {J.~M.}\ \bibnamefont {Marmolejo-Tejada}}, \bibinfo {author} {\bibfnamefont {K.}~\bibnamefont {Dolui}}, \bibinfo {author} {\bibfnamefont {P.}~\bibnamefont {Lazi{\ifmmode\acute{c}\else\'{c}\fi}}}, \bibinfo {author} {\bibfnamefont {P.-H.}\ \bibnamefont {Chang}}, \bibinfo {author} {\bibfnamefont {S.}~\bibnamefont {Smidstrup}}, \bibinfo {author} {\bibfnamefont {D.}~\bibnamefont {Stradi}}, \bibinfo {author} {\bibfnamefont {K.}~\bibnamefont {Stokbro}}, \ and\ \bibinfo {author} {\bibfnamefont {B.~K.}\ \bibnamefont {Nikoli{\ifmmode\acute{c}\else\'{c}\fi}}},\ }\href {\doibase 10.1021/acs.nanolett.7b02511} {\bibfield  {journal} {\bibinfo  {journal} {Nano Lett.}\ }\textbf {\bibinfo {volume} {17}},\ \bibinfo {pages} {5626} (\bibinfo {year} {2017})}\BibitemShut {NoStop}%
\bibitem [{\citenamefont {Mellnik}\ \emph {et~al.}(2014)\citenamefont {Mellnik}, \citenamefont {Lee}, \citenamefont {Richardella}, \citenamefont {Grab}, \citenamefont {Mintun}, \citenamefont {Fischer}, \citenamefont {Vaezi}, \citenamefont {Manchon}, \citenamefont {Kim}, \citenamefont {Samarth},\ and\ \citenamefont {Ralph}}]{ti_ralph}%
  \BibitemOpen
  \bibfield  {author} {\bibinfo {author} {\bibfnamefont {A.~R.}\ \bibnamefont {Mellnik}}, \bibinfo {author} {\bibfnamefont {J.~S.}\ \bibnamefont {Lee}}, \bibinfo {author} {\bibfnamefont {A.}~\bibnamefont {Richardella}}, \bibinfo {author} {\bibfnamefont {J.~L.}\ \bibnamefont {Grab}}, \bibinfo {author} {\bibfnamefont {P.~J.}\ \bibnamefont {Mintun}}, \bibinfo {author} {\bibfnamefont {M.~H.}\ \bibnamefont {Fischer}}, \bibinfo {author} {\bibfnamefont {A.}~\bibnamefont {Vaezi}}, \bibinfo {author} {\bibfnamefont {A.}~\bibnamefont {Manchon}}, \bibinfo {author} {\bibfnamefont {E.-A.}\ \bibnamefont {Kim}}, \bibinfo {author} {\bibfnamefont {N.}~\bibnamefont {Samarth}}, \ and\ \bibinfo {author} {\bibfnamefont {D.~C.}\ \bibnamefont {Ralph}},\ }\href {\doibase 10.1038/nature13534} {\bibfield  {journal} {\bibinfo  {journal} {Nature}\ }\textbf {\bibinfo {volume} {511}},\ \bibinfo {pages} {449} (\bibinfo {year} {2014})}\BibitemShut {NoStop}%
\bibitem [{\citenamefont {Fan}\ \emph {et~al.}(2014)\citenamefont {Fan}, \citenamefont {Upadhyaya}, \citenamefont {Kou}, \citenamefont {Lang}, \citenamefont {Takei}, \citenamefont {Wang}, \citenamefont {Tang}, \citenamefont {He}, \citenamefont {Chang}, \citenamefont {Montazeri}, \citenamefont {Yu}, \citenamefont {Jiang}, \citenamefont {Nie}, \citenamefont {Schwartz}, \citenamefont {Tserkovnyak},\ and\ \citenamefont {Wang}}]{mag_doped_ti_kang_wang}%
  \BibitemOpen
  \bibfield  {author} {\bibinfo {author} {\bibfnamefont {Y.}~\bibnamefont {Fan}}, \bibinfo {author} {\bibfnamefont {P.}~\bibnamefont {Upadhyaya}}, \bibinfo {author} {\bibfnamefont {X.}~\bibnamefont {Kou}}, \bibinfo {author} {\bibfnamefont {M.}~\bibnamefont {Lang}}, \bibinfo {author} {\bibfnamefont {S.}~\bibnamefont {Takei}}, \bibinfo {author} {\bibfnamefont {Z.}~\bibnamefont {Wang}}, \bibinfo {author} {\bibfnamefont {J.}~\bibnamefont {Tang}}, \bibinfo {author} {\bibfnamefont {L.}~\bibnamefont {He}}, \bibinfo {author} {\bibfnamefont {L.-T.}\ \bibnamefont {Chang}}, \bibinfo {author} {\bibfnamefont {M.}~\bibnamefont {Montazeri}}, \bibinfo {author} {\bibfnamefont {G.}~\bibnamefont {Yu}}, \bibinfo {author} {\bibfnamefont {W.}~\bibnamefont {Jiang}}, \bibinfo {author} {\bibfnamefont {T.}~\bibnamefont {Nie}}, \bibinfo {author} {\bibfnamefont {R.~N.}\ \bibnamefont {Schwartz}}, \bibinfo {author} {\bibfnamefont {Y.}~\bibnamefont {Tserkovnyak}}, \ and\ \bibinfo {author} {\bibfnamefont {K.~L.}\ \bibnamefont {Wang}},\
  }\href {\doibase 10.1038/nmat3973} {\bibfield  {journal} {\bibinfo  {journal} {Nat. Mater.}\ }\textbf {\bibinfo {volume} {13}},\ \bibinfo {pages} {699} (\bibinfo {year} {2014})}\BibitemShut {NoStop}%
\bibitem [{\citenamefont {Han}\ \emph {et~al.}(2017)\citenamefont {Han}, \citenamefont {Richardella}, \citenamefont {Siddiqui}, \citenamefont {Finley}, \citenamefont {Samarth},\ and\ \citenamefont {Liu}}]{ti_TbCo_Liu}%
  \BibitemOpen
  \bibfield  {author} {\bibinfo {author} {\bibfnamefont {J.}~\bibnamefont {Han}}, \bibinfo {author} {\bibfnamefont {A.}~\bibnamefont {Richardella}}, \bibinfo {author} {\bibfnamefont {S.~A.}\ \bibnamefont {Siddiqui}}, \bibinfo {author} {\bibfnamefont {J.}~\bibnamefont {Finley}}, \bibinfo {author} {\bibfnamefont {N.}~\bibnamefont {Samarth}}, \ and\ \bibinfo {author} {\bibfnamefont {L.}~\bibnamefont {Liu}},\ }\href {\doibase 10.1103/PhysRevLett.119.077702} {\bibfield  {journal} {\bibinfo  {journal} {Phys. Rev. Lett.}\ }\textbf {\bibinfo {volume} {119}},\ \bibinfo {pages} {077702} (\bibinfo {year} {2017})}\BibitemShut {NoStop}%
\bibitem [{\citenamefont {Khang}\ \emph {et~al.}(2018)\citenamefont {Khang}, \citenamefont {Ueda},\ and\ \citenamefont {Hai}}]{Pham1}%
  \BibitemOpen
  \bibfield  {author} {\bibinfo {author} {\bibfnamefont {N.~H.~D.}\ \bibnamefont {Khang}}, \bibinfo {author} {\bibfnamefont {Y.}~\bibnamefont {Ueda}}, \ and\ \bibinfo {author} {\bibfnamefont {P.~N.}\ \bibnamefont {Hai}},\ }\href {\doibase 10.1038/s41563-018-0137-y} {\bibfield  {journal} {\bibinfo  {journal} {Nat. Mater.}\ }\textbf {\bibinfo {volume} {17}},\ \bibinfo {pages} {808} (\bibinfo {year} {2018})}\BibitemShut {NoStop}%
\bibitem [{\citenamefont {Wulf}\ and\ \citenamefont {McKee}(1995)}]{memory_wall}%
  \BibitemOpen
  \bibfield  {author} {\bibinfo {author} {\bibfnamefont {W.~A.}\ \bibnamefont {Wulf}}\ and\ \bibinfo {author} {\bibfnamefont {S.~A.}\ \bibnamefont {McKee}},\ }\href@noop {} {\bibfield  {journal} {\bibinfo  {journal} {ACM SIGARCH computer architecture news}\ }\textbf {\bibinfo {volume} {23}},\ \bibinfo {pages} {20} (\bibinfo {year} {1995})}\BibitemShut {NoStop}%
\bibitem [{\citenamefont {Zou}\ \emph {et~al.}(2021)\citenamefont {Zou}, \citenamefont {Xu}, \citenamefont {Chen}, \citenamefont {Yan},\ and\ \citenamefont {Han}}]{zou2021breaking}%
  \BibitemOpen
  \bibfield  {author} {\bibinfo {author} {\bibfnamefont {X.}~\bibnamefont {Zou}}, \bibinfo {author} {\bibfnamefont {S.}~\bibnamefont {Xu}}, \bibinfo {author} {\bibfnamefont {X.}~\bibnamefont {Chen}}, \bibinfo {author} {\bibfnamefont {L.}~\bibnamefont {Yan}}, \ and\ \bibinfo {author} {\bibfnamefont {Y.}~\bibnamefont {Han}},\ }\href@noop {} {\bibfield  {journal} {\bibinfo  {journal} {Science China Information Sciences}\ }\textbf {\bibinfo {volume} {64}},\ \bibinfo {pages} {160404} (\bibinfo {year} {2021})}\BibitemShut {NoStop}%
\bibitem [{\citenamefont {Patterson}\ \emph {et~al.}(1997)\citenamefont {Patterson}, \citenamefont {Anderson}, \citenamefont {Cardwell}, \citenamefont {Fromm}, \citenamefont {Keeton}, \citenamefont {Kozyrakis}, \citenamefont {Thomas},\ and\ \citenamefont {Yelick}}]{early_pim}%
  \BibitemOpen
  \bibfield  {author} {\bibinfo {author} {\bibfnamefont {D.}~\bibnamefont {Patterson}}, \bibinfo {author} {\bibfnamefont {T.}~\bibnamefont {Anderson}}, \bibinfo {author} {\bibfnamefont {N.}~\bibnamefont {Cardwell}}, \bibinfo {author} {\bibfnamefont {R.}~\bibnamefont {Fromm}}, \bibinfo {author} {\bibfnamefont {K.}~\bibnamefont {Keeton}}, \bibinfo {author} {\bibfnamefont {C.}~\bibnamefont {Kozyrakis}}, \bibinfo {author} {\bibfnamefont {R.}~\bibnamefont {Thomas}}, \ and\ \bibinfo {author} {\bibfnamefont {K.}~\bibnamefont {Yelick}},\ }\href {\doibase 10.1109/40.592312} {\bibfield  {journal} {\bibinfo  {journal} {IEEE Micro}\ }\textbf {\bibinfo {volume} {17}},\ \bibinfo {pages} {34} (\bibinfo {year} {1997})}\BibitemShut {NoStop}%
\bibitem [{\citenamefont {Ghose}\ \emph {et~al.}(2019)\citenamefont {Ghose}, \citenamefont {Boroumand}, \citenamefont {Kim}, \citenamefont {G{\ifmmode\acute{o}\else\'{o}\fi}mez-Luna},\ and\ \citenamefont {Mutlu}}]{pim_mutlu}%
  \BibitemOpen
  \bibfield  {author} {\bibinfo {author} {\bibfnamefont {S.}~\bibnamefont {Ghose}}, \bibinfo {author} {\bibfnamefont {A.}~\bibnamefont {Boroumand}}, \bibinfo {author} {\bibfnamefont {J.~S.}\ \bibnamefont {Kim}}, \bibinfo {author} {\bibfnamefont {J.}~\bibnamefont {G{\ifmmode\acute{o}\else\'{o}\fi}mez-Luna}}, \ and\ \bibinfo {author} {\bibfnamefont {O.}~\bibnamefont {Mutlu}},\ }\href {\doibase 10.1147/JRD.2019.2934048} {\bibfield  {journal} {\bibinfo  {journal} {IBM J. Res. Dev.}\ }\textbf {\bibinfo {volume} {63}},\ \bibinfo {pages} {1} (\bibinfo {year} {2019})}\BibitemShut {NoStop}%
\bibitem [{\citenamefont {Verma}\ \emph {et~al.}(2019)\citenamefont {Verma}, \citenamefont {Jia}, \citenamefont {Valavi}, \citenamefont {Tang}, \citenamefont {Ozatay}, \citenamefont {Chen}, \citenamefont {Zhang},\ and\ \citenamefont {Deaville}}]{pim_princeton}%
  \BibitemOpen
  \bibfield  {author} {\bibinfo {author} {\bibfnamefont {N.}~\bibnamefont {Verma}}, \bibinfo {author} {\bibfnamefont {H.}~\bibnamefont {Jia}}, \bibinfo {author} {\bibfnamefont {H.}~\bibnamefont {Valavi}}, \bibinfo {author} {\bibfnamefont {Y.}~\bibnamefont {Tang}}, \bibinfo {author} {\bibfnamefont {M.}~\bibnamefont {Ozatay}}, \bibinfo {author} {\bibfnamefont {L.-Y.}\ \bibnamefont {Chen}}, \bibinfo {author} {\bibfnamefont {B.}~\bibnamefont {Zhang}}, \ and\ \bibinfo {author} {\bibfnamefont {P.}~\bibnamefont {Deaville}},\ }\href {\doibase 10.1109/MSSC.2019.2922889} {\bibfield  {journal} {\bibinfo  {journal} {IEEE Solid-State Circuits Mag.}\ }\textbf {\bibinfo {volume} {11}},\ \bibinfo {pages} {43} (\bibinfo {year} {2019})}\BibitemShut {NoStop}%
\bibitem [{\citenamefont {Mannocci}\ \emph {et~al.}(2023)\citenamefont {Mannocci}, \citenamefont {Farronato}, \citenamefont {Lepri}, \citenamefont {Cattaneo}, \citenamefont {Glukhov}, \citenamefont {Sun},\ and\ \citenamefont {Ielmini}}]{pim_survey}%
  \BibitemOpen
  \bibfield  {author} {\bibinfo {author} {\bibfnamefont {P.}~\bibnamefont {Mannocci}}, \bibinfo {author} {\bibfnamefont {M.}~\bibnamefont {Farronato}}, \bibinfo {author} {\bibfnamefont {N.}~\bibnamefont {Lepri}}, \bibinfo {author} {\bibfnamefont {L.}~\bibnamefont {Cattaneo}}, \bibinfo {author} {\bibfnamefont {A.}~\bibnamefont {Glukhov}}, \bibinfo {author} {\bibfnamefont {Z.}~\bibnamefont {Sun}}, \ and\ \bibinfo {author} {\bibfnamefont {D.}~\bibnamefont {Ielmini}},\ }\href {\doibase 10.1063/5.0136403} {\bibfield  {journal} {\bibinfo  {journal} {APL Mach. Learn.}\ }\textbf {\bibinfo {volume} {1}} (\bibinfo {year} {2023}),\ 10.1063/5.0136403}\BibitemShut {NoStop}%
\bibitem [{\citenamefont {Mutlu}\ \emph {et~al.}(2019)\citenamefont {Mutlu}, \citenamefont {Ghose}, \citenamefont {G{\ifmmode\acute{o}\else\'{o}\fi}mez-Luna},\ and\ \citenamefont {Ausavarungnirun}}]{Mutlu2019Jun}%
  \BibitemOpen
  \bibfield  {author} {\bibinfo {author} {\bibfnamefont {O.}~\bibnamefont {Mutlu}}, \bibinfo {author} {\bibfnamefont {S.}~\bibnamefont {Ghose}}, \bibinfo {author} {\bibfnamefont {J.}~\bibnamefont {G{\ifmmode\acute{o}\else\'{o}\fi}mez-Luna}}, \ and\ \bibinfo {author} {\bibfnamefont {R.}~\bibnamefont {Ausavarungnirun}},\ }\href {\doibase 10.1016/j.micpro.2019.01.009} {\bibfield  {journal} {\bibinfo  {journal} {Microprocess. Microsyst.}\ }\textbf {\bibinfo {volume} {67}},\ \bibinfo {pages} {28} (\bibinfo {year} {2019})}\BibitemShut {NoStop}%
\bibitem [{\citenamefont {Parveen}\ \emph {et~al.}(2017)\citenamefont {Parveen}, \citenamefont {Angizi}, \citenamefont {He},\ and\ \citenamefont {Fan}}]{Parveen}%
  \BibitemOpen
  \bibfield  {author} {\bibinfo {author} {\bibfnamefont {F.}~\bibnamefont {Parveen}}, \bibinfo {author} {\bibfnamefont {S.}~\bibnamefont {Angizi}}, \bibinfo {author} {\bibfnamefont {Z.}~\bibnamefont {He}}, \ and\ \bibinfo {author} {\bibfnamefont {D.}~\bibnamefont {Fan}},\ }in\ \href {\doibase 10.1109/ISLPED.2017.8009200} {\emph {\bibinfo {booktitle} {2017 IEEE/ACM International Symposium on Low Power Electronics and Design (ISLPED)}}}\ (\bibinfo {year} {2017})\ pp.\ \bibinfo {pages} {1--6}\BibitemShut {NoStop}%
\bibitem [{\citenamefont {Lee}\ \emph {et~al.}(2019)\citenamefont {Lee}, \citenamefont {Park},\ and\ \citenamefont {Kim}}]{Boolean_1}%
  \BibitemOpen
  \bibfield  {author} {\bibinfo {author} {\bibfnamefont {J.}~\bibnamefont {Lee}}, \bibinfo {author} {\bibfnamefont {B.-G.}\ \bibnamefont {Park}}, \ and\ \bibinfo {author} {\bibfnamefont {Y.}~\bibnamefont {Kim}},\ }\href {\doibase 10.1109/LED.2019.2928335} {\bibfield  {journal} {\bibinfo  {journal} {IEEE Electron Device Lett.}\ }\textbf {\bibinfo {volume} {40}},\ \bibinfo {pages} {1358} (\bibinfo {year} {2019})}\BibitemShut {NoStop}%
\bibitem [{\citenamefont {Xie}\ \emph {et~al.}(2021)\citenamefont {Xie}, \citenamefont {Liang}, \citenamefont {Gu}, \citenamefont {Basak}, \citenamefont {Deng}, \citenamefont {Liang}, \citenamefont {Hu},\ and\ \citenamefont {Xie}}]{sparse_matrix}%
  \BibitemOpen
  \bibfield  {author} {\bibinfo {author} {\bibfnamefont {X.}~\bibnamefont {Xie}}, \bibinfo {author} {\bibfnamefont {Z.}~\bibnamefont {Liang}}, \bibinfo {author} {\bibfnamefont {P.}~\bibnamefont {Gu}}, \bibinfo {author} {\bibfnamefont {A.}~\bibnamefont {Basak}}, \bibinfo {author} {\bibfnamefont {L.}~\bibnamefont {Deng}}, \bibinfo {author} {\bibfnamefont {L.}~\bibnamefont {Liang}}, \bibinfo {author} {\bibfnamefont {X.}~\bibnamefont {Hu}}, \ and\ \bibinfo {author} {\bibfnamefont {Y.}~\bibnamefont {Xie}},\ }in\ \href {\doibase 10.1109/HPCA51647.2021.00055} {\emph {\bibinfo {booktitle} {2021 IEEE International Symposium on High-Performance Computer Architecture (HPCA)}}}\ (\bibinfo {year} {2021})\ pp.\ \bibinfo {pages} {570--583}\BibitemShut {NoStop}%
\bibitem [{\citenamefont {Fan}\ \emph {et~al.}(2017)\citenamefont {Fan}, \citenamefont {Angizi},\ and\ \citenamefont {He}}]{pim_spintronics}%
  \BibitemOpen
  \bibfield  {author} {\bibinfo {author} {\bibfnamefont {D.}~\bibnamefont {Fan}}, \bibinfo {author} {\bibfnamefont {S.}~\bibnamefont {Angizi}}, \ and\ \bibinfo {author} {\bibfnamefont {Z.}~\bibnamefont {He}},\ }in\ \href {\doibase 10.1109/ISVLSI.2017.116} {\emph {\bibinfo {booktitle} {2017 IEEE Computer Society Annual Symposium on VLSI (ISVLSI)}}}\ (\bibinfo {year} {2017})\ pp.\ \bibinfo {pages} {683--688}\BibitemShut {NoStop}%
\bibitem [{\citenamefont {He}\ \emph {et~al.}(2018)\citenamefont {He}, \citenamefont {Zhang}, \citenamefont {Angizi}, \citenamefont {Gong},\ and\ \citenamefont {Fan}}]{sot_pim_fan}%
  \BibitemOpen
  \bibfield  {author} {\bibinfo {author} {\bibfnamefont {Z.}~\bibnamefont {He}}, \bibinfo {author} {\bibfnamefont {Y.}~\bibnamefont {Zhang}}, \bibinfo {author} {\bibfnamefont {S.}~\bibnamefont {Angizi}}, \bibinfo {author} {\bibfnamefont {B.}~\bibnamefont {Gong}}, \ and\ \bibinfo {author} {\bibfnamefont {D.}~\bibnamefont {Fan}},\ }\href {\doibase 10.1109/TMSCS.2018.2836967} {\bibfield  {journal} {\bibinfo  {journal} {IEEE Trans. Multi-Scale Comput. Syst.}\ }\textbf {\bibinfo {volume} {4}},\ \bibinfo {pages} {676} (\bibinfo {year} {2018})}\BibitemShut {NoStop}%
\bibitem [{\citenamefont {Liu}\ \emph {et~al.}(2012{\natexlab{a}})\citenamefont {Liu}, \citenamefont {Pai}, \citenamefont {Li}, \citenamefont {Tseng}, \citenamefont {Ralph},\ and\ \citenamefont {Buhrman}}]{Liu_sot_1}%
  \BibitemOpen
  \bibfield  {author} {\bibinfo {author} {\bibfnamefont {L.}~\bibnamefont {Liu}}, \bibinfo {author} {\bibfnamefont {C.-F.}\ \bibnamefont {Pai}}, \bibinfo {author} {\bibfnamefont {Y.}~\bibnamefont {Li}}, \bibinfo {author} {\bibfnamefont {H.~W.}\ \bibnamefont {Tseng}}, \bibinfo {author} {\bibfnamefont {D.~C.}\ \bibnamefont {Ralph}}, \ and\ \bibinfo {author} {\bibfnamefont {R.~A.}\ \bibnamefont {Buhrman}},\ }\href {\doibase 10.1126/science.1218197} {\bibfield  {journal} {\bibinfo  {journal} {Science}\ }\textbf {\bibinfo {volume} {336}},\ \bibinfo {pages} {555} (\bibinfo {year} {2012}{\natexlab{a}})}\BibitemShut {NoStop}%
\bibitem [{\citenamefont {Liu}\ \emph {et~al.}(2012{\natexlab{b}})\citenamefont {Liu}, \citenamefont {Lee}, \citenamefont {Gudmundsen}, \citenamefont {Ralph},\ and\ \citenamefont {Buhrman}}]{Liu_sot_2}%
  \BibitemOpen
  \bibfield  {author} {\bibinfo {author} {\bibfnamefont {L.}~\bibnamefont {Liu}}, \bibinfo {author} {\bibfnamefont {O.~J.}\ \bibnamefont {Lee}}, \bibinfo {author} {\bibfnamefont {T.~J.}\ \bibnamefont {Gudmundsen}}, \bibinfo {author} {\bibfnamefont {D.~C.}\ \bibnamefont {Ralph}}, \ and\ \bibinfo {author} {\bibfnamefont {R.~A.}\ \bibnamefont {Buhrman}},\ }\href {\doibase 10.1103/PhysRevLett.109.096602} {\bibfield  {journal} {\bibinfo  {journal} {Phys. Rev. Lett.}\ }\textbf {\bibinfo {volume} {109}},\ \bibinfo {pages} {096602} (\bibinfo {year} {2012}{\natexlab{b}})}\BibitemShut {NoStop}%
\bibitem [{\citenamefont {Miron}\ \emph {et~al.}(2011)\citenamefont {Miron}, \citenamefont {Garello}, \citenamefont {Gaudin}, \citenamefont {Zermatten}, \citenamefont {Costache}, \citenamefont {Auffret}, \citenamefont {Bandiera}, \citenamefont {Rodmacq}, \citenamefont {Schuhl},\ and\ \citenamefont {Gambardella}}]{Miron2011Aug}%
  \BibitemOpen
  \bibfield  {author} {\bibinfo {author} {\bibfnamefont {I.~M.}\ \bibnamefont {Miron}}, \bibinfo {author} {\bibfnamefont {K.}~\bibnamefont {Garello}}, \bibinfo {author} {\bibfnamefont {G.}~\bibnamefont {Gaudin}}, \bibinfo {author} {\bibfnamefont {P.-J.}\ \bibnamefont {Zermatten}}, \bibinfo {author} {\bibfnamefont {M.~V.}\ \bibnamefont {Costache}}, \bibinfo {author} {\bibfnamefont {S.}~\bibnamefont {Auffret}}, \bibinfo {author} {\bibfnamefont {S.}~\bibnamefont {Bandiera}}, \bibinfo {author} {\bibfnamefont {B.}~\bibnamefont {Rodmacq}}, \bibinfo {author} {\bibfnamefont {A.}~\bibnamefont {Schuhl}}, \ and\ \bibinfo {author} {\bibfnamefont {P.}~\bibnamefont {Gambardella}},\ }\href {\doibase 10.1038/nature10309} {\bibfield  {journal} {\bibinfo  {journal} {Nature}\ }\textbf {\bibinfo {volume} {476}},\ \bibinfo {pages} {189} (\bibinfo {year} {2011})}\BibitemShut {NoStop}%
\bibitem [{\citenamefont {Prenat}\ \emph {et~al.}(2015)\citenamefont {Prenat}, \citenamefont {Jabeur}, \citenamefont {Di~Pendina}, \citenamefont {Boulle},\ and\ \citenamefont {Gaudin}}]{Prenat2015}%
  \BibitemOpen
  \bibfield  {author} {\bibinfo {author} {\bibfnamefont {G.}~\bibnamefont {Prenat}}, \bibinfo {author} {\bibfnamefont {K.}~\bibnamefont {Jabeur}}, \bibinfo {author} {\bibfnamefont {G.}~\bibnamefont {Di~Pendina}}, \bibinfo {author} {\bibfnamefont {O.}~\bibnamefont {Boulle}}, \ and\ \bibinfo {author} {\bibfnamefont {G.}~\bibnamefont {Gaudin}},\ }in\ \href {\doibase 10.1007/978-3-319-15180-9_4} {\emph {\bibinfo {booktitle} {{Spintronics-based Computing}}}}\ (\bibinfo  {publisher} {Springer},\ \bibinfo {address} {Cham, Switzerland},\ \bibinfo {year} {2015})\ pp.\ \bibinfo {pages} {145--157}\BibitemShut {NoStop}%
\bibitem [{\citenamefont {Ahmed}\ \emph {et~al.}(2017)\citenamefont {Ahmed}, \citenamefont {Zhao}, \citenamefont {Mankalale}, \citenamefont {Sapatnekar}, \citenamefont {Wang},\ and\ \citenamefont {Kim}}]{Ahmed2017Oct}%
  \BibitemOpen
  \bibfield  {author} {\bibinfo {author} {\bibfnamefont {I.}~\bibnamefont {Ahmed}}, \bibinfo {author} {\bibfnamefont {Z.}~\bibnamefont {Zhao}}, \bibinfo {author} {\bibfnamefont {M.~G.}\ \bibnamefont {Mankalale}}, \bibinfo {author} {\bibfnamefont {S.~S.}\ \bibnamefont {Sapatnekar}}, \bibinfo {author} {\bibfnamefont {J.-P.}\ \bibnamefont {Wang}}, \ and\ \bibinfo {author} {\bibfnamefont {C.~H.}\ \bibnamefont {Kim}},\ }\href {\doibase 10.1109/JXCDC.2017.2762699} {\bibfield  {journal} {\bibinfo  {journal} {IEEE J. Explor. Solid-State Comput. Devices Circuits}\ }\textbf {\bibinfo {volume} {3}},\ \bibinfo {pages} {74} (\bibinfo {year} {2017})}\BibitemShut {NoStop}%
\bibitem [{\citenamefont {Shao}\ \emph {et~al.}(2021)\citenamefont {Shao}, \citenamefont {Li}, \citenamefont {Liu}, \citenamefont {Yang}, \citenamefont {Fukami}, \citenamefont {Razavi}, \citenamefont {Wu}, \citenamefont {Wang}, \citenamefont {Freimuth}, \citenamefont {Mokrousov}, \citenamefont {Stiles}, \citenamefont {Emori}, \citenamefont {Hoffmann}, \citenamefont {{\AA}kerman}, \citenamefont {Roy}, \citenamefont {Wang}, \citenamefont {Yang}, \citenamefont {Garello},\ and\ \citenamefont {Zhang}}]{Shao2021May}%
  \BibitemOpen
  \bibfield  {author} {\bibinfo {author} {\bibfnamefont {Q.}~\bibnamefont {Shao}}, \bibinfo {author} {\bibfnamefont {P.}~\bibnamefont {Li}}, \bibinfo {author} {\bibfnamefont {L.}~\bibnamefont {Liu}}, \bibinfo {author} {\bibfnamefont {H.}~\bibnamefont {Yang}}, \bibinfo {author} {\bibfnamefont {S.}~\bibnamefont {Fukami}}, \bibinfo {author} {\bibfnamefont {A.}~\bibnamefont {Razavi}}, \bibinfo {author} {\bibfnamefont {H.}~\bibnamefont {Wu}}, \bibinfo {author} {\bibfnamefont {K.}~\bibnamefont {Wang}}, \bibinfo {author} {\bibfnamefont {F.}~\bibnamefont {Freimuth}}, \bibinfo {author} {\bibfnamefont {Y.}~\bibnamefont {Mokrousov}}, \bibinfo {author} {\bibfnamefont {M.~D.}\ \bibnamefont {Stiles}}, \bibinfo {author} {\bibfnamefont {S.}~\bibnamefont {Emori}}, \bibinfo {author} {\bibfnamefont {A.}~\bibnamefont {Hoffmann}}, \bibinfo {author} {\bibfnamefont {J.}~\bibnamefont {{\AA}kerman}}, \bibinfo {author} {\bibfnamefont {K.}~\bibnamefont {Roy}}, \bibinfo {author} {\bibfnamefont {J.-P.}\ \bibnamefont {Wang}}, \bibinfo
  {author} {\bibfnamefont {S.-H.}\ \bibnamefont {Yang}}, \bibinfo {author} {\bibfnamefont {K.}~\bibnamefont {Garello}}, \ and\ \bibinfo {author} {\bibfnamefont {W.}~\bibnamefont {Zhang}},\ }\href {\doibase 10.1109/TMAG.2021.3078583} {\bibfield  {journal} {\bibinfo  {journal} {IEEE Trans. Magn.}\ }\textbf {\bibinfo {volume} {57}},\ \bibinfo {pages} {ArticleSequenceNumber:800439} (\bibinfo {year} {2021})}\BibitemShut {NoStop}%
\bibitem [{\citenamefont {Garello}\ \emph {et~al.}(2018)\citenamefont {Garello}, \citenamefont {Yasin}, \citenamefont {Couet}, \citenamefont {Souriau}, \citenamefont {Swerts}, \citenamefont {Rao}, \citenamefont {Van~Beek}, \citenamefont {Kim}, \citenamefont {Liu}, \citenamefont {Kundu}, \citenamefont {Tsvetanova}, \citenamefont {Croes}, \citenamefont {Jossart}, \citenamefont {Grimaldi}, \citenamefont {Baumgartner}, \citenamefont {Crotti}, \citenamefont {Fumémont}, \citenamefont {Gambardella},\ and\ \citenamefont {Kar}}]{SOT_cache}%
  \BibitemOpen
  \bibfield  {author} {\bibinfo {author} {\bibfnamefont {K.}~\bibnamefont {Garello}}, \bibinfo {author} {\bibfnamefont {F.}~\bibnamefont {Yasin}}, \bibinfo {author} {\bibfnamefont {S.}~\bibnamefont {Couet}}, \bibinfo {author} {\bibfnamefont {L.}~\bibnamefont {Souriau}}, \bibinfo {author} {\bibfnamefont {J.}~\bibnamefont {Swerts}}, \bibinfo {author} {\bibfnamefont {S.}~\bibnamefont {Rao}}, \bibinfo {author} {\bibfnamefont {S.}~\bibnamefont {Van~Beek}}, \bibinfo {author} {\bibfnamefont {W.}~\bibnamefont {Kim}}, \bibinfo {author} {\bibfnamefont {E.}~\bibnamefont {Liu}}, \bibinfo {author} {\bibfnamefont {S.}~\bibnamefont {Kundu}}, \bibinfo {author} {\bibfnamefont {D.}~\bibnamefont {Tsvetanova}}, \bibinfo {author} {\bibfnamefont {K.}~\bibnamefont {Croes}}, \bibinfo {author} {\bibfnamefont {N.}~\bibnamefont {Jossart}}, \bibinfo {author} {\bibfnamefont {E.}~\bibnamefont {Grimaldi}}, \bibinfo {author} {\bibfnamefont {M.}~\bibnamefont {Baumgartner}}, \bibinfo {author} {\bibfnamefont {D.}~\bibnamefont {Crotti}},
  \bibinfo {author} {\bibfnamefont {A.}~\bibnamefont {Fumémont}}, \bibinfo {author} {\bibfnamefont {P.}~\bibnamefont {Gambardella}}, \ and\ \bibinfo {author} {\bibfnamefont {G.}~\bibnamefont {Kar}},\ }in\ \href {\doibase 10.1109/VLSIC.2018.8502269} {\emph {\bibinfo {booktitle} {2018 IEEE Symposium on VLSI Circuits}}}\ (\bibinfo {year} {2018})\ pp.\ \bibinfo {pages} {81--82}\BibitemShut {NoStop}%
\bibitem [{\citenamefont {Oboril}\ \emph {et~al.}(2015)\citenamefont {Oboril}, \citenamefont {Bishnoi}, \citenamefont {Ebrahimi},\ and\ \citenamefont {Tahoori}}]{Oboril2015Jan}%
  \BibitemOpen
  \bibfield  {author} {\bibinfo {author} {\bibfnamefont {F.}~\bibnamefont {Oboril}}, \bibinfo {author} {\bibfnamefont {R.}~\bibnamefont {Bishnoi}}, \bibinfo {author} {\bibfnamefont {M.}~\bibnamefont {Ebrahimi}}, \ and\ \bibinfo {author} {\bibfnamefont {M.~B.}\ \bibnamefont {Tahoori}},\ }\href {\doibase 10.1109/TCAD.2015.2391254} {\bibfield  {journal} {\bibinfo  {journal} {IEEE Trans. Comput. Aided Des. Integr. Circuits Syst.}\ }\textbf {\bibinfo {volume} {34}},\ \bibinfo {pages} {367} (\bibinfo {year} {2015})}\BibitemShut {NoStop}%
\bibitem [{\citenamefont {Kim}\ \emph {et~al.}(2022)\citenamefont {Kim}, \citenamefont {Jang}, \citenamefont {Kang}, \citenamefont {Park}, \citenamefont {Lee},\ and\ \citenamefont {Park}}]{sot_matrix}%
  \BibitemOpen
  \bibfield  {author} {\bibinfo {author} {\bibfnamefont {T.}~\bibnamefont {Kim}}, \bibinfo {author} {\bibfnamefont {Y.}~\bibnamefont {Jang}}, \bibinfo {author} {\bibfnamefont {M.-G.}\ \bibnamefont {Kang}}, \bibinfo {author} {\bibfnamefont {B.-G.}\ \bibnamefont {Park}}, \bibinfo {author} {\bibfnamefont {K.-J.}\ \bibnamefont {Lee}}, \ and\ \bibinfo {author} {\bibfnamefont {J.}~\bibnamefont {Park}},\ }\href {\doibase 10.1109/TC.2022.3155277} {\bibfield  {journal} {\bibinfo  {journal} {IEEE Transactions on Computers}\ }\textbf {\bibinfo {volume} {71}},\ \bibinfo {pages} {2816} (\bibinfo {year} {2022})}\BibitemShut {NoStop}%
\bibitem [{\citenamefont {Ostwal}\ \emph {et~al.}(2017)\citenamefont {Ostwal}, \citenamefont {Penumatcha}, \citenamefont {Hung}, \citenamefont {Kent},\ and\ \citenamefont {Appenzeller}}]{Pt_sot}%
  \BibitemOpen
  \bibfield  {author} {\bibinfo {author} {\bibfnamefont {V.}~\bibnamefont {Ostwal}}, \bibinfo {author} {\bibfnamefont {A.}~\bibnamefont {Penumatcha}}, \bibinfo {author} {\bibfnamefont {Y.-M.}\ \bibnamefont {Hung}}, \bibinfo {author} {\bibfnamefont {A.~D.}\ \bibnamefont {Kent}}, \ and\ \bibinfo {author} {\bibfnamefont {J.}~\bibnamefont {Appenzeller}},\ }\href {\doibase 10.1063/1.4994711} {\bibfield  {journal} {\bibinfo  {journal} {J. Appl. Phys.}\ }\textbf {\bibinfo {volume} {122}} (\bibinfo {year} {2017}),\ 10.1063/1.4994711}\BibitemShut {NoStop}%
\bibitem [{\citenamefont {Pai}\ \emph {et~al.}(2012)\citenamefont {Pai}, \citenamefont {Liu}, \citenamefont {Li}, \citenamefont {Tseng}, \citenamefont {Ralph},\ and\ \citenamefont {Buhrman}}]{sot_w}%
  \BibitemOpen
  \bibfield  {author} {\bibinfo {author} {\bibfnamefont {C.-F.}\ \bibnamefont {Pai}}, \bibinfo {author} {\bibfnamefont {L.}~\bibnamefont {Liu}}, \bibinfo {author} {\bibfnamefont {Y.}~\bibnamefont {Li}}, \bibinfo {author} {\bibfnamefont {H.~W.}\ \bibnamefont {Tseng}}, \bibinfo {author} {\bibfnamefont {D.~C.}\ \bibnamefont {Ralph}}, \ and\ \bibinfo {author} {\bibfnamefont {R.~A.}\ \bibnamefont {Buhrman}},\ }\href {\doibase 10.1063/1.4753947} {\bibfield  {journal} {\bibinfo  {journal} {Appl. Phys. Lett.}\ }\textbf {\bibinfo {volume} {101}} (\bibinfo {year} {2012}),\ 10.1063/1.4753947}\BibitemShut {NoStop}%
\bibitem [{\citenamefont {Fan}\ \emph {et~al.}(2022)\citenamefont {Fan}, \citenamefont {Khang}, \citenamefont {Nakano},\ and\ \citenamefont {Hai}}]{Pham2}%
  \BibitemOpen
  \bibfield  {author} {\bibinfo {author} {\bibfnamefont {T.}~\bibnamefont {Fan}}, \bibinfo {author} {\bibfnamefont {N.~H.~D.}\ \bibnamefont {Khang}}, \bibinfo {author} {\bibfnamefont {S.}~\bibnamefont {Nakano}}, \ and\ \bibinfo {author} {\bibfnamefont {P.~N.}\ \bibnamefont {Hai}},\ }\href {\doibase 10.1038/s41598-022-06779-3} {\bibfield  {journal} {\bibinfo  {journal} {Sci. Rep.}\ }\textbf {\bibinfo {volume} {12}},\ \bibinfo {pages} {1} (\bibinfo {year} {2022})}\BibitemShut {NoStop}%
\bibitem [{\citenamefont {Wang}\ \emph {et~al.}(2017)\citenamefont {Wang}, \citenamefont {Zhu}, \citenamefont {Wu}, \citenamefont {Yang}, \citenamefont {Yu}, \citenamefont {Ramaswamy}, \citenamefont {Mishra}, \citenamefont {Shi}, \citenamefont {Elyasi}, \citenamefont {Teo}, \citenamefont {Wu},\ and\ \citenamefont {Yang}}]{ti_current_ratio}%
  \BibitemOpen
  \bibfield  {author} {\bibinfo {author} {\bibfnamefont {Y.}~\bibnamefont {Wang}}, \bibinfo {author} {\bibfnamefont {D.}~\bibnamefont {Zhu}}, \bibinfo {author} {\bibfnamefont {Y.}~\bibnamefont {Wu}}, \bibinfo {author} {\bibfnamefont {Y.}~\bibnamefont {Yang}}, \bibinfo {author} {\bibfnamefont {J.}~\bibnamefont {Yu}}, \bibinfo {author} {\bibfnamefont {R.}~\bibnamefont {Ramaswamy}}, \bibinfo {author} {\bibfnamefont {R.}~\bibnamefont {Mishra}}, \bibinfo {author} {\bibfnamefont {S.}~\bibnamefont {Shi}}, \bibinfo {author} {\bibfnamefont {M.}~\bibnamefont {Elyasi}}, \bibinfo {author} {\bibfnamefont {K.-L.}\ \bibnamefont {Teo}}, \bibinfo {author} {\bibfnamefont {Y.}~\bibnamefont {Wu}}, \ and\ \bibinfo {author} {\bibfnamefont {H.}~\bibnamefont {Yang}},\ }\href {\doibase 10.1038/s41467-017-01583-4} {\bibfield  {journal} {\bibinfo  {journal} {Nat. Commun.}\ }\textbf {\bibinfo {volume} {8}},\ \bibinfo {pages} {1} (\bibinfo {year} {2017})}\BibitemShut {NoStop}%
\bibitem [{\citenamefont {Dc}\ \emph {et~al.}(2018)\citenamefont {Dc}, \citenamefont {Grassi}, \citenamefont {Chen}, \citenamefont {Jamali}, \citenamefont {Reifsnyder~Hickey}, \citenamefont {Zhang}, \citenamefont {Zhao}, \citenamefont {Li}, \citenamefont {Quarterman}, \citenamefont {Lv}, \citenamefont {Li}, \citenamefont {Manchon}, \citenamefont {Mkhoyan}, \citenamefont {Low},\ and\ \citenamefont {Wang}}]{ti_Jian_Ping}%
  \BibitemOpen
  \bibfield  {author} {\bibinfo {author} {\bibfnamefont {M.}~\bibnamefont {Dc}}, \bibinfo {author} {\bibfnamefont {R.}~\bibnamefont {Grassi}}, \bibinfo {author} {\bibfnamefont {J.-Y.}\ \bibnamefont {Chen}}, \bibinfo {author} {\bibfnamefont {M.}~\bibnamefont {Jamali}}, \bibinfo {author} {\bibfnamefont {D.}~\bibnamefont {Reifsnyder~Hickey}}, \bibinfo {author} {\bibfnamefont {D.}~\bibnamefont {Zhang}}, \bibinfo {author} {\bibfnamefont {Z.}~\bibnamefont {Zhao}}, \bibinfo {author} {\bibfnamefont {H.}~\bibnamefont {Li}}, \bibinfo {author} {\bibfnamefont {P.}~\bibnamefont {Quarterman}}, \bibinfo {author} {\bibfnamefont {Y.}~\bibnamefont {Lv}}, \bibinfo {author} {\bibfnamefont {M.}~\bibnamefont {Li}}, \bibinfo {author} {\bibfnamefont {A.}~\bibnamefont {Manchon}}, \bibinfo {author} {\bibfnamefont {K.~A.}\ \bibnamefont {Mkhoyan}}, \bibinfo {author} {\bibfnamefont {T.}~\bibnamefont {Low}}, \ and\ \bibinfo {author} {\bibfnamefont {J.-P.}\ \bibnamefont {Wang}},\ }\href {\doibase 10.1038/s41563-018-0136-z} {\bibfield
  {journal} {\bibinfo  {journal} {Nat. Mater.}\ }\textbf {\bibinfo {volume} {17}},\ \bibinfo {pages} {800} (\bibinfo {year} {2018})}\BibitemShut {NoStop}%
\bibitem [{\citenamefont {Hasan}\ and\ \citenamefont {Kane}(2010)}]{zahid_hasan_princeton}%
  \BibitemOpen
  \bibfield  {author} {\bibinfo {author} {\bibfnamefont {M.~Z.}\ \bibnamefont {Hasan}}\ and\ \bibinfo {author} {\bibfnamefont {C.~L.}\ \bibnamefont {Kane}},\ }\href {\doibase 10.1103/RevModPhys.82.3045} {\bibfield  {journal} {\bibinfo  {journal} {Rev. Mod. Phys.}\ }\textbf {\bibinfo {volume} {82}},\ \bibinfo {pages} {3045} (\bibinfo {year} {2010})}\BibitemShut {NoStop}%
\bibitem [{\citenamefont {Ando}(2013)}]{ti_review}%
  \BibitemOpen
  \bibfield  {author} {\bibinfo {author} {\bibfnamefont {Y.}~\bibnamefont {Ando}},\ }\href {\doibase 10.7566/JPSJ.82.102001} {\bibfield  {journal} {\bibinfo  {journal} {J. Phys. Soc. Jpn.}\ }\textbf {\bibinfo {volume} {82}},\ \bibinfo {pages} {102001} (\bibinfo {year} {2013})}\BibitemShut {NoStop}%
\bibitem [{\citenamefont {Tokura}\ \emph {et~al.}(2019)\citenamefont {Tokura}, \citenamefont {Yasuda},\ and\ \citenamefont {Tsukazaki}}]{ti_review_2}%
  \BibitemOpen
  \bibfield  {author} {\bibinfo {author} {\bibfnamefont {Y.}~\bibnamefont {Tokura}}, \bibinfo {author} {\bibfnamefont {K.}~\bibnamefont {Yasuda}}, \ and\ \bibinfo {author} {\bibfnamefont {A.}~\bibnamefont {Tsukazaki}},\ }\href {\doibase 10.1038/s42254-018-0011-5} {\bibfield  {journal} {\bibinfo  {journal} {Nat. Rev. Phys.}\ }\textbf {\bibinfo {volume} {1}},\ \bibinfo {pages} {126} (\bibinfo {year} {2019})}\BibitemShut {NoStop}%
\bibitem [{\citenamefont {Bahramy}\ \emph {et~al.}(2012)\citenamefont {Bahramy}, \citenamefont {King}, \citenamefont {de~la Torre}, \citenamefont {Chang}, \citenamefont {Shi}, \citenamefont {Patthey}, \citenamefont {Balakrishnan}, \citenamefont {Hofmann}, \citenamefont {Arita}, \citenamefont {Nagaosa},\ and\ \citenamefont {Baumberger}}]{ti_tss_1}%
  \BibitemOpen
  \bibfield  {author} {\bibinfo {author} {\bibfnamefont {M.~S.}\ \bibnamefont {Bahramy}}, \bibinfo {author} {\bibfnamefont {P.~D.~C.}\ \bibnamefont {King}}, \bibinfo {author} {\bibfnamefont {A.}~\bibnamefont {de~la Torre}}, \bibinfo {author} {\bibfnamefont {J.}~\bibnamefont {Chang}}, \bibinfo {author} {\bibfnamefont {M.}~\bibnamefont {Shi}}, \bibinfo {author} {\bibfnamefont {L.}~\bibnamefont {Patthey}}, \bibinfo {author} {\bibfnamefont {G.}~\bibnamefont {Balakrishnan}}, \bibinfo {author} {\bibfnamefont {{\relax Ph}.}~\bibnamefont {Hofmann}}, \bibinfo {author} {\bibfnamefont {R.}~\bibnamefont {Arita}}, \bibinfo {author} {\bibfnamefont {N.}~\bibnamefont {Nagaosa}}, \ and\ \bibinfo {author} {\bibfnamefont {F.}~\bibnamefont {Baumberger}},\ }\href {\doibase 10.1038/ncomms2162} {\bibfield  {journal} {\bibinfo  {journal} {Nat. Commun.}\ }\textbf {\bibinfo {volume} {3}},\ \bibinfo {pages} {1} (\bibinfo {year} {2012})}\BibitemShut {NoStop}%
\bibitem [{\citenamefont {Barua}\ \emph {et~al.}(2014)\citenamefont {Barua}, \citenamefont {Rajeev},\ and\ \citenamefont {Gupta}}]{ti_tss_2}%
  \BibitemOpen
  \bibfield  {author} {\bibinfo {author} {\bibfnamefont {S.}~\bibnamefont {Barua}}, \bibinfo {author} {\bibfnamefont {K.~P.}\ \bibnamefont {Rajeev}}, \ and\ \bibinfo {author} {\bibfnamefont {A.~K.}\ \bibnamefont {Gupta}},\ }\href {\doibase 10.1088/0953-8984/27/1/015601} {\bibfield  {journal} {\bibinfo  {journal} {J. Phys.: Condens. Matter}\ }\textbf {\bibinfo {volume} {27}},\ \bibinfo {pages} {015601} (\bibinfo {year} {2014})}\BibitemShut {NoStop}%
\bibitem [{\citenamefont {Wang}\ \emph {et~al.}(2015)\citenamefont {Wang}, \citenamefont {Deorani}, \citenamefont {Banerjee}, \citenamefont {Koirala}, \citenamefont {Brahlek}, \citenamefont {Oh},\ and\ \citenamefont {Yang}}]{ti_tss_3}%
  \BibitemOpen
  \bibfield  {author} {\bibinfo {author} {\bibfnamefont {Y.}~\bibnamefont {Wang}}, \bibinfo {author} {\bibfnamefont {P.}~\bibnamefont {Deorani}}, \bibinfo {author} {\bibfnamefont {K.}~\bibnamefont {Banerjee}}, \bibinfo {author} {\bibfnamefont {N.}~\bibnamefont {Koirala}}, \bibinfo {author} {\bibfnamefont {M.}~\bibnamefont {Brahlek}}, \bibinfo {author} {\bibfnamefont {S.}~\bibnamefont {Oh}}, \ and\ \bibinfo {author} {\bibfnamefont {H.}~\bibnamefont {Yang}},\ }\href {\doibase 10.1103/PhysRevLett.114.257202} {\bibfield  {journal} {\bibinfo  {journal} {Phys. Rev. Lett.}\ }\textbf {\bibinfo {volume} {114}},\ \bibinfo {pages} {257202} (\bibinfo {year} {2015})}\BibitemShut {NoStop}%
\bibitem [{\citenamefont {Wu}\ \emph {et~al.}(2021)\citenamefont {Wu}, \citenamefont {Chen}, \citenamefont {Zhang}, \citenamefont {He}, \citenamefont {Nance}, \citenamefont {Guo}, \citenamefont {Sasaki}, \citenamefont {Shirokura}, \citenamefont {Hai}, \citenamefont {Fang}, \citenamefont {Razavi}, \citenamefont {Wong}, \citenamefont {Wen}, \citenamefont {Ma}, \citenamefont {Yu}, \citenamefont {Carman}, \citenamefont {Han}, \citenamefont {Zhang},\ and\ \citenamefont {Wang}}]{ti_memory}%
  \BibitemOpen
  \bibfield  {author} {\bibinfo {author} {\bibfnamefont {H.}~\bibnamefont {Wu}}, \bibinfo {author} {\bibfnamefont {A.}~\bibnamefont {Chen}}, \bibinfo {author} {\bibfnamefont {P.}~\bibnamefont {Zhang}}, \bibinfo {author} {\bibfnamefont {H.}~\bibnamefont {He}}, \bibinfo {author} {\bibfnamefont {J.}~\bibnamefont {Nance}}, \bibinfo {author} {\bibfnamefont {C.}~\bibnamefont {Guo}}, \bibinfo {author} {\bibfnamefont {J.}~\bibnamefont {Sasaki}}, \bibinfo {author} {\bibfnamefont {T.}~\bibnamefont {Shirokura}}, \bibinfo {author} {\bibfnamefont {P.~N.}\ \bibnamefont {Hai}}, \bibinfo {author} {\bibfnamefont {B.}~\bibnamefont {Fang}}, \bibinfo {author} {\bibfnamefont {S.~A.}\ \bibnamefont {Razavi}}, \bibinfo {author} {\bibfnamefont {K.}~\bibnamefont {Wong}}, \bibinfo {author} {\bibfnamefont {Y.}~\bibnamefont {Wen}}, \bibinfo {author} {\bibfnamefont {Y.}~\bibnamefont {Ma}}, \bibinfo {author} {\bibfnamefont {G.}~\bibnamefont {Yu}}, \bibinfo {author} {\bibfnamefont {G.~P.}\ \bibnamefont {Carman}}, \bibinfo {author}
  {\bibfnamefont {X.}~\bibnamefont {Han}}, \bibinfo {author} {\bibfnamefont {X.}~\bibnamefont {Zhang}}, \ and\ \bibinfo {author} {\bibfnamefont {K.~L.}\ \bibnamefont {Wang}},\ }\href {\doibase 10.1038/s41467-021-26478-3} {\bibfield  {journal} {\bibinfo  {journal} {Nat. Commun.}\ }\textbf {\bibinfo {volume} {12}},\ \bibinfo {pages} {1} (\bibinfo {year} {2021})}\BibitemShut {NoStop}%
\bibitem [{\citenamefont {Vakili}\ \emph {et~al.}(2022)\citenamefont {Vakili}, \citenamefont {Ganguly}, \citenamefont {de~Coster}, \citenamefont {Neupane},\ and\ \citenamefont {Ghosh}}]{hamed_fm_ti}%
  \BibitemOpen
  \bibfield  {author} {\bibinfo {author} {\bibfnamefont {H.}~\bibnamefont {Vakili}}, \bibinfo {author} {\bibfnamefont {S.}~\bibnamefont {Ganguly}}, \bibinfo {author} {\bibfnamefont {G.~J.}\ \bibnamefont {de~Coster}}, \bibinfo {author} {\bibfnamefont {M.~R.}\ \bibnamefont {Neupane}}, \ and\ \bibinfo {author} {\bibfnamefont {A.~W.}\ \bibnamefont {Ghosh}},\ }\href {\doibase 10.1021/acsnano.2c05645} {\bibfield  {journal} {\bibinfo  {journal} {ACS Nano}\ }\textbf {\bibinfo {volume} {16}},\ \bibinfo {pages} {20222} (\bibinfo {year} {2022})}\BibitemShut {NoStop}%
\bibitem [{\citenamefont {Atulasimha}\ and\ \citenamefont {Bandyopadhyay}(2010)}]{strain_2010}%
  \BibitemOpen
  \bibfield  {author} {\bibinfo {author} {\bibfnamefont {J.}~\bibnamefont {Atulasimha}}\ and\ \bibinfo {author} {\bibfnamefont {S.}~\bibnamefont {Bandyopadhyay}},\ }\href {\doibase 10.1063/1.3506690} {\bibfield  {journal} {\bibinfo  {journal} {Appl. Phys. Lett.}\ }\textbf {\bibinfo {volume} {97}} (\bibinfo {year} {2010}),\ 10.1063/1.3506690}\BibitemShut {NoStop}%
\bibitem [{\citenamefont {Roy}\ \emph {et~al.}(2011{\natexlab{b}})\citenamefont {Roy}, \citenamefont {Bandyopadhyay},\ and\ \citenamefont {Atulasimha}}]{strain_2011}%
  \BibitemOpen
  \bibfield  {author} {\bibinfo {author} {\bibfnamefont {K.}~\bibnamefont {Roy}}, \bibinfo {author} {\bibfnamefont {S.}~\bibnamefont {Bandyopadhyay}}, \ and\ \bibinfo {author} {\bibfnamefont {J.}~\bibnamefont {Atulasimha}},\ }\href {\doibase 10.1103/PhysRevB.83.224412} {\bibfield  {journal} {\bibinfo  {journal} {Phys. Rev. B}\ }\textbf {\bibinfo {volume} {83}},\ \bibinfo {pages} {224412} (\bibinfo {year} {2011}{\natexlab{b}})}\BibitemShut {NoStop}%
\bibitem [{\citenamefont {Trassin}(2015)}]{strain_2016}%
  \BibitemOpen
  \bibfield  {author} {\bibinfo {author} {\bibfnamefont {M.}~\bibnamefont {Trassin}},\ }\href {\doibase 10.1088/0953-8984/28/3/033001} {\bibfield  {journal} {\bibinfo  {journal} {J. Phys.: Condens. Matter}\ }\textbf {\bibinfo {volume} {28}},\ \bibinfo {pages} {033001} (\bibinfo {year} {2015})}\BibitemShut {NoStop}%
\bibitem [{\citenamefont {Bandyopadhyay}\ \emph {et~al.}(2021)\citenamefont {Bandyopadhyay}, \citenamefont {Atulasimha},\ and\ \citenamefont {Barman}}]{sb_review}%
  \BibitemOpen
  \bibfield  {author} {\bibinfo {author} {\bibfnamefont {S.}~\bibnamefont {Bandyopadhyay}}, \bibinfo {author} {\bibfnamefont {J.}~\bibnamefont {Atulasimha}}, \ and\ \bibinfo {author} {\bibfnamefont {A.}~\bibnamefont {Barman}},\ }\href {\doibase 10.1063/5.0062993} {\bibfield  {journal} {\bibinfo  {journal} {Appl. Phys. Rev.}\ }\textbf {\bibinfo {volume} {8}} (\bibinfo {year} {2021}),\ 10.1063/5.0062993}\BibitemShut {NoStop}%
\bibitem [{\citenamefont {Manipatruni}\ \emph {et~al.}(2018)\citenamefont {Manipatruni}, \citenamefont {Nikonov}, \citenamefont {Lin}, \citenamefont {Prasad}, \citenamefont {Huang}, \citenamefont {Damodaran}, \citenamefont {Chen}, \citenamefont {Ramesh},\ and\ \citenamefont {Young}}]{vcma}%
  \BibitemOpen
  \bibfield  {author} {\bibinfo {author} {\bibfnamefont {S.}~\bibnamefont {Manipatruni}}, \bibinfo {author} {\bibfnamefont {D.~E.}\ \bibnamefont {Nikonov}}, \bibinfo {author} {\bibfnamefont {C.-C.}\ \bibnamefont {Lin}}, \bibinfo {author} {\bibfnamefont {B.}~\bibnamefont {Prasad}}, \bibinfo {author} {\bibfnamefont {Y.-L.}\ \bibnamefont {Huang}}, \bibinfo {author} {\bibfnamefont {A.~R.}\ \bibnamefont {Damodaran}}, \bibinfo {author} {\bibfnamefont {Z.}~\bibnamefont {Chen}}, \bibinfo {author} {\bibfnamefont {R.}~\bibnamefont {Ramesh}}, \ and\ \bibinfo {author} {\bibfnamefont {I.~A.}\ \bibnamefont {Young}},\ }\href {\doibase 10.1126/sciadv.aat4229} {\bibfield  {journal} {\bibinfo  {journal} {Sci. Adv.}\ }\textbf {\bibinfo {volume} {4}} (\bibinfo {year} {2018}),\ 10.1126/sciadv.aat4229}\BibitemShut {NoStop}%
\bibitem [{\citenamefont {Semenov}\ \emph {et~al.}(2012)\citenamefont {Semenov}, \citenamefont {Duan},\ and\ \citenamefont {Kim}}]{voltage}%
  \BibitemOpen
  \bibfield  {author} {\bibinfo {author} {\bibfnamefont {Y.~G.}\ \bibnamefont {Semenov}}, \bibinfo {author} {\bibfnamefont {X.}~\bibnamefont {Duan}}, \ and\ \bibinfo {author} {\bibfnamefont {K.~W.}\ \bibnamefont {Kim}},\ }\href {\doibase 10.1103/PhysRevB.86.161406} {\bibfield  {journal} {\bibinfo  {journal} {Phys. Rev. B}\ }\textbf {\bibinfo {volume} {86}},\ \bibinfo {pages} {161406} (\bibinfo {year} {2012})}\BibitemShut {NoStop}%
\bibitem [{\citenamefont {Roy}\ \emph {et~al.}(2012)\citenamefont {Roy}, \citenamefont {Bandyopadhyay},\ and\ \citenamefont {Atulasimha}}]{Roy2012Jul}%
  \BibitemOpen
  \bibfield  {author} {\bibinfo {author} {\bibfnamefont {K.}~\bibnamefont {Roy}}, \bibinfo {author} {\bibfnamefont {S.}~\bibnamefont {Bandyopadhyay}}, \ and\ \bibinfo {author} {\bibfnamefont {J.}~\bibnamefont {Atulasimha}},\ }\href {\doibase 10.1063/1.4737792} {\bibfield  {journal} {\bibinfo  {journal} {J. Appl. Phys.}\ }\textbf {\bibinfo {volume} {112}} (\bibinfo {year} {2012}),\ 10.1063/1.4737792}\BibitemShut {NoStop}%
\bibitem [{\citenamefont {Fashami}\ \emph {et~al.}(2011)\citenamefont {Fashami}, \citenamefont {Roy}, \citenamefont {Atulasimha},\ and\ \citenamefont {Bandyopadhyay}}]{Fashami2011Mar}%
  \BibitemOpen
  \bibfield  {author} {\bibinfo {author} {\bibfnamefont {M.~S.}\ \bibnamefont {Fashami}}, \bibinfo {author} {\bibfnamefont {K.}~\bibnamefont {Roy}}, \bibinfo {author} {\bibfnamefont {J.}~\bibnamefont {Atulasimha}}, \ and\ \bibinfo {author} {\bibfnamefont {S.}~\bibnamefont {Bandyopadhyay}},\ }\href {\doibase 10.1088/0957-4484/22/15/155201} {\bibfield  {journal} {\bibinfo  {journal} {Nanotechnology}\ }\textbf {\bibinfo {volume} {22}},\ \bibinfo {pages} {155201} (\bibinfo {year} {2011})}\BibitemShut {NoStop}%
\bibitem [{\citenamefont {Winters}\ \emph {et~al.}(2019)\citenamefont {Winters}, \citenamefont {Abeed}, \citenamefont {Sahoo}, \citenamefont {Barman},\ and\ \citenamefont {Bandyopadhyay}}]{Winters2019Sep}%
  \BibitemOpen
  \bibfield  {author} {\bibinfo {author} {\bibfnamefont {D.}~\bibnamefont {Winters}}, \bibinfo {author} {\bibfnamefont {M.~A.}\ \bibnamefont {Abeed}}, \bibinfo {author} {\bibfnamefont {S.}~\bibnamefont {Sahoo}}, \bibinfo {author} {\bibfnamefont {A.}~\bibnamefont {Barman}}, \ and\ \bibinfo {author} {\bibfnamefont {S.}~\bibnamefont {Bandyopadhyay}},\ }\href {\doibase 10.1103/PhysRevApplied.12.034010} {\bibfield  {journal} {\bibinfo  {journal} {Phys. Rev. Appl.}\ }\textbf {\bibinfo {volume} {12}},\ \bibinfo {pages} {034010} (\bibinfo {year} {2019})}\BibitemShut {NoStop}%
\bibitem [{\citenamefont {Mishra}\ \emph {et~al.}(2023)\citenamefont {Mishra}, \citenamefont {Halavath},\ and\ \citenamefont {Bhuktare}}]{Mishra2023Sep}%
  \BibitemOpen
  \bibfield  {author} {\bibinfo {author} {\bibfnamefont {P.~K.}\ \bibnamefont {Mishra}}, \bibinfo {author} {\bibfnamefont {N.}~\bibnamefont {Halavath}}, \ and\ \bibinfo {author} {\bibfnamefont {S.}~\bibnamefont {Bhuktare}},\ }\href {\doibase 10.1063/5.0161990} {\bibfield  {journal} {\bibinfo  {journal} {J. Appl. Phys.}\ }\textbf {\bibinfo {volume} {134}} (\bibinfo {year} {2023}),\ 10.1063/5.0161990}\BibitemShut {NoStop}%
\bibitem [{\citenamefont {Yu}\ \emph {et~al.}(2010)\citenamefont {Yu}, \citenamefont {Zhang}, \citenamefont {Zhang}, \citenamefont {Zhang}, \citenamefont {Dai},\ and\ \citenamefont {Fang}}]{sci_dirac}%
  \BibitemOpen
  \bibfield  {author} {\bibinfo {author} {\bibfnamefont {R.}~\bibnamefont {Yu}}, \bibinfo {author} {\bibfnamefont {W.}~\bibnamefont {Zhang}}, \bibinfo {author} {\bibfnamefont {H.-J.}\ \bibnamefont {Zhang}}, \bibinfo {author} {\bibfnamefont {S.-C.}\ \bibnamefont {Zhang}}, \bibinfo {author} {\bibfnamefont {X.}~\bibnamefont {Dai}}, \ and\ \bibinfo {author} {\bibfnamefont {Z.}~\bibnamefont {Fang}},\ }\href {\doibase 10.1126/science.1187485} {\bibfield  {journal} {\bibinfo  {journal} {Science}\ }\textbf {\bibinfo {volume} {329}},\ \bibinfo {pages} {61} (\bibinfo {year} {2010})}\BibitemShut {NoStop}%
\bibitem [{\citenamefont {Cho}\ \emph {et~al.}(2011)\citenamefont {Cho}, \citenamefont {Butch}, \citenamefont {Paglione},\ and\ \citenamefont {Fuhrer}}]{Cho2011May}%
  \BibitemOpen
  \bibfield  {author} {\bibinfo {author} {\bibfnamefont {S.}~\bibnamefont {Cho}}, \bibinfo {author} {\bibfnamefont {N.~P.}\ \bibnamefont {Butch}}, \bibinfo {author} {\bibfnamefont {J.}~\bibnamefont {Paglione}}, \ and\ \bibinfo {author} {\bibfnamefont {M.~S.}\ \bibnamefont {Fuhrer}},\ }\href {\doibase 10.1021/nl200017f} {\bibfield  {journal} {\bibinfo  {journal} {Nano Lett.}\ }\textbf {\bibinfo {volume} {11}},\ \bibinfo {pages} {1925} (\bibinfo {year} {2011})}\BibitemShut {NoStop}%
\bibitem [{\citenamefont {Xie}\ \emph {et~al.}(2023)\citenamefont {Xie}, \citenamefont {Vakili}, \citenamefont {Ganguly},\ and\ \citenamefont {Ghosh}}]{yunkun_ti}%
  \BibitemOpen
  \bibfield  {author} {\bibinfo {author} {\bibfnamefont {Y.}~\bibnamefont {Xie}}, \bibinfo {author} {\bibfnamefont {H.}~\bibnamefont {Vakili}}, \bibinfo {author} {\bibfnamefont {S.}~\bibnamefont {Ganguly}}, \ and\ \bibinfo {author} {\bibfnamefont {A.~W.}\ \bibnamefont {Ghosh}},\ }\href {\doibase 10.1038/s41598-023-35623-5} {\bibfield  {journal} {\bibinfo  {journal} {Sci. Rep.}\ }\textbf {\bibinfo {volume} {13}},\ \bibinfo {pages} {1} (\bibinfo {year} {2023})}\BibitemShut {NoStop}%
\bibitem [{\citenamefont {Fukami}\ \emph {et~al.}(2016)\citenamefont {Fukami}, \citenamefont {Anekawa}, \citenamefont {Zhang},\ and\ \citenamefont {Ohno}}]{Fukami2016Jul}%
  \BibitemOpen
  \bibfield  {author} {\bibinfo {author} {\bibfnamefont {S.}~\bibnamefont {Fukami}}, \bibinfo {author} {\bibfnamefont {T.}~\bibnamefont {Anekawa}}, \bibinfo {author} {\bibfnamefont {C.}~\bibnamefont {Zhang}}, \ and\ \bibinfo {author} {\bibfnamefont {H.}~\bibnamefont {Ohno}},\ }\href {\doibase 10.1038/nnano.2016.29} {\bibfield  {journal} {\bibinfo  {journal} {Nat. Nanotechnol.}\ }\textbf {\bibinfo {volume} {11}},\ \bibinfo {pages} {621} (\bibinfo {year} {2016})}\BibitemShut {NoStop}%
\bibitem [{\citenamefont {Yu}\ \emph {et~al.}(2014{\natexlab{a}})\citenamefont {Yu}, \citenamefont {Upadhyaya}, \citenamefont {Fan}, \citenamefont {Alzate}, \citenamefont {Jiang}, \citenamefont {Wong}, \citenamefont {Takei}, \citenamefont {Bender}, \citenamefont {Chang}, \citenamefont {Jiang}, \citenamefont {Lang}, \citenamefont {Tang}, \citenamefont {Wang}, \citenamefont {Tserkovnyak}, \citenamefont {Amiri},\ and\ \citenamefont {Wang}}]{field-free1}%
  \BibitemOpen
  \bibfield  {author} {\bibinfo {author} {\bibfnamefont {G.}~\bibnamefont {Yu}}, \bibinfo {author} {\bibfnamefont {P.}~\bibnamefont {Upadhyaya}}, \bibinfo {author} {\bibfnamefont {Y.}~\bibnamefont {Fan}}, \bibinfo {author} {\bibfnamefont {J.~G.}\ \bibnamefont {Alzate}}, \bibinfo {author} {\bibfnamefont {W.}~\bibnamefont {Jiang}}, \bibinfo {author} {\bibfnamefont {K.~L.}\ \bibnamefont {Wong}}, \bibinfo {author} {\bibfnamefont {S.}~\bibnamefont {Takei}}, \bibinfo {author} {\bibfnamefont {S.~A.}\ \bibnamefont {Bender}}, \bibinfo {author} {\bibfnamefont {L.-T.}\ \bibnamefont {Chang}}, \bibinfo {author} {\bibfnamefont {Y.}~\bibnamefont {Jiang}}, \bibinfo {author} {\bibfnamefont {M.}~\bibnamefont {Lang}}, \bibinfo {author} {\bibfnamefont {J.}~\bibnamefont {Tang}}, \bibinfo {author} {\bibfnamefont {Y.}~\bibnamefont {Wang}}, \bibinfo {author} {\bibfnamefont {Y.}~\bibnamefont {Tserkovnyak}}, \bibinfo {author} {\bibfnamefont {P.~K.}\ \bibnamefont {Amiri}}, \ and\ \bibinfo {author} {\bibfnamefont {K.~L.}\ \bibnamefont
  {Wang}},\ }\href {\doibase 10.1038/nnano.2014.94} {\bibfield  {journal} {\bibinfo  {journal} {Nat. Nanotechnol.}\ }\textbf {\bibinfo {volume} {9}},\ \bibinfo {pages} {548} (\bibinfo {year} {2014}{\natexlab{a}})}\BibitemShut {NoStop}%
\bibitem [{\citenamefont {Yu}\ \emph {et~al.}(2014{\natexlab{b}})\citenamefont {Yu}, \citenamefont {Chang}, \citenamefont {Akyol}, \citenamefont {Upadhyaya}, \citenamefont {He}, \citenamefont {Li}, \citenamefont {Wong}, \citenamefont {Amiri},\ and\ \citenamefont {Wang}}]{field-free2}%
  \BibitemOpen
  \bibfield  {author} {\bibinfo {author} {\bibfnamefont {G.}~\bibnamefont {Yu}}, \bibinfo {author} {\bibfnamefont {L.-T.}\ \bibnamefont {Chang}}, \bibinfo {author} {\bibfnamefont {M.}~\bibnamefont {Akyol}}, \bibinfo {author} {\bibfnamefont {P.}~\bibnamefont {Upadhyaya}}, \bibinfo {author} {\bibfnamefont {C.}~\bibnamefont {He}}, \bibinfo {author} {\bibfnamefont {X.}~\bibnamefont {Li}}, \bibinfo {author} {\bibfnamefont {K.~L.}\ \bibnamefont {Wong}}, \bibinfo {author} {\bibfnamefont {P.~K.}\ \bibnamefont {Amiri}}, \ and\ \bibinfo {author} {\bibfnamefont {K.~L.}\ \bibnamefont {Wang}},\ }\href {\doibase 10.1063/1.4895735} {\bibfield  {journal} {\bibinfo  {journal} {Appl. Phys. Lett.}\ }\textbf {\bibinfo {volume} {105}} (\bibinfo {year} {2014}{\natexlab{b}}),\ 10.1063/1.4895735}\BibitemShut {NoStop}%
\bibitem [{\citenamefont {Katine}\ \emph {et~al.}(2000)\citenamefont {Katine}, \citenamefont {Albert}, \citenamefont {Buhrman}, \citenamefont {Myers},\ and\ \citenamefont {Ralph}}]{Katine2000Apr}%
  \BibitemOpen
  \bibfield  {author} {\bibinfo {author} {\bibfnamefont {J.~A.}\ \bibnamefont {Katine}}, \bibinfo {author} {\bibfnamefont {F.~J.}\ \bibnamefont {Albert}}, \bibinfo {author} {\bibfnamefont {R.~A.}\ \bibnamefont {Buhrman}}, \bibinfo {author} {\bibfnamefont {E.~B.}\ \bibnamefont {Myers}}, \ and\ \bibinfo {author} {\bibfnamefont {D.~C.}\ \bibnamefont {Ralph}},\ }\href {\doibase 10.1103/PhysRevLett.84.3149} {\bibfield  {journal} {\bibinfo  {journal} {Phys. Rev. Lett.}\ }\textbf {\bibinfo {volume} {84}},\ \bibinfo {pages} {3149} (\bibinfo {year} {2000})}\BibitemShut {NoStop}%
\bibitem [{\citenamefont {Lee}\ \emph {et~al.}(2013)\citenamefont {Lee}, \citenamefont {Lee}, \citenamefont {Min},\ and\ \citenamefont {Lee}}]{Lee2013Mar}%
  \BibitemOpen
  \bibfield  {author} {\bibinfo {author} {\bibfnamefont {K.-S.}\ \bibnamefont {Lee}}, \bibinfo {author} {\bibfnamefont {S.-W.}\ \bibnamefont {Lee}}, \bibinfo {author} {\bibfnamefont {B.-C.}\ \bibnamefont {Min}}, \ and\ \bibinfo {author} {\bibfnamefont {K.-J.}\ \bibnamefont {Lee}},\ }\href {\doibase 10.1063/1.4798288} {\bibfield  {journal} {\bibinfo  {journal} {Appl. Phys. Lett.}\ }\textbf {\bibinfo {volume} {102}} (\bibinfo {year} {2013}),\ 10.1063/1.4798288}\BibitemShut {NoStop}%
\bibitem [{\citenamefont {Yoshino}\ \emph {et~al.}(1984)\citenamefont {Yoshino}, \citenamefont {Takagi}, \citenamefont {Tsunashima}, \citenamefont {Masuda},\ and\ \citenamefont {Uchiyama}}]{Youngs_mod}%
  \BibitemOpen
  \bibfield  {author} {\bibinfo {author} {\bibfnamefont {S.}~\bibnamefont {Yoshino}}, \bibinfo {author} {\bibfnamefont {H.}~\bibnamefont {Takagi}}, \bibinfo {author} {\bibfnamefont {S.}~\bibnamefont {Tsunashima}}, \bibinfo {author} {\bibfnamefont {M.}~\bibnamefont {Masuda}}, \ and\ \bibinfo {author} {\bibfnamefont {S.}~\bibnamefont {Uchiyama}},\ }\href {\doibase 10.1143/JJAP.23.188} {\bibfield  {journal} {\bibinfo  {journal} {Jpn. J. Appl. Phys.}\ }\textbf {\bibinfo {volume} {23}},\ \bibinfo {pages} {188} (\bibinfo {year} {1984})}\BibitemShut {NoStop}%
\bibitem [{\citenamefont {Betz}\ \emph {et~al.}(1999)\citenamefont {Betz}, \citenamefont {Mackay},\ and\ \citenamefont {Givord}}]{magnetostriction}%
  \BibitemOpen
  \bibfield  {author} {\bibinfo {author} {\bibfnamefont {J.}~\bibnamefont {Betz}}, \bibinfo {author} {\bibfnamefont {K.}~\bibnamefont {Mackay}}, \ and\ \bibinfo {author} {\bibfnamefont {D.}~\bibnamefont {Givord}},\ }\href {\doibase 10.1016/S0304-8853(99)00457-6} {\bibfield  {journal} {\bibinfo  {journal} {J. Magn. Magn. Mater.}\ }\textbf {\bibinfo {volume} {207}},\ \bibinfo {pages} {180} (\bibinfo {year} {1999})}\BibitemShut {NoStop}%
\bibitem [{\citenamefont {B{\ifmmode\ddot{u}\else\"{u}\fi}ttner}\ \emph {et~al.}(2018)\citenamefont {B{\ifmmode\ddot{u}\else\"{u}\fi}ttner}, \citenamefont {Lemesh},\ and\ \citenamefont {Beach}}]{TbCo_ku_ms}%
  \BibitemOpen
  \bibfield  {author} {\bibinfo {author} {\bibfnamefont {F.}~\bibnamefont {B{\ifmmode\ddot{u}\else\"{u}\fi}ttner}}, \bibinfo {author} {\bibfnamefont {I.}~\bibnamefont {Lemesh}}, \ and\ \bibinfo {author} {\bibfnamefont {G.~S.~D.}\ \bibnamefont {Beach}},\ }\href {\doibase 10.1038/s41598-018-22242-8} {\bibfield  {journal} {\bibinfo  {journal} {Sci. Rep.}\ }\textbf {\bibinfo {volume} {8}},\ \bibinfo {pages} {1} (\bibinfo {year} {2018})}\BibitemShut {NoStop}%
\bibitem [{\citenamefont {Shukla}\ \emph {et~al.}(2023)\citenamefont {Shukla}, \citenamefont {Heller}, \citenamefont {Morshed}, \citenamefont {Rehm}, \citenamefont {Ghosh}, \citenamefont {Kent},\ and\ \citenamefont {Rakheja}}]{ankit}%
  \BibitemOpen
  \bibfield  {author} {\bibinfo {author} {\bibfnamefont {A.}~\bibnamefont {Shukla}}, \bibinfo {author} {\bibfnamefont {L.}~\bibnamefont {Heller}}, \bibinfo {author} {\bibfnamefont {M.~G.}\ \bibnamefont {Morshed}}, \bibinfo {author} {\bibfnamefont {L.}~\bibnamefont {Rehm}}, \bibinfo {author} {\bibfnamefont {A.~W.}\ \bibnamefont {Ghosh}}, \bibinfo {author} {\bibfnamefont {A.~D.}\ \bibnamefont {Kent}}, \ and\ \bibinfo {author} {\bibfnamefont {S.}~\bibnamefont {Rakheja}},\ }in\ \href {\doibase 10.1109/ISQED57927.2023.10129319} {\emph {\bibinfo {booktitle} {2023 24th International Symposium on Quality Electronic Design (ISQED)}}}\ (\bibinfo {year} {2023})\ pp.\ \bibinfo {pages} {1--10}\BibitemShut {NoStop}%
\bibitem [{\citenamefont {Zhao}\ \emph {et~al.}(2015)\citenamefont {Zhao}, \citenamefont {Jamali}, \citenamefont {Smith},\ and\ \citenamefont {Wang}}]{Zhao2015Mar}%
  \BibitemOpen
  \bibfield  {author} {\bibinfo {author} {\bibfnamefont {Z.}~\bibnamefont {Zhao}}, \bibinfo {author} {\bibfnamefont {M.}~\bibnamefont {Jamali}}, \bibinfo {author} {\bibfnamefont {A.~K.}\ \bibnamefont {Smith}}, \ and\ \bibinfo {author} {\bibfnamefont {J.-P.}\ \bibnamefont {Wang}},\ }\href {\doibase 10.1063/1.4916665} {\bibfield  {journal} {\bibinfo  {journal} {Appl. Phys. Lett.}\ }\textbf {\bibinfo {volume} {106}} (\bibinfo {year} {2015}),\ 10.1063/1.4916665}\BibitemShut {NoStop}%
\bibitem [{\citenamefont {Finley}\ and\ \citenamefont {Liu}(2016)}]{TbCo_param_Liu}%
  \BibitemOpen
  \bibfield  {author} {\bibinfo {author} {\bibfnamefont {J.}~\bibnamefont {Finley}}\ and\ \bibinfo {author} {\bibfnamefont {L.}~\bibnamefont {Liu}},\ }\href {\doibase 10.1103/PhysRevApplied.6.054001} {\bibfield  {journal} {\bibinfo  {journal} {Phys. Rev. Appl.}\ }\textbf {\bibinfo {volume} {6}},\ \bibinfo {pages} {054001} (\bibinfo {year} {2016})}\BibitemShut {NoStop}%
\bibitem [{\citenamefont {Bedau}\ \emph {et~al.}(2010)\citenamefont {Bedau}, \citenamefont {Liu}, \citenamefont {Bouzaglou}, \citenamefont {Kent}, \citenamefont {Sun}, \citenamefont {Katine}, \citenamefont {Fullerton},\ and\ \citenamefont {Mangin}}]{Bedau2010Jan}%
  \BibitemOpen
  \bibfield  {author} {\bibinfo {author} {\bibfnamefont {D.}~\bibnamefont {Bedau}}, \bibinfo {author} {\bibfnamefont {H.}~\bibnamefont {Liu}}, \bibinfo {author} {\bibfnamefont {J.-J.}\ \bibnamefont {Bouzaglou}}, \bibinfo {author} {\bibfnamefont {A.~D.}\ \bibnamefont {Kent}}, \bibinfo {author} {\bibfnamefont {J.~Z.}\ \bibnamefont {Sun}}, \bibinfo {author} {\bibfnamefont {J.~A.}\ \bibnamefont {Katine}}, \bibinfo {author} {\bibfnamefont {E.~E.}\ \bibnamefont {Fullerton}}, \ and\ \bibinfo {author} {\bibfnamefont {S.}~\bibnamefont {Mangin}},\ }\href {\doibase 10.1063/1.3284515} {\bibfield  {journal} {\bibinfo  {journal} {Appl. Phys. Lett.}\ }\textbf {\bibinfo {volume} {96}} (\bibinfo {year} {2010}),\ 10.1063/1.3284515}\BibitemShut {NoStop}%
\bibitem [{\citenamefont {Wang}\ \emph {et~al.}(2018)\citenamefont {Wang}, \citenamefont {Domann}, \citenamefont {Yu}, \citenamefont {Barra}, \citenamefont {Wang},\ and\ \citenamefont {Carman}}]{strain_sot}%
  \BibitemOpen
  \bibfield  {author} {\bibinfo {author} {\bibfnamefont {Q.}~\bibnamefont {Wang}}, \bibinfo {author} {\bibfnamefont {J.}~\bibnamefont {Domann}}, \bibinfo {author} {\bibfnamefont {G.}~\bibnamefont {Yu}}, \bibinfo {author} {\bibfnamefont {A.}~\bibnamefont {Barra}}, \bibinfo {author} {\bibfnamefont {K.~L.}\ \bibnamefont {Wang}}, \ and\ \bibinfo {author} {\bibfnamefont {G.~P.}\ \bibnamefont {Carman}},\ }\href {\doibase 10.1103/PhysRevApplied.10.034052} {\bibfield  {journal} {\bibinfo  {journal} {Phys. Rev. Appl.}\ }\textbf {\bibinfo {volume} {10}},\ \bibinfo {pages} {034052} (\bibinfo {year} {2018})}\BibitemShut {NoStop}%
\bibitem [{\citenamefont {Morshed}\ \emph {et~al.}(2023)\citenamefont {Morshed}, \citenamefont {Rehm}, \citenamefont {Shukla}, \citenamefont {Xie}, \citenamefont {Ganguly}, \citenamefont {Rakheja}, \citenamefont {Kent},\ and\ \citenamefont {Ghosh}}]{golam_mbm}%
  \BibitemOpen
  \bibfield  {author} {\bibinfo {author} {\bibfnamefont {M.~G.}\ \bibnamefont {Morshed}}, \bibinfo {author} {\bibfnamefont {L.}~\bibnamefont {Rehm}}, \bibinfo {author} {\bibfnamefont {A.}~\bibnamefont {Shukla}}, \bibinfo {author} {\bibfnamefont {Y.}~\bibnamefont {Xie}}, \bibinfo {author} {\bibfnamefont {S.}~\bibnamefont {Ganguly}}, \bibinfo {author} {\bibfnamefont {S.}~\bibnamefont {Rakheja}}, \bibinfo {author} {\bibfnamefont {A.~D.}\ \bibnamefont {Kent}}, \ and\ \bibinfo {author} {\bibfnamefont {A.~W.}\ \bibnamefont {Ghosh}},\ }\href {\doibase 10.48550/arXiv.2310.18781} {\bibfield  {journal} {\bibinfo  {journal} {arXiv}\ } (\bibinfo {year} {2023}),\ 10.48550/arXiv.2310.18781},\ \Eprint {http://arxiv.org/abs/2310.18781} {2310.18781} \BibitemShut {NoStop}%
\bibitem [{\citenamefont {Roy}\ \emph {et~al.}(2016)\citenamefont {Roy}, \citenamefont {Pramanik}, \citenamefont {Register},\ and\ \citenamefont {Banerjee}}]{Roy2016Jun}%
  \BibitemOpen
  \bibfield  {author} {\bibinfo {author} {\bibfnamefont {U.}~\bibnamefont {Roy}}, \bibinfo {author} {\bibfnamefont {T.}~\bibnamefont {Pramanik}}, \bibinfo {author} {\bibfnamefont {L.~F.}\ \bibnamefont {Register}}, \ and\ \bibinfo {author} {\bibfnamefont {S.~K.}\ \bibnamefont {Banerjee}},\ }\href {\doibase 10.1109/TMAG.2016.2580532} {\bibfield  {journal} {\bibinfo  {journal} {IEEE Trans. Magn.}\ }\textbf {\bibinfo {volume} {52}},\ \bibinfo {pages} {ArticleSequenceNumber:3402106} (\bibinfo {year} {2016})}\BibitemShut {NoStop}%
\bibitem [{\citenamefont {Yu}\ and\ \citenamefont {Chen}(2016)}]{Yu2016Jun}%
  \BibitemOpen
  \bibfield  {author} {\bibinfo {author} {\bibfnamefont {S.}~\bibnamefont {Yu}}\ and\ \bibinfo {author} {\bibfnamefont {P.-Y.}\ \bibnamefont {Chen}},\ }\href {\doibase 10.1109/MSSC.2016.2546199} {\bibfield  {journal} {\bibinfo  {journal} {IEEE Solid-State Circuits Mag.}\ }\textbf {\bibinfo {volume} {8}},\ \bibinfo {pages} {43} (\bibinfo {year} {2016})}\BibitemShut {NoStop}%
\bibitem [{\citenamefont {Honda}\ \emph {et~al.}(1987)\citenamefont {Honda}, \citenamefont {Nawate}, \citenamefont {Yoshiyama},\ and\ \citenamefont {Kusuda}}]{TbCo_highMs}%
  \BibitemOpen
  \bibfield  {author} {\bibinfo {author} {\bibfnamefont {S.}~\bibnamefont {Honda}}, \bibinfo {author} {\bibfnamefont {M.}~\bibnamefont {Nawate}}, \bibinfo {author} {\bibfnamefont {M.}~\bibnamefont {Yoshiyama}}, \ and\ \bibinfo {author} {\bibfnamefont {T.}~\bibnamefont {Kusuda}},\ }\href {\doibase 10.1109/TMAG.1987.1065727} {\bibfield  {journal} {\bibinfo  {journal} {IEEE Trans. Magn.}\ }\textbf {\bibinfo {volume} {23}},\ \bibinfo {pages} {2952} (\bibinfo {year} {1987})}\BibitemShut {NoStop}%
\bibitem [{\citenamefont {Deorani}\ \emph {et~al.}(2014)\citenamefont {Deorani}, \citenamefont {Son}, \citenamefont {Banerjee}, \citenamefont {Koirala}, \citenamefont {Brahlek}, \citenamefont {Oh},\ and\ \citenamefont {Yang}}]{spin_diffusion}%
  \BibitemOpen
  \bibfield  {author} {\bibinfo {author} {\bibfnamefont {P.}~\bibnamefont {Deorani}}, \bibinfo {author} {\bibfnamefont {J.}~\bibnamefont {Son}}, \bibinfo {author} {\bibfnamefont {K.}~\bibnamefont {Banerjee}}, \bibinfo {author} {\bibfnamefont {N.}~\bibnamefont {Koirala}}, \bibinfo {author} {\bibfnamefont {M.}~\bibnamefont {Brahlek}}, \bibinfo {author} {\bibfnamefont {S.}~\bibnamefont {Oh}}, \ and\ \bibinfo {author} {\bibfnamefont {H.}~\bibnamefont {Yang}},\ }\href {\doibase 10.1103/PhysRevB.90.094403} {\bibfield  {journal} {\bibinfo  {journal} {Phys. Rev. B}\ }\textbf {\bibinfo {volume} {90}},\ \bibinfo {pages} {094403} (\bibinfo {year} {2014})}\BibitemShut {NoStop}%
\bibitem [{\citenamefont {Rehm}\ \emph {et~al.}(2023)\citenamefont {Rehm}, \citenamefont {Capriata}, \citenamefont {Misra}, \citenamefont {Smith}, \citenamefont {Pinarbasi}, \citenamefont {Malm},\ and\ \citenamefont {Kent}}]{laura_smart}%
  \BibitemOpen
  \bibfield  {author} {\bibinfo {author} {\bibfnamefont {L.}~\bibnamefont {Rehm}}, \bibinfo {author} {\bibfnamefont {C.~C.~M.}\ \bibnamefont {Capriata}}, \bibinfo {author} {\bibfnamefont {S.}~\bibnamefont {Misra}}, \bibinfo {author} {\bibfnamefont {J.~D.}\ \bibnamefont {Smith}}, \bibinfo {author} {\bibfnamefont {M.}~\bibnamefont {Pinarbasi}}, \bibinfo {author} {\bibfnamefont {B.~G.}\ \bibnamefont {Malm}}, \ and\ \bibinfo {author} {\bibfnamefont {A.~D.}\ \bibnamefont {Kent}},\ }\href {\doibase 10.1103/PhysRevApplied.19.024035} {\bibfield  {journal} {\bibinfo  {journal} {Phys. Rev. Appl.}\ }\textbf {\bibinfo {volume} {19}},\ \bibinfo {pages} {024035} (\bibinfo {year} {2023})}\BibitemShut {NoStop}%
\bibitem [{\citenamefont {Wei}\ \emph {et~al.}(2019)\citenamefont {Wei}, \citenamefont {Alzate}, \citenamefont {Arslan}, \citenamefont {Brockman}, \citenamefont {Das}, \citenamefont {Fischer}, \citenamefont {Ghani}, \citenamefont {Golonzka}, \citenamefont {Hentges}, \citenamefont {Jahan}, \citenamefont {Jain}, \citenamefont {Lin}, \citenamefont {Meterelliyoz}, \citenamefont {O’Donnell}, \citenamefont {Puls}, \citenamefont {Quintero}, \citenamefont {Sahu}, \citenamefont {Sekhar}, \citenamefont {Vangapaty}, \citenamefont {Wiegand},\ and\ \citenamefont {Hamzaoglu}}]{Wei}%
  \BibitemOpen
  \bibfield  {author} {\bibinfo {author} {\bibfnamefont {L.}~\bibnamefont {Wei}}, \bibinfo {author} {\bibfnamefont {J.~G.}\ \bibnamefont {Alzate}}, \bibinfo {author} {\bibfnamefont {U.}~\bibnamefont {Arslan}}, \bibinfo {author} {\bibfnamefont {J.}~\bibnamefont {Brockman}}, \bibinfo {author} {\bibfnamefont {N.}~\bibnamefont {Das}}, \bibinfo {author} {\bibfnamefont {K.}~\bibnamefont {Fischer}}, \bibinfo {author} {\bibfnamefont {T.}~\bibnamefont {Ghani}}, \bibinfo {author} {\bibfnamefont {O.}~\bibnamefont {Golonzka}}, \bibinfo {author} {\bibfnamefont {P.}~\bibnamefont {Hentges}}, \bibinfo {author} {\bibfnamefont {R.}~\bibnamefont {Jahan}}, \bibinfo {author} {\bibfnamefont {P.}~\bibnamefont {Jain}}, \bibinfo {author} {\bibfnamefont {B.}~\bibnamefont {Lin}}, \bibinfo {author} {\bibfnamefont {M.}~\bibnamefont {Meterelliyoz}}, \bibinfo {author} {\bibfnamefont {J.}~\bibnamefont {O’Donnell}}, \bibinfo {author} {\bibfnamefont {C.}~\bibnamefont {Puls}}, \bibinfo {author} {\bibfnamefont {P.}~\bibnamefont {Quintero}},
  \bibinfo {author} {\bibfnamefont {T.}~\bibnamefont {Sahu}}, \bibinfo {author} {\bibfnamefont {M.}~\bibnamefont {Sekhar}}, \bibinfo {author} {\bibfnamefont {A.}~\bibnamefont {Vangapaty}}, \bibinfo {author} {\bibfnamefont {C.}~\bibnamefont {Wiegand}}, \ and\ \bibinfo {author} {\bibfnamefont {F.}~\bibnamefont {Hamzaoglu}},\ }in\ \href {\doibase 10.1109/ISSCC.2019.8662444} {\emph {\bibinfo {booktitle} {2019 IEEE International Solid-State Circuits Conference - (ISSCC)}}}\ (\bibinfo {year} {2019})\ pp.\ \bibinfo {pages} {214--216}\BibitemShut {NoStop}%
\bibitem [{\citenamefont {Song}\ \emph {et~al.}(2022)\citenamefont {Song}, \citenamefont {Lee}, \citenamefont {Yang}, \citenamefont {Chen}, \citenamefont {Chen}, \citenamefont {Wang}, \citenamefont {Hsin}, \citenamefont {Chang}, \citenamefont {Hsu}, \citenamefont {Li}, \citenamefont {Wei}, \citenamefont {Lee}, \citenamefont {Chang}, \citenamefont {Bao}, \citenamefont {Diaz},\ and\ \citenamefont {Lin}}]{symmetric_current}%
  \BibitemOpen
  \bibfield  {author} {\bibinfo {author} {\bibfnamefont {M.~Y.}\ \bibnamefont {Song}}, \bibinfo {author} {\bibfnamefont {C.~M.}\ \bibnamefont {Lee}}, \bibinfo {author} {\bibfnamefont {S.~Y.}\ \bibnamefont {Yang}}, \bibinfo {author} {\bibfnamefont {G.~L.}\ \bibnamefont {Chen}}, \bibinfo {author} {\bibfnamefont {K.~M.}\ \bibnamefont {Chen}}, \bibinfo {author} {\bibfnamefont {I.~J.}\ \bibnamefont {Wang}}, \bibinfo {author} {\bibfnamefont {Y.~C.}\ \bibnamefont {Hsin}}, \bibinfo {author} {\bibfnamefont {K.~T.}\ \bibnamefont {Chang}}, \bibinfo {author} {\bibfnamefont {C.~F.}\ \bibnamefont {Hsu}}, \bibinfo {author} {\bibfnamefont {S.~H.}\ \bibnamefont {Li}}, \bibinfo {author} {\bibfnamefont {J.~H.}\ \bibnamefont {Wei}}, \bibinfo {author} {\bibfnamefont {T.~Y.}\ \bibnamefont {Lee}}, \bibinfo {author} {\bibfnamefont {M.~F.}\ \bibnamefont {Chang}}, \bibinfo {author} {\bibfnamefont {X.~Y.}\ \bibnamefont {Bao}}, \bibinfo {author} {\bibfnamefont {C.~H.}\ \bibnamefont {Diaz}}, \ and\ \bibinfo {author} {\bibfnamefont
  {S.~J.}\ \bibnamefont {Lin}},\ }in\ \href {\doibase 10.1109/VLSITechnologyandCir46769.2022.9830149} {\emph {\bibinfo {booktitle} {2022 IEEE Symposium on VLSI Technology and Circuits (VLSI Technology and Circuits)}}}\ (\bibinfo {year} {2022})\ pp.\ \bibinfo {pages} {377--378}\BibitemShut {NoStop}%
\bibitem [{\citenamefont {Tozer}(1992)}]{Tozer1992May}%
  \BibitemOpen
  \bibfield  {author} {\bibinfo {author} {\bibfnamefont {R.~C.}\ \bibnamefont {Tozer}},\ }\href {\doibase 10.1088/0957-0233/3/5/011} {\bibfield  {journal} {\bibinfo  {journal} {Meas. Sci. Technol.}\ }\textbf {\bibinfo {volume} {3}},\ \bibinfo {pages} {508} (\bibinfo {year} {1992})}\BibitemShut {NoStop}%
\bibitem [{\citenamefont {Conte}\ \emph {et~al.}(2005)\citenamefont {Conte}, \citenamefont {Giudice}, \citenamefont {Palumbo},\ and\ \citenamefont {Signorello}}]{Conte2005Jan}%
  \BibitemOpen
  \bibfield  {author} {\bibinfo {author} {\bibfnamefont {A.}~\bibnamefont {Conte}}, \bibinfo {author} {\bibfnamefont {G.~L.}\ \bibnamefont {Giudice}}, \bibinfo {author} {\bibfnamefont {G.}~\bibnamefont {Palumbo}}, \ and\ \bibinfo {author} {\bibfnamefont {A.}~\bibnamefont {Signorello}},\ }\href {\doibase 10.1109/JSSC.2004.840985} {\bibfield  {journal} {\bibinfo  {journal} {IEEE J. Solid-State Circuits}\ }\textbf {\bibinfo {volume} {40}},\ \bibinfo {pages} {507} (\bibinfo {year} {2005})}\BibitemShut {NoStop}%
\bibitem [{\citenamefont {Wolf}\ \emph {et~al.}(2010)\citenamefont {Wolf}, \citenamefont {Lu}, \citenamefont {Stan}, \citenamefont {Chen},\ and\ \citenamefont {Treger}}]{sptlgt_chat_1}%
  \BibitemOpen
  \bibfield  {author} {\bibinfo {author} {\bibfnamefont {S.~A.}\ \bibnamefont {Wolf}}, \bibinfo {author} {\bibfnamefont {J.}~\bibnamefont {Lu}}, \bibinfo {author} {\bibfnamefont {M.~R.}\ \bibnamefont {Stan}}, \bibinfo {author} {\bibfnamefont {E.}~\bibnamefont {Chen}}, \ and\ \bibinfo {author} {\bibfnamefont {D.~M.}\ \bibnamefont {Treger}},\ }\href {\doibase 10.1109/JPROC.2010.2064150} {\bibfield  {journal} {\bibinfo  {journal} {Proceedings of the IEEE}\ }\textbf {\bibinfo {volume} {98}},\ \bibinfo {pages} {2155} (\bibinfo {year} {2010})}\BibitemShut {NoStop}%
\bibitem [{\citenamefont {Endoh}\ \emph {et~al.}(2016)\citenamefont {Endoh}, \citenamefont {Koike}, \citenamefont {Ikeda}, \citenamefont {Hanyu},\ and\ \citenamefont {Ohno}}]{sptlgt_chat_2}%
  \BibitemOpen
  \bibfield  {author} {\bibinfo {author} {\bibfnamefont {T.}~\bibnamefont {Endoh}}, \bibinfo {author} {\bibfnamefont {H.}~\bibnamefont {Koike}}, \bibinfo {author} {\bibfnamefont {S.}~\bibnamefont {Ikeda}}, \bibinfo {author} {\bibfnamefont {T.}~\bibnamefont {Hanyu}}, \ and\ \bibinfo {author} {\bibfnamefont {H.}~\bibnamefont {Ohno}},\ }\href {\doibase 10.1109/JETCAS.2016.2547704} {\bibfield  {journal} {\bibinfo  {journal} {IEEE Journal on Emerging and Selected Topics in Circuits and Systems}\ }\textbf {\bibinfo {volume} {6}},\ \bibinfo {pages} {109} (\bibinfo {year} {2016})}\BibitemShut {NoStop}%
\bibitem [{\citenamefont {Sebastian}\ \emph {et~al.}(2020)\citenamefont {Sebastian}, \citenamefont {Le~Gallo}, \citenamefont {Khaddam-Aljameh},\ and\ \citenamefont {Eleftheriou}}]{sptlgt_chat_3}%
  \BibitemOpen
  \bibfield  {author} {\bibinfo {author} {\bibfnamefont {A.}~\bibnamefont {Sebastian}}, \bibinfo {author} {\bibfnamefont {M.}~\bibnamefont {Le~Gallo}}, \bibinfo {author} {\bibfnamefont {R.}~\bibnamefont {Khaddam-Aljameh}}, \ and\ \bibinfo {author} {\bibfnamefont {E.}~\bibnamefont {Eleftheriou}},\ }\href {\doibase 10.1038/s41565-020-0655-z} {\bibfield  {journal} {\bibinfo  {journal} {Nat. Nanotechnol.}\ }\textbf {\bibinfo {volume} {15}},\ \bibinfo {pages} {529} (\bibinfo {year} {2020})}\BibitemShut {NoStop}%
\bibitem [{\citenamefont {Schenk}\ \emph {et~al.}(2020)\citenamefont {Schenk}, \citenamefont {Pe{\ifmmode\check{s}\else\v{s}\fi}i{\ifmmode\acute{c}\else\'{c}\fi}}, \citenamefont {Slesazeck}, \citenamefont {Schroeder},\ and\ \citenamefont {Mikolajick}}]{Schenk2020Jun}%
  \BibitemOpen
  \bibfield  {author} {\bibinfo {author} {\bibfnamefont {T.}~\bibnamefont {Schenk}}, \bibinfo {author} {\bibfnamefont {M.}~\bibnamefont {Pe{\ifmmode\check{s}\else\v{s}\fi}i{\ifmmode\acute{c}\else\'{c}\fi}}}, \bibinfo {author} {\bibfnamefont {S.}~\bibnamefont {Slesazeck}}, \bibinfo {author} {\bibfnamefont {U.}~\bibnamefont {Schroeder}}, \ and\ \bibinfo {author} {\bibfnamefont {T.}~\bibnamefont {Mikolajick}},\ }\href {\doibase 10.1088/1361-6633/ab8f86} {\bibfield  {journal} {\bibinfo  {journal} {Rep. Prog. Phys.}\ }\textbf {\bibinfo {volume} {83}},\ \bibinfo {pages} {086501} (\bibinfo {year} {2020})}\BibitemShut {NoStop}%
\bibitem [{\citenamefont {Anuniwat}\ \emph {et~al.}(2013)\citenamefont {Anuniwat}, \citenamefont {Ding}, \citenamefont {Poon}, \citenamefont {Wolf},\ and\ \citenamefont {Lu}}]{Anuniwat2013Jan}%
  \BibitemOpen
  \bibfield  {author} {\bibinfo {author} {\bibfnamefont {N.}~\bibnamefont {Anuniwat}}, \bibinfo {author} {\bibfnamefont {M.}~\bibnamefont {Ding}}, \bibinfo {author} {\bibfnamefont {S.~J.}\ \bibnamefont {Poon}}, \bibinfo {author} {\bibfnamefont {S.~A.}\ \bibnamefont {Wolf}}, \ and\ \bibinfo {author} {\bibfnamefont {J.}~\bibnamefont {Lu}},\ }\href {\doibase 10.1063/1.4788807} {\bibfield  {journal} {\bibinfo  {journal} {J. Appl. Phys.}\ }\textbf {\bibinfo {volume} {113}} (\bibinfo {year} {2013}),\ 10.1063/1.4788807}\BibitemShut {NoStop}%
\bibitem [{\citenamefont {Mandal}\ \emph {et~al.}(2018)\citenamefont {Mandal}, \citenamefont {Jung}, \citenamefont {Masuda}, \citenamefont {Takahashi}, \citenamefont {Sakuraba}, \citenamefont {Kasai}, \citenamefont {Miura}, \citenamefont {Ohkubo},\ and\ \citenamefont {Hono}}]{Mandal2018Dec}%
  \BibitemOpen
  \bibfield  {author} {\bibinfo {author} {\bibfnamefont {R.}~\bibnamefont {Mandal}}, \bibinfo {author} {\bibfnamefont {J.~W.}\ \bibnamefont {Jung}}, \bibinfo {author} {\bibfnamefont {K.}~\bibnamefont {Masuda}}, \bibinfo {author} {\bibfnamefont {Y.~K.}\ \bibnamefont {Takahashi}}, \bibinfo {author} {\bibfnamefont {Y.}~\bibnamefont {Sakuraba}}, \bibinfo {author} {\bibfnamefont {S.}~\bibnamefont {Kasai}}, \bibinfo {author} {\bibfnamefont {Y.}~\bibnamefont {Miura}}, \bibinfo {author} {\bibfnamefont {T.}~\bibnamefont {Ohkubo}}, \ and\ \bibinfo {author} {\bibfnamefont {K.}~\bibnamefont {Hono}},\ }\href {\doibase 10.1063/1.5052721} {\bibfield  {journal} {\bibinfo  {journal} {Appl. Phys. Lett.}\ }\textbf {\bibinfo {volume} {113}} (\bibinfo {year} {2018}),\ 10.1063/1.5052721}\BibitemShut {NoStop}%
\bibitem [{\citenamefont {Serizawa}\ \emph {et~al.}(2019)\citenamefont {Serizawa}, \citenamefont {Ohtake}, \citenamefont {Kawai}, \citenamefont {Futamoto}, \citenamefont {Kirino},\ and\ \citenamefont {Inaba}}]{Serizawa2019May}%
  \BibitemOpen
  \bibfield  {author} {\bibinfo {author} {\bibfnamefont {K.}~\bibnamefont {Serizawa}}, \bibinfo {author} {\bibfnamefont {M.}~\bibnamefont {Ohtake}}, \bibinfo {author} {\bibfnamefont {T.}~\bibnamefont {Kawai}}, \bibinfo {author} {\bibfnamefont {M.}~\bibnamefont {Futamoto}}, \bibinfo {author} {\bibfnamefont {F.}~\bibnamefont {Kirino}}, \ and\ \bibinfo {author} {\bibfnamefont {N.}~\bibnamefont {Inaba}},\ }\href {\doibase 10.3379/msjmag.1905R003} {\bibfield  {journal} {\bibinfo  {journal} {J. Magn. Soc. Jpn.}\ }\textbf {\bibinfo {volume} {43}},\ \bibinfo {pages} {50} (\bibinfo {year} {2019})}\BibitemShut {NoStop}%
\bibitem [{\citenamefont {Fu}\ \emph {et~al.}(2016)\citenamefont {Fu}, \citenamefont {Hanaguri}, \citenamefont {Igarashi}, \citenamefont {Kawamura}, \citenamefont {Bahramy},\ and\ \citenamefont {Sasagawa}}]{Fu2016Feb}%
  \BibitemOpen
  \bibfield  {author} {\bibinfo {author} {\bibfnamefont {Y.-S.}\ \bibnamefont {Fu}}, \bibinfo {author} {\bibfnamefont {T.}~\bibnamefont {Hanaguri}}, \bibinfo {author} {\bibfnamefont {K.}~\bibnamefont {Igarashi}}, \bibinfo {author} {\bibfnamefont {M.}~\bibnamefont {Kawamura}}, \bibinfo {author} {\bibfnamefont {M.~S.}\ \bibnamefont {Bahramy}}, \ and\ \bibinfo {author} {\bibfnamefont {T.}~\bibnamefont {Sasagawa}},\ }\href {\doibase 10.1038/ncomms10829} {\bibfield  {journal} {\bibinfo  {journal} {Nat. Commun.}\ }\textbf {\bibinfo {volume} {7}},\ \bibinfo {pages} {1} (\bibinfo {year} {2016})}\BibitemShut {NoStop}%
\bibitem [{\citenamefont {Wang}\ \emph {et~al.}(2023)\citenamefont {Wang}, \citenamefont {Wang}, \citenamefont {Ozerov}, \citenamefont {Zhang}, \citenamefont {Bermejo-Ortiz}, \citenamefont {Bac}, \citenamefont {Trinh}, \citenamefont {Zhukovskyi}, \citenamefont {Orlova}, \citenamefont {Ambaye}, \citenamefont {Keum}, \citenamefont {de~Vaulchier}, \citenamefont {Guldner}, \citenamefont {Smirnov}, \citenamefont {Lauter}, \citenamefont {Liu},\ and\ \citenamefont {Assaf}}]{Wang2023Aug}%
  \BibitemOpen
  \bibfield  {author} {\bibinfo {author} {\bibfnamefont {J.}~\bibnamefont {Wang}}, \bibinfo {author} {\bibfnamefont {T.}~\bibnamefont {Wang}}, \bibinfo {author} {\bibfnamefont {M.}~\bibnamefont {Ozerov}}, \bibinfo {author} {\bibfnamefont {Z.}~\bibnamefont {Zhang}}, \bibinfo {author} {\bibfnamefont {J.}~\bibnamefont {Bermejo-Ortiz}}, \bibinfo {author} {\bibfnamefont {S.-K.}\ \bibnamefont {Bac}}, \bibinfo {author} {\bibfnamefont {H.}~\bibnamefont {Trinh}}, \bibinfo {author} {\bibfnamefont {M.}~\bibnamefont {Zhukovskyi}}, \bibinfo {author} {\bibfnamefont {T.}~\bibnamefont {Orlova}}, \bibinfo {author} {\bibfnamefont {H.}~\bibnamefont {Ambaye}}, \bibinfo {author} {\bibfnamefont {J.}~\bibnamefont {Keum}}, \bibinfo {author} {\bibfnamefont {L.-A.}\ \bibnamefont {de~Vaulchier}}, \bibinfo {author} {\bibfnamefont {Y.}~\bibnamefont {Guldner}}, \bibinfo {author} {\bibfnamefont {D.}~\bibnamefont {Smirnov}}, \bibinfo {author} {\bibfnamefont {V.}~\bibnamefont {Lauter}}, \bibinfo {author} {\bibfnamefont {X.}~\bibnamefont
  {Liu}}, \ and\ \bibinfo {author} {\bibfnamefont {B.~A.}\ \bibnamefont {Assaf}},\ }\href {\doibase 10.1038/s42005-023-01327-5} {\bibfield  {journal} {\bibinfo  {journal} {Commun. Phys.}\ }\textbf {\bibinfo {volume} {6}},\ \bibinfo {pages} {1} (\bibinfo {year} {2023})}\BibitemShut {NoStop}%
\bibitem [{\citenamefont {Kaveev}\ \emph {et~al.}(2021)\citenamefont {Kaveev}, \citenamefont {Suturin}, \citenamefont {Golyashov}, \citenamefont {Kokh}, \citenamefont {Eremeev}, \citenamefont {Estyunin}, \citenamefont {Shikin}, \citenamefont {Okotrub}, \citenamefont {Lavrov}, \citenamefont {Schwier},\ and\ \citenamefont {Tereshchenko}}]{Kaveev2021Dec}%
  \BibitemOpen
  \bibfield  {author} {\bibinfo {author} {\bibfnamefont {A.~K.}\ \bibnamefont {Kaveev}}, \bibinfo {author} {\bibfnamefont {S.~M.}\ \bibnamefont {Suturin}}, \bibinfo {author} {\bibfnamefont {V.~A.}\ \bibnamefont {Golyashov}}, \bibinfo {author} {\bibfnamefont {K.~A.}\ \bibnamefont {Kokh}}, \bibinfo {author} {\bibfnamefont {S.~V.}\ \bibnamefont {Eremeev}}, \bibinfo {author} {\bibfnamefont {D.~A.}\ \bibnamefont {Estyunin}}, \bibinfo {author} {\bibfnamefont {A.~M.}\ \bibnamefont {Shikin}}, \bibinfo {author} {\bibfnamefont {A.~V.}\ \bibnamefont {Okotrub}}, \bibinfo {author} {\bibfnamefont {A.~N.}\ \bibnamefont {Lavrov}}, \bibinfo {author} {\bibfnamefont {E.~F.}\ \bibnamefont {Schwier}}, \ and\ \bibinfo {author} {\bibfnamefont {O.~E.}\ \bibnamefont {Tereshchenko}},\ }\href {\doibase 10.1103/PhysRevMaterials.5.124204} {\bibfield  {journal} {\bibinfo  {journal} {Phys. Rev. Mater.}\ }\textbf {\bibinfo {volume} {5}},\ \bibinfo {pages} {124204} (\bibinfo {year} {2021})}\BibitemShut {NoStop}%
\bibitem [{\citenamefont {Khang}\ \emph {et~al.}(2022)\citenamefont {Khang}, \citenamefont {Shirokura}, \citenamefont {Fan}, \citenamefont {Takahashi}, \citenamefont {Nakatani}, \citenamefont {Kato}, \citenamefont {Miyamoto},\ and\ \citenamefont {Hai}}]{Pham3}%
  \BibitemOpen
  \bibfield  {author} {\bibinfo {author} {\bibfnamefont {N.~H.~D.}\ \bibnamefont {Khang}}, \bibinfo {author} {\bibfnamefont {T.}~\bibnamefont {Shirokura}}, \bibinfo {author} {\bibfnamefont {T.}~\bibnamefont {Fan}}, \bibinfo {author} {\bibfnamefont {M.}~\bibnamefont {Takahashi}}, \bibinfo {author} {\bibfnamefont {N.}~\bibnamefont {Nakatani}}, \bibinfo {author} {\bibfnamefont {D.}~\bibnamefont {Kato}}, \bibinfo {author} {\bibfnamefont {Y.}~\bibnamefont {Miyamoto}}, \ and\ \bibinfo {author} {\bibfnamefont {P.~N.}\ \bibnamefont {Hai}},\ }\href {\doibase 10.1063/5.0084927} {\bibfield  {journal} {\bibinfo  {journal} {Appl. Phys. Lett.}\ }\textbf {\bibinfo {volume} {120}} (\bibinfo {year} {2022}),\ 10.1063/5.0084927}\BibitemShut {NoStop}%
\bibitem [{\citenamefont {Rehm}\ \emph {et~al.}(2024)\citenamefont {Rehm}, \citenamefont {Morshed}, \citenamefont {Misra}, \citenamefont {Shukla}, \citenamefont {Rakheja}, \citenamefont {Pinarbasi}, \citenamefont {Ghosh},\ and\ \citenamefont {Kent}}]{Rehm2024Jan}%
  \BibitemOpen
  \bibfield  {author} {\bibinfo {author} {\bibfnamefont {L.}~\bibnamefont {Rehm}}, \bibinfo {author} {\bibfnamefont {M.~G.}\ \bibnamefont {Morshed}}, \bibinfo {author} {\bibfnamefont {S.}~\bibnamefont {Misra}}, \bibinfo {author} {\bibfnamefont {A.}~\bibnamefont {Shukla}}, \bibinfo {author} {\bibfnamefont {S.}~\bibnamefont {Rakheja}}, \bibinfo {author} {\bibfnamefont {M.}~\bibnamefont {Pinarbasi}}, \bibinfo {author} {\bibfnamefont {A.~W.}\ \bibnamefont {Ghosh}}, \ and\ \bibinfo {author} {\bibfnamefont {A.~D.}\ \bibnamefont {Kent}},\ }\href {\doibase 10.1063/5.0186810} {\bibfield  {journal} {\bibinfo  {journal} {Appl. Phys. Lett.}\ }\textbf {\bibinfo {volume} {124}} (\bibinfo {year} {2024}),\ 10.1063/5.0186810}\BibitemShut {NoStop}%
\bibitem [{\citenamefont {Chang}\ \emph {et~al.}(2012)\citenamefont {Chang}, \citenamefont {Register},\ and\ \citenamefont {Banerjee}}]{Chang2012Dec}%
  \BibitemOpen
  \bibfield  {author} {\bibinfo {author} {\bibfnamefont {J.}~\bibnamefont {Chang}}, \bibinfo {author} {\bibfnamefont {L.~F.}\ \bibnamefont {Register}}, \ and\ \bibinfo {author} {\bibfnamefont {S.~K.}\ \bibnamefont {Banerjee}},\ }\href {\doibase 10.1063/1.4770324} {\bibfield  {journal} {\bibinfo  {journal} {J. Appl. Phys.}\ }\textbf {\bibinfo {volume} {112}} (\bibinfo {year} {2012}),\ 10.1063/1.4770324}\BibitemShut {NoStop}%
\bibitem [{\citenamefont {Wang}\ \emph {et~al.}(2019)\citenamefont {Wang}, \citenamefont {Zhu}, \citenamefont {Yang}, \citenamefont {Lee}, \citenamefont {Mishra}, \citenamefont {Go}, \citenamefont {Oh}, \citenamefont {Kim}, \citenamefont {Cai}, \citenamefont {Liu}, \citenamefont {Pollard}, \citenamefont {Shi}, \citenamefont {Lee}, \citenamefont {Teo}, \citenamefont {Wu}, \citenamefont {Lee},\ and\ \citenamefont {Yang}}]{Wang2019Nov}%
  \BibitemOpen
  \bibfield  {author} {\bibinfo {author} {\bibfnamefont {Y.}~\bibnamefont {Wang}}, \bibinfo {author} {\bibfnamefont {D.}~\bibnamefont {Zhu}}, \bibinfo {author} {\bibfnamefont {Y.}~\bibnamefont {Yang}}, \bibinfo {author} {\bibfnamefont {K.}~\bibnamefont {Lee}}, \bibinfo {author} {\bibfnamefont {R.}~\bibnamefont {Mishra}}, \bibinfo {author} {\bibfnamefont {G.}~\bibnamefont {Go}}, \bibinfo {author} {\bibfnamefont {S.-H.}\ \bibnamefont {Oh}}, \bibinfo {author} {\bibfnamefont {D.-H.}\ \bibnamefont {Kim}}, \bibinfo {author} {\bibfnamefont {K.}~\bibnamefont {Cai}}, \bibinfo {author} {\bibfnamefont {E.}~\bibnamefont {Liu}}, \bibinfo {author} {\bibfnamefont {S.~D.}\ \bibnamefont {Pollard}}, \bibinfo {author} {\bibfnamefont {S.}~\bibnamefont {Shi}}, \bibinfo {author} {\bibfnamefont {J.}~\bibnamefont {Lee}}, \bibinfo {author} {\bibfnamefont {K.~L.}\ \bibnamefont {Teo}}, \bibinfo {author} {\bibfnamefont {Y.}~\bibnamefont {Wu}}, \bibinfo {author} {\bibfnamefont {K.-J.}\ \bibnamefont {Lee}}, \ and\ \bibinfo {author}
  {\bibfnamefont {H.}~\bibnamefont {Yang}},\ }\href {\doibase 10.1126/science.aav8076} {\bibfield  {journal} {\bibinfo  {journal} {Science}\ }\textbf {\bibinfo {volume} {366}},\ \bibinfo {pages} {1125} (\bibinfo {year} {2019})}\BibitemShut {NoStop}%
\bibitem [{\citenamefont {Khanal}\ \emph {et~al.}(2021)\citenamefont {Khanal}, \citenamefont {Zhou}, \citenamefont {Andrade}, \citenamefont {Dang}, \citenamefont {Davydov}, \citenamefont {Habiboglu}, \citenamefont {Saidian}, \citenamefont {Laurie}, \citenamefont {Wang}, \citenamefont {Gopman},\ and\ \citenamefont {Wang}}]{Khanal2021Dec}%
  \BibitemOpen
  \bibfield  {author} {\bibinfo {author} {\bibfnamefont {P.}~\bibnamefont {Khanal}}, \bibinfo {author} {\bibfnamefont {B.}~\bibnamefont {Zhou}}, \bibinfo {author} {\bibfnamefont {M.}~\bibnamefont {Andrade}}, \bibinfo {author} {\bibfnamefont {Y.}~\bibnamefont {Dang}}, \bibinfo {author} {\bibfnamefont {A.}~\bibnamefont {Davydov}}, \bibinfo {author} {\bibfnamefont {A.}~\bibnamefont {Habiboglu}}, \bibinfo {author} {\bibfnamefont {J.}~\bibnamefont {Saidian}}, \bibinfo {author} {\bibfnamefont {A.}~\bibnamefont {Laurie}}, \bibinfo {author} {\bibfnamefont {J.-P.}\ \bibnamefont {Wang}}, \bibinfo {author} {\bibfnamefont {D.~B.}\ \bibnamefont {Gopman}}, \ and\ \bibinfo {author} {\bibfnamefont {W.}~\bibnamefont {Wang}},\ }\href {\doibase 10.1063/5.0066782} {\bibfield  {journal} {\bibinfo  {journal} {Appl. Phys. Lett.}\ }\textbf {\bibinfo {volume} {119}} (\bibinfo {year} {2021}),\ 10.1063/5.0066782}\BibitemShut {NoStop}%
\bibitem [{\citenamefont {Li}\ \emph {et~al.}(2019)\citenamefont {Li}, \citenamefont {Kally}, \citenamefont {Zhang}, \citenamefont {Pillsbury}, \citenamefont {Ding}, \citenamefont {Csaba}, \citenamefont {Ding}, \citenamefont {Jiang}, \citenamefont {Liu}, \citenamefont {Sinclair}, \citenamefont {Bi}, \citenamefont {DeMann}, \citenamefont {Rimal}, \citenamefont {Zhang}, \citenamefont {Field}, \citenamefont {Tang}, \citenamefont {Wang}, \citenamefont {Heinonen}, \citenamefont {Novosad}, \citenamefont {Hoffmann}, \citenamefont {Samarth},\ and\ \citenamefont {Wu}}]{Li2019Aug}%
  \BibitemOpen
  \bibfield  {author} {\bibinfo {author} {\bibfnamefont {P.}~\bibnamefont {Li}}, \bibinfo {author} {\bibfnamefont {J.}~\bibnamefont {Kally}}, \bibinfo {author} {\bibfnamefont {S.~S.-L.}\ \bibnamefont {Zhang}}, \bibinfo {author} {\bibfnamefont {T.}~\bibnamefont {Pillsbury}}, \bibinfo {author} {\bibfnamefont {J.}~\bibnamefont {Ding}}, \bibinfo {author} {\bibfnamefont {G.}~\bibnamefont {Csaba}}, \bibinfo {author} {\bibfnamefont {J.}~\bibnamefont {Ding}}, \bibinfo {author} {\bibfnamefont {J.~S.}\ \bibnamefont {Jiang}}, \bibinfo {author} {\bibfnamefont {Y.}~\bibnamefont {Liu}}, \bibinfo {author} {\bibfnamefont {R.}~\bibnamefont {Sinclair}}, \bibinfo {author} {\bibfnamefont {C.}~\bibnamefont {Bi}}, \bibinfo {author} {\bibfnamefont {A.}~\bibnamefont {DeMann}}, \bibinfo {author} {\bibfnamefont {G.}~\bibnamefont {Rimal}}, \bibinfo {author} {\bibfnamefont {W.}~\bibnamefont {Zhang}}, \bibinfo {author} {\bibfnamefont {S.~B.}\ \bibnamefont {Field}}, \bibinfo {author} {\bibfnamefont {J.}~\bibnamefont {Tang}}, \bibinfo
  {author} {\bibfnamefont {W.}~\bibnamefont {Wang}}, \bibinfo {author} {\bibfnamefont {O.~G.}\ \bibnamefont {Heinonen}}, \bibinfo {author} {\bibfnamefont {V.}~\bibnamefont {Novosad}}, \bibinfo {author} {\bibfnamefont {A.}~\bibnamefont {Hoffmann}}, \bibinfo {author} {\bibfnamefont {N.}~\bibnamefont {Samarth}}, \ and\ \bibinfo {author} {\bibfnamefont {M.}~\bibnamefont {Wu}},\ }\href {\doibase 10.1126/sciadv.aaw3415} {\bibfield  {journal} {\bibinfo  {journal} {Sci. Adv.}\ }\textbf {\bibinfo {volume} {5}} (\bibinfo {year} {2019}),\ 10.1126/sciadv.aaw3415}\BibitemShut {NoStop}%
\bibitem [{\citenamefont {Dc}\ \emph {et~al.}(2019)\citenamefont {Dc}, \citenamefont {Chen}, \citenamefont {Peterson}, \citenamefont {Sahu}, \citenamefont {Ma}, \citenamefont {Mousavi}, \citenamefont {Harjani},\ and\ \citenamefont {Wang}}]{Dc2019Aug}%
  \BibitemOpen
  \bibfield  {author} {\bibinfo {author} {\bibfnamefont {M.}~\bibnamefont {Dc}}, \bibinfo {author} {\bibfnamefont {J.-Y.}\ \bibnamefont {Chen}}, \bibinfo {author} {\bibfnamefont {T.}~\bibnamefont {Peterson}}, \bibinfo {author} {\bibfnamefont {P.}~\bibnamefont {Sahu}}, \bibinfo {author} {\bibfnamefont {B.}~\bibnamefont {Ma}}, \bibinfo {author} {\bibfnamefont {N.}~\bibnamefont {Mousavi}}, \bibinfo {author} {\bibfnamefont {R.}~\bibnamefont {Harjani}}, \ and\ \bibinfo {author} {\bibfnamefont {J.-P.}\ \bibnamefont {Wang}},\ }\href {\doibase 10.1021/acs.nanolett.8b05011} {\bibfield  {journal} {\bibinfo  {journal} {Nano Lett.}\ }\textbf {\bibinfo {volume} {19}},\ \bibinfo {pages} {4836} (\bibinfo {year} {2019})}\BibitemShut {NoStop}%
\bibitem [{\citenamefont {Lu}\ \emph {et~al.}(2022)\citenamefont {Lu}, \citenamefont {Li}, \citenamefont {Guo}, \citenamefont {Dong}, \citenamefont {Peng}, \citenamefont {Zha}, \citenamefont {Min}, \citenamefont {Zhou},\ and\ \citenamefont {Liu}}]{Lu2022Mar}%
  \BibitemOpen
  \bibfield  {author} {\bibinfo {author} {\bibfnamefont {Q.}~\bibnamefont {Lu}}, \bibinfo {author} {\bibfnamefont {P.}~\bibnamefont {Li}}, \bibinfo {author} {\bibfnamefont {Z.}~\bibnamefont {Guo}}, \bibinfo {author} {\bibfnamefont {G.}~\bibnamefont {Dong}}, \bibinfo {author} {\bibfnamefont {B.}~\bibnamefont {Peng}}, \bibinfo {author} {\bibfnamefont {X.}~\bibnamefont {Zha}}, \bibinfo {author} {\bibfnamefont {T.}~\bibnamefont {Min}}, \bibinfo {author} {\bibfnamefont {Z.}~\bibnamefont {Zhou}}, \ and\ \bibinfo {author} {\bibfnamefont {M.}~\bibnamefont {Liu}},\ }\href {\doibase 10.1038/s41467-022-29281-w} {\bibfield  {journal} {\bibinfo  {journal} {Nat. Commun.}\ }\textbf {\bibinfo {volume} {13}},\ \bibinfo {pages} {1} (\bibinfo {year} {2022})}\BibitemShut {NoStop}%
\bibitem [{\citenamefont {Barton}\ \emph {et~al.}(2019)\citenamefont {Barton}, \citenamefont {Walsh}, \citenamefont {Smyth}, \citenamefont {Qin}, \citenamefont {Addou}, \citenamefont {Cormier}, \citenamefont {Hurley}, \citenamefont {Wallace},\ and\ \citenamefont {Hinkle}}]{Barton2019Sep}%
  \BibitemOpen
  \bibfield  {author} {\bibinfo {author} {\bibfnamefont {A.~T.}\ \bibnamefont {Barton}}, \bibinfo {author} {\bibfnamefont {L.~A.}\ \bibnamefont {Walsh}}, \bibinfo {author} {\bibfnamefont {C.~M.}\ \bibnamefont {Smyth}}, \bibinfo {author} {\bibfnamefont {X.}~\bibnamefont {Qin}}, \bibinfo {author} {\bibfnamefont {R.}~\bibnamefont {Addou}}, \bibinfo {author} {\bibfnamefont {C.}~\bibnamefont {Cormier}}, \bibinfo {author} {\bibfnamefont {P.~K.}\ \bibnamefont {Hurley}}, \bibinfo {author} {\bibfnamefont {R.~M.}\ \bibnamefont {Wallace}}, \ and\ \bibinfo {author} {\bibfnamefont {C.~L.}\ \bibnamefont {Hinkle}},\ }\href {\doibase 10.1021/acsami.9b10625} {\bibfield  {journal} {\bibinfo  {journal} {ACS Appl. Mater. Interfaces}\ }\textbf {\bibinfo {volume} {11}},\ \bibinfo {pages} {32144} (\bibinfo {year} {2019})}\BibitemShut {NoStop}%
\bibitem [{\citenamefont {Karki}\ \emph {et~al.}(2023)\citenamefont {Karki}, \citenamefont {Kwon}, \citenamefont {Davies}, \citenamefont {Fabiha}, \citenamefont {Rogers}, \citenamefont {Leonard}, \citenamefont {Bandyopadhyay},\ and\ \citenamefont {Incorvia}}]{Karki2023Nov}%
  \BibitemOpen
  \bibfield  {author} {\bibinfo {author} {\bibfnamefont {S.}~\bibnamefont {Karki}}, \bibinfo {author} {\bibfnamefont {J.}~\bibnamefont {Kwon}}, \bibinfo {author} {\bibfnamefont {J.}~\bibnamefont {Davies}}, \bibinfo {author} {\bibfnamefont {R.}~\bibnamefont {Fabiha}}, \bibinfo {author} {\bibfnamefont {V.}~\bibnamefont {Rogers}}, \bibinfo {author} {\bibfnamefont {T.}~\bibnamefont {Leonard}}, \bibinfo {author} {\bibfnamefont {S.}~\bibnamefont {Bandyopadhyay}}, \ and\ \bibinfo {author} {\bibfnamefont {J.~A.~C.}\ \bibnamefont {Incorvia}},\ }\href {\doibase 10.48550/arXiv.2311.08984} {\bibfield  {journal} {\bibinfo  {journal} {arXiv}\ } (\bibinfo {year} {2023}),\ 10.48550/arXiv.2311.08984},\ \Eprint {http://arxiv.org/abs/2311.08984} {2311.08984} \BibitemShut {NoStop}%
\bibitem [{\citenamefont {Zhu}\ \emph {et~al.}(2013)\citenamefont {Zhu}, \citenamefont {Richter}, \citenamefont {Zhao}, \citenamefont {Bonevich}, \citenamefont {Kimes}, \citenamefont {Jang}, \citenamefont {Yuan}, \citenamefont {Li}, \citenamefont {Arab}, \citenamefont {Kirillov}, \citenamefont {Maslar}, \citenamefont {Ioannou},\ and\ \citenamefont {Li}}]{Zhu2013Apr}%
  \BibitemOpen
  \bibfield  {author} {\bibinfo {author} {\bibfnamefont {H.}~\bibnamefont {Zhu}}, \bibinfo {author} {\bibfnamefont {C.~A.}\ \bibnamefont {Richter}}, \bibinfo {author} {\bibfnamefont {E.}~\bibnamefont {Zhao}}, \bibinfo {author} {\bibfnamefont {J.~E.}\ \bibnamefont {Bonevich}}, \bibinfo {author} {\bibfnamefont {W.~A.}\ \bibnamefont {Kimes}}, \bibinfo {author} {\bibfnamefont {H.-J.}\ \bibnamefont {Jang}}, \bibinfo {author} {\bibfnamefont {H.}~\bibnamefont {Yuan}}, \bibinfo {author} {\bibfnamefont {H.}~\bibnamefont {Li}}, \bibinfo {author} {\bibfnamefont {A.}~\bibnamefont {Arab}}, \bibinfo {author} {\bibfnamefont {O.}~\bibnamefont {Kirillov}}, \bibinfo {author} {\bibfnamefont {J.~E.}\ \bibnamefont {Maslar}}, \bibinfo {author} {\bibfnamefont {D.~E.}\ \bibnamefont {Ioannou}}, \ and\ \bibinfo {author} {\bibfnamefont {Q.}~\bibnamefont {Li}},\ }\href {\doibase 10.1038/srep01757} {\bibfield  {journal} {\bibinfo  {journal} {Sci. Rep.}\ }\textbf {\bibinfo {volume} {3}},\ \bibinfo {pages} {1} (\bibinfo {year}
  {2013})}\BibitemShut {NoStop}%
\bibitem [{\citenamefont {Jang}\ \emph {et~al.}(2020)\citenamefont {Jang}, \citenamefont {Hwang}, \citenamefont {Min}, \citenamefont {Kim}, \citenamefont {Ahn}, \citenamefont {Choi}, \citenamefont {Hahn}, \citenamefont {Choi}, \citenamefont {Park}, \citenamefont {Jung},\ and\ \citenamefont {Yoon}}]{Jang2020Nov}%
  \BibitemOpen
  \bibfield  {author} {\bibinfo {author} {\bibfnamefont {J.}~\bibnamefont {Jang}}, \bibinfo {author} {\bibfnamefont {G.-T.}\ \bibnamefont {Hwang}}, \bibinfo {author} {\bibfnamefont {Y.}~\bibnamefont {Min}}, \bibinfo {author} {\bibfnamefont {J.-W.}\ \bibnamefont {Kim}}, \bibinfo {author} {\bibfnamefont {C.-W.}\ \bibnamefont {Ahn}}, \bibinfo {author} {\bibfnamefont {J.-J.}\ \bibnamefont {Choi}}, \bibinfo {author} {\bibfnamefont {B.-D.}\ \bibnamefont {Hahn}}, \bibinfo {author} {\bibfnamefont {J.-H.}\ \bibnamefont {Choi}}, \bibinfo {author} {\bibfnamefont {D.-S.}\ \bibnamefont {Park}}, \bibinfo {author} {\bibfnamefont {Y.}~\bibnamefont {Jung}}, \ and\ \bibinfo {author} {\bibfnamefont {W.-H.}\ \bibnamefont {Yoon}},\ }\href {\doibase 10.1007/s43207-020-00062-9} {\bibfield  {journal} {\bibinfo  {journal} {J. Korean Ceram. Soc.}\ }\textbf {\bibinfo {volume} {57}},\ \bibinfo {pages} {645} (\bibinfo {year} {2020})}\BibitemShut {NoStop}%
\bibitem [{\citenamefont {Hey}(2023)}]{Hey2023May}%
  \BibitemOpen
  \bibfield  {author} {\bibinfo {author} {\bibfnamefont {T.}~\bibnamefont {Hey}},\ }\href {https://www.taylorfrancis.com/books/edit/10.1201/9781003358817/feynman-lectures-computation-tony-hey} {\emph {\bibinfo {title} {{Feynman Lectures on Computation:Anniversary Edition}}}}\ (\bibinfo  {publisher} {Taylor {\&} Francis},\ \bibinfo {address} {Andover, England, UK},\ \bibinfo {year} {2023})\BibitemShut {NoStop}%
\bibitem [{\citenamefont {Liu}\ \emph {et~al.}(2023)\citenamefont {Liu}, \citenamefont {Shi}, \citenamefont {Kumar}, \citenamefont {Kim}, \citenamefont {Shi}, \citenamefont {Yang}, \citenamefont {Zhang}, \citenamefont {Zhang}, \citenamefont {Wang}, \citenamefont {Yang}, \citenamefont {Pu}, \citenamefont {Yu}, \citenamefont {Cai},\ and\ \citenamefont {Yang}}]{WSM1}%
  \BibitemOpen
  \bibfield  {author} {\bibinfo {author} {\bibfnamefont {Y.}~\bibnamefont {Liu}}, \bibinfo {author} {\bibfnamefont {G.}~\bibnamefont {Shi}}, \bibinfo {author} {\bibfnamefont {D.}~\bibnamefont {Kumar}}, \bibinfo {author} {\bibfnamefont {T.}~\bibnamefont {Kim}}, \bibinfo {author} {\bibfnamefont {S.}~\bibnamefont {Shi}}, \bibinfo {author} {\bibfnamefont {D.}~\bibnamefont {Yang}}, \bibinfo {author} {\bibfnamefont {J.}~\bibnamefont {Zhang}}, \bibinfo {author} {\bibfnamefont {C.}~\bibnamefont {Zhang}}, \bibinfo {author} {\bibfnamefont {F.}~\bibnamefont {Wang}}, \bibinfo {author} {\bibfnamefont {S.}~\bibnamefont {Yang}}, \bibinfo {author} {\bibfnamefont {Y.}~\bibnamefont {Pu}}, \bibinfo {author} {\bibfnamefont {P.}~\bibnamefont {Yu}}, \bibinfo {author} {\bibfnamefont {K.}~\bibnamefont {Cai}}, \ and\ \bibinfo {author} {\bibfnamefont {H.}~\bibnamefont {Yang}},\ }\href {\doibase 10.1038/s41928-023-01039-2} {\bibfield  {journal} {\bibinfo  {journal} {Nat. Electron.}\ }\textbf {\bibinfo {volume} {6}},\ \bibinfo {pages}
  {732} (\bibinfo {year} {2023})}\BibitemShut {NoStop}%
\bibitem [{\citenamefont {Zhang}\ \emph {et~al.}(2023)\citenamefont {Zhang}, \citenamefont {Xu}, \citenamefont {Jia}, \citenamefont {Lan}, \citenamefont {Huang}, \citenamefont {He}, \citenamefont {He}, \citenamefont {Shao}, \citenamefont {Wang}, \citenamefont {Zhao}, \citenamefont {Ma}, \citenamefont {Dong}, \citenamefont {Guo}, \citenamefont {Cheng}, \citenamefont {Feng}, \citenamefont {Wan}, \citenamefont {Wei}, \citenamefont {Shi}, \citenamefont {Zhang}, \citenamefont {Han},\ and\ \citenamefont {Yu}}]{WSM2}%
  \BibitemOpen
  \bibfield  {author} {\bibinfo {author} {\bibfnamefont {Y.}~\bibnamefont {Zhang}}, \bibinfo {author} {\bibfnamefont {H.}~\bibnamefont {Xu}}, \bibinfo {author} {\bibfnamefont {K.}~\bibnamefont {Jia}}, \bibinfo {author} {\bibfnamefont {G.}~\bibnamefont {Lan}}, \bibinfo {author} {\bibfnamefont {Z.}~\bibnamefont {Huang}}, \bibinfo {author} {\bibfnamefont {B.}~\bibnamefont {He}}, \bibinfo {author} {\bibfnamefont {C.}~\bibnamefont {He}}, \bibinfo {author} {\bibfnamefont {Q.}~\bibnamefont {Shao}}, \bibinfo {author} {\bibfnamefont {Y.}~\bibnamefont {Wang}}, \bibinfo {author} {\bibfnamefont {M.}~\bibnamefont {Zhao}}, \bibinfo {author} {\bibfnamefont {T.}~\bibnamefont {Ma}}, \bibinfo {author} {\bibfnamefont {J.}~\bibnamefont {Dong}}, \bibinfo {author} {\bibfnamefont {C.}~\bibnamefont {Guo}}, \bibinfo {author} {\bibfnamefont {C.}~\bibnamefont {Cheng}}, \bibinfo {author} {\bibfnamefont {J.}~\bibnamefont {Feng}}, \bibinfo {author} {\bibfnamefont {C.}~\bibnamefont {Wan}}, \bibinfo {author} {\bibfnamefont {H.}~\bibnamefont
  {Wei}}, \bibinfo {author} {\bibfnamefont {Y.}~\bibnamefont {Shi}}, \bibinfo {author} {\bibfnamefont {G.}~\bibnamefont {Zhang}}, \bibinfo {author} {\bibfnamefont {X.}~\bibnamefont {Han}}, \ and\ \bibinfo {author} {\bibfnamefont {G.}~\bibnamefont {Yu}},\ }\href {\doibase 10.1126/sciadv.adg9819} {\bibfield  {journal} {\bibinfo  {journal} {Sci. Adv.}\ }\textbf {\bibinfo {volume} {9}} (\bibinfo {year} {2023}),\ 10.1126/sciadv.adg9819}\BibitemShut {NoStop}%
\bibitem [{\citenamefont {Bainsla}\ \emph {et~al.}(2024)\citenamefont {Bainsla}, \citenamefont {Zhao}, \citenamefont {Behera}, \citenamefont {Hoque}, \citenamefont {Sj{\ifmmode\ddot{o}\else\"{o}\fi}str{\ifmmode\ddot{o}\else\"{o}\fi}m}, \citenamefont {Martinelli}, \citenamefont {Abdel-Hafiez}, \citenamefont {{\AA}kerman},\ and\ \citenamefont {Dash}}]{WSM3}%
  \BibitemOpen
  \bibfield  {author} {\bibinfo {author} {\bibfnamefont {L.}~\bibnamefont {Bainsla}}, \bibinfo {author} {\bibfnamefont {B.}~\bibnamefont {Zhao}}, \bibinfo {author} {\bibfnamefont {N.}~\bibnamefont {Behera}}, \bibinfo {author} {\bibfnamefont {A.~{\relax Md}.}\ \bibnamefont {Hoque}}, \bibinfo {author} {\bibfnamefont {L.}~\bibnamefont {Sj{\ifmmode\ddot{o}\else\"{o}\fi}str{\ifmmode\ddot{o}\else\"{o}\fi}m}}, \bibinfo {author} {\bibfnamefont {A.}~\bibnamefont {Martinelli}}, \bibinfo {author} {\bibfnamefont {M.}~\bibnamefont {Abdel-Hafiez}}, \bibinfo {author} {\bibfnamefont {J.}~\bibnamefont {{\AA}kerman}}, \ and\ \bibinfo {author} {\bibfnamefont {S.~P.}\ \bibnamefont {Dash}},\ }\href {\doibase 10.1038/s41467-024-48872-3} {\bibfield  {journal} {\bibinfo  {journal} {Nat. Commun.}\ }\textbf {\bibinfo {volume} {15}},\ \bibinfo {pages} {1} (\bibinfo {year} {2024})}\BibitemShut {NoStop}%
\end{thebibliography}%
\bibliographystyle{apsrev4-1}
\end{document}